\documentclass[12pt,a4paper,twoside]{my_report}

\usepackage[tbtags]{amsmath}
\usepackage{amssymb}
\usepackage{cite}
\usepackage[usenames]{color}
\usepackage{float}
\usepackage{graphicx}
\usepackage{graphics}
\usepackage{minitoc}
\usepackage[pdf]{optional}
\usepackage{picins}
\usepackage{tabularx}
\usepackage{times}
\usepackage{theorem}
\usepackage[T1]{url}
\usepackage{fancyheadings}
\usepackage[Sonny]{fncychap}

\opt{print}{%
\usepackage[ps2pdf,colorlinks=false,linkcolor=black, citecolor=black,%
  pagecolor=black,urlcolor=black,anchorcolor=black,pdfborder={0 0 0},%
  linktocpage=true,plainpages=false,pdfpagelabels=true,%
  bookmarksopen=true,pagebackref=false,breaklinks=true,%
  bookmarksopenlevel=1,a4paper=true,draft=false,%
  pdftitle={Quantum Chaos: Spectral Analysis of Floquet Operators},%
  pdfauthor={James Matthew McCaw},%
  pdfsubject={PhD Thesis/School of Physics/Uni Melbourne/Australia},%
  pdfkeywords={spectral analysis, floquet operator, time evolution,%
    periodic kick, quantum dynamics, subspace, spectrum, pure point,%
    singularly continuous, absolutely continuous, chaos, quantum chaos,%
    unitary, number theory, Weyl sum, rank N, rank 1, projection operator,%
    unitary operator, non-perturbative, spectral decomposition,%
    functional analysis, cyclicity, kicked rotor,%
    kicked harmonic oscillator, chaotic dynamics}
  pdfcreator = {LaTeX with hyperref package},
  pdfproducer = {dvips, ps2pdf}]{hyperref}
}
\opt{pdf}{%
\usepackage[ps2pdf,colorlinks=true,linkcolor=darkblue, citecolor=darkgreen,%
  pagecolor=darkblue, urlcolor=blue,anchorcolor=black,pdfborder={0 0 0},%
  linktocpage=true,plainpages=false,pdfpagelabels=true,%
  bookmarksopen=true,pagebackref=true,breaklinks=true,%
  bookmarksopenlevel=1,a4paper=true,draft=false,%
  pdftitle={Quantum Chaos: Spectral Analysis of Floquet Operators},%
  pdfauthor={James Matthew McCaw},%
  pdfsubject={PhD Thesis/School of Physics/Uni Melbourne/Australia},%
  pdfkeywords={spectral analysis, floquet operator, time evolution,%
    periodic kick, quantum dynamics, subspace, spectrum, pure point,%
    singularly continuous, absolutely continuous, chaos, quantum chaos,%
    unitary, number theory, Weyl sum, rank N, rank 1, projection operator,%
    unitary operator, non-perturbative, spectral decomposition,%
    functional analysis, cyclicity, kicked rotor,%
    kicked harmonic oscillator, chaotic dynamics}
  pdfcreator = {LaTeX with hyperref package},
  pdfproducer = {dvips, ps2pdf}]{hyperref}
}

\definecolor{darkblue}{rgb}{0,0,0.5}
\definecolor{darkgreen}{rgb}{0,0.5,0}
\definecolor{darkred}{rgb}{0.5,0,0}

\renewcommand{\chaptermark}[1]{\markboth{#1}{}}

\theorembodyfont{\rmfamily \itshape} \newtheorem{conj}{Conjecture}[chapter]
\theorembodyfont{\rmfamily \itshape} 
\theorembodyfont{\rmfamily \itshape} \newtheorem{defn}[conj]{Definition}
\theorembodyfont{\rmfamily \itshape} \newtheorem{lemma}[conj]{Lemma}
\theorembodyfont{\rmfamily \itshape} \newtheorem{prop}[conj]{Proposition}
\theorembodyfont{\rmfamily \itshape} \newtheorem{thrm}[conj]{Theorem}
\theoremheaderfont{\bf\scshape}
\renewcommand{\theenumi}{\alph{enumi}}

\newcommand{\Arctan}{\operatorname{Arctan}}
\newcommand{\Arg}{\operatorname{Arg}}
\newcommand{\cotg}{\operatorname{cotg}}

\newcommand{\tr}{\operatorname{Tr}}
\newcommand{\refeq}[1]{(\ref{#1})}
\newcommand{\reffig}[1]{Figure~\ref{#1}}
\newcommand{\reftab}[1]{Table~\ref{#1}}
\newcommand{\refchap}[1]{Chapter~\ref{#1}}
\newcommand{\refsec}[1]{Section~\ref{#1}}
\newcommand{\etal}{\emph{et.\ al.}}
\newcommand{\pp}{\text{pp}}
\newcommand{\ac}{\text{ac}}
\newcommand{\cs}{\text{sc}}
\newcommand{\cont}{\text{cont}}
\newcommand{\point}{\text{p}}
\newcommand{\sing}{\text{s}}
\newcommand{\lpar}{\parallel \negthickspace}
\newcommand{\rpar}{\negthickspace \parallel}
\newcommand{\longpage}[1][1]{\enlargethispage{#1\baselineskip}}
\newcommand{\shortpage}[1][1]{\enlargethispage{-#1\baselineskip}}
\newcommand{\proofend}{\hfill$\Box$\newline}

\newcommand{\clearemptydoublepage}{\newpage{\pagestyle{empty}\cleardoublepage}}
\newcommand{\emptyline}[1][1]{\vspace*{#1\baselineskip}}

\renewcommand{\abstractname}{A\lowercase{bstract}}
\renewcommand{\appendixname}{A\lowercase{ppendix}}
\renewcommand{\bibname}{R\lowercase{eferences}}
\renewcommand{\chaptername}{C\lowercase{hapter}}
\renewcommand{\contentsname}{C\lowercase{ontents}}
\renewcommand{\listfigurename}{L\lowercase{ist of} F\lowercase{igures}}
\renewcommand{\listtablename}{L\lowercase{ist of} T\lowercase{ables}}
\renewcommand{\listtheoremname}{L\lowercase{ist of} D\lowercase{efinitions,} T\lowercase{heorems and} C\lowercase{onjectures}}

\begin{document}

\raggedbottom


\pagestyle{empty}

\emptyline[5]
\noindent
\Large
{\sf\scshape Quantum Chaos: Spectral \\ Analysis of Floquet Operators}

\clearemptydoublepage


\pagenumbering{roman}

\begin{titlepage}

\begin{figure}[ht]
\begin{center}
	\scalebox{1.0}{\includegraphics{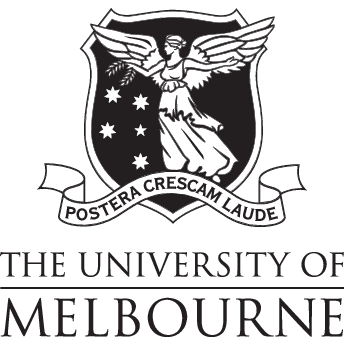}}
\end{center}
\end{figure}

\noindent
\huge
{\sf\scshape Quantum Chaos: Spectral \\ Analysis of Floquet Operators}

\emptyline[2]
\noindent
\large
James Matthew McCaw

\emptyline[2]
\noindent
\normalsize
Submitted in total fulfilment of the requirements\\
of the degree of Doctor of Philosophy.

\emptyline[15]
\noindent
\large
\begin{tabular}{l c r}
{\sf\scshape School of Physics} & {\sf\scshape $\bullet$} &
{\sf\scshape The University of Melbourne} \\
{\sf\scshape December 2004} & & {\sf\scshape Australia}
\end{tabular}

\end{titlepage}


\noindent
\normalsize
Copyright \copyright\ James Matthew McCaw 2004--2005.

\emptyline
\noindent
This work is licensed under the Creative Commons \emph{Attribution-NoDerivs
2.0 \\ 
License (Australia)}
\href{http://creativecommons.org/licenses/by-nd/2.0/au/}
{http://creativecommons.org/licenses/by-nd/2.0/au/}

\emptyline
\noindent
You are free to copy, distribute and display this work, and to make
commercial use of this work, provided that you give credit to the author
James McCaw. You may not alter, transform, or build upon this work. For any
reuse or distribution, you must make clear to others the license terms of
this work. Any of these conditions can be waived if you get permission from
the copyright holder, James McCaw. Your fair use and other rights are in no
way affected by the above.

\emptyline[2]
\noindent
\opt{print}{%
Hardback print edition.
}
\opt{pdf}{%
Online edition.
}

\emptyline[2]
\noindent
\emph{National Library of Australia Cataloguing-in-Publication entry:} \\
McCaw, James Matthew. \\
Quantum Chaos: Spectral Analysis of Floquet Operators.

\emptyline
\noindent
Bibliography. \\
\opt{print}{%
ISBN 0 7340 3068 1. \\
ISBN 0 7340 3071 1. (online edition)
}
\opt{pdf}{%
ISBN 0 7340 3071 1. \\
ISBN 0 7340 3068 1. (hardback print edition)
}

\emptyline
\noindent
1.\ Quantum chaos. 2.\ Differential operators. 3.\ Floquet theory. \\
I.\ University of Melbourne. School of Physics. II.\ Title.

\emptyline
\noindent
530.12

\emptyline[2]
\noindent
The online version of this thesis is archived at the University of
Melbourne ePrints Repository and is part of the Australian Digital Theses
project. It has the Open Archives Initiative identifier
oai:unimelb.edu.au:848, and is accessible at \\
\href{http://errol.oclc.org/oai:unimelb.edu.au:848.resource}
{http://errol.oclc.org/oai:unimelb.edu.au:848.resource}

\clearemptydoublepage


\pagestyle{empty}

\emptyline[5]
\begin{flushright}
\noindent
\emph{To Victoria.}
\end{flushright}

\clearemptydoublepage

\setcounter{page}{5}


\renewcommand{\baselinestretch}{1.5}
\normalsize
\pagestyle{fancyplain}

\chapter*{Abstract}
\chaptermark{Abstract}

The \emph{Floquet operator}, defined as the time-evolution operator over
one period, plays a central role in the work presented in this thesis on
periodically perturbed quantum systems. Knowledge of the spectral nature of
the Floquet operator gives us information on the dynamics of such systems.
The work presented here on the spectrum of the Floquet operator gives
further insight into the nature of chaos in quantum mechanics. After
discussing the links between the spectrum, dynamics and chaos and pointing
out an ambiguity in the physics literature, I present a
number of new mathematical results on the existence of different types of
spectra of the Floquet operator. I characterise the conditions for which
the spectrum remains \emph{pure point} and then, on relaxing these
conditions, show the emergence of a continuous spectral component. The
nature of the continuous spectrum is further analysed, and shown to be
\emph{singularly continuous}. Thus, the dynamics of these systems are a
candidate for classification as chaotic.  A conjecture on the emergence of
a continuous spectral component is linked to a long standing
number-theoretic conjecture on the estimation of finite exponential sums.

\clearemptydoublepage
\chapter*{Declaration}
\chaptermark{Declaration}

This is to certify that

(i) the thesis comprises only my original work towards the PhD, except
where indicated in the Preface,

(ii) due acknowledgement has been made in the text to all other material used,

(iii) the thesis is less than 100,000 words in length, exclusive of tables,
maps, bibliographies and appendices.

\emptyline[2]

{\raggedleft
\hspace{\stretch{0.6}}James McCaw \hrulefill

\emptyline
\hspace{\stretch{0.84}}Date \hrulefill}

\clearemptydoublepage
\chapter*{Preface}
\chaptermark{Preface}

The original work in this thesis has been prepared for publication as
follows:

\begin{itemize}
 
\item James~McCaw and B.~H.~J.~McKellar. Pure point spectrum for the time
evolution of a periodically rank-N kicked Hamiltonian.
\emph{J. Math. Phys.}, 46:032108, 2005. Also available at
\href{http://arxiv.org/abs/math-ph/0404006}
{http://arxiv.org/abs/math-ph/0404006}. This covers the work presented in
\refchap{chap:rankN}.

\item James McCaw and B.~H.~J.~McKellar. On the continuous spectral component
of the Floquet operator for a periodically kicked quantum system. In
preparation. This covers the work presented in \refchap{chap:comb}.

\item J.~M.~McCaw and B.~H.~J.~McKellar. An analysis of the spectrum for
the time evolution of a periodically rank-N kicked Hamiltonian.
To appear in the \emph{Proceedings of the AIP National Congress 2005}. 
This covers the work presented in \refchap{chap:rankN} and
\refchap{chap:comb}.
\end{itemize}

Except where explicitly mentioned in the text, this thesis comprises my
own work.

\clearemptydoublepage
\chapter*{Acknowledgements}
\chaptermark{Acknowledgements}

Firstly, I must thank my supervisor, Professor Bruce McKellar. Without your 
experience and guidance, and never ending commitment to provide ideas and
feedback, I simply could not have produced this work. Our initial ideas
have lead us into an unexpected field---the journey may not have been
expected, but it is one which has been very rewarding!

To Bruce's colleagues and other students, especially Sasha Ignatiev, Kristian
McDonald and Ivona Okuniewicz, thank-you for the many stimulating discussions,
distractions and advice. Thank-you also to Simon Devitt for providing so
much entertainment (and some physics too!) in our office.

For the uncountably many discussions exploring the concepts, mathematics
and peculiarities of chaos in quantum mechanics, thank-you Peter. Your
insightful questions helped me clarify ideas and provide improved
explanations of much of the work I have done. Without your help, this
thesis would not be what it is.

Thank-you Cath and Gaby---our daily coffee trips for the past five years
have been essential in keeping me sane. We did a lot of physics over those
coffees too. Here's to a life long friendship to remember our time in the
School of Physics together.

To mum and dad---thank-you for always supporting me throughout my studies.
Your belief in my abilities and potential has always been a fantastic
motivator. Thank-you Chris for always reminding me of the alternatives to
``being at uni'' and just for being a great brother. I must also thank
Stephen and Stephanie West for showing such a strong interest in my studies
and life---you have been a significant influence and I can't thank-you
enough for that. I also thank Malcolm and Judy for their support and
friendship over the past five years.

Finally, and most importantly, thank-you Vic for your never ending support
and love.  Through honours and my PhD you have always been there and I
can't imagine having got through it without you. We have learnt so much
from each other.  With the completion of my PhD, we are starting a new
chapter in our life together. Thank-you!

\emptyline
\noindent
James McCaw \\
18$^{\text{th}}$ October 2004

\clearemptydoublepage


\setcounter{tocdepth}{1}
\setcounter{minitocdepth}{2}
\dominitoc
\tableofcontents
\clearemptydoublepage
\listoffigures
\clearemptydoublepage
\listoftables
\clearemptydoublepage
\listoftheorems
\clearemptydoublepage


\pagenumbering{arabic}
\setcounter{page}{1}

\chapter{\label{chap:intro}Introduction}
\minitoc

The main work presented in this thesis is an extension of a number of
mathematical results concerning the spectral analysis of time-evolution
operators in quantum mechanics for a class of Hamiltonian systems. While
the results are of a technical nature and do not lend themselves to a
simple ``physical'' interpretation, they do have potential impacts on the
broad field of quantum chaos; specifically the link between the
spectrum of an operator\footnote{The operator of interest is either the
Hamiltonian or the time-evolution operator over one period, depending upon
the type of system under consideration.} and the dynamics of the system.

While this link to quantum chaos turns out to have little impact on the core
results presented, it sets the context for the research and drives the
course of work presented in the two main chapters. I will begin with an
investigation of the solutions to the Schr\"odinger equation when the
Hamiltonian is some simple system (e.g.,\ the harmonic oscillator) which is
perturbed in a periodic fashion. I will then characterise the spectrum and
extend a number of known results in the field of spectral analysis.  Along
the way, I will build unitary equivalents of a number of well known
self-adjoint theorems from functional analysis and discover links between
physical conjectures in the physics literature and deep number-theoretic
conjectures.  The results presented in \refchap{chap:rankN} and
\refchap{chap:comb} constitute the main body of research undertaken.

\refchap{chap:chaos} provides an introduction to both classical and quantum
chaos, outlining definitions, techniques and model systems widely used in
the literature. \refchap{chap:floquet} introduces the \emph{Floquet
operator}, an essential tool in the investigation of time-periodic quantum
mechanical systems.

To discuss the links between dynamics and spectra presented in
\refchap{chap:spectrum} properly, an understanding of functional analysis
and the mathematics of Hilbert spaces is required. \refchap{chap:maths}
provides this background, and pointers into the mathematics literature
where required. This chapter also develops the background theory and
notation used in \refchap{chap:rankN} and \refchap{chap:comb}.

As already mentioned, \refchap{chap:spectrum} constitutes a more detailed
examination of the literature that motivated my study of the spectrum of
operators. I discuss the link between quantum dynamics and spectral
analysis of the Hamiltonian operator (time-independent systems) or the
time-evolution operator (time-dependent systems). I will also identify a
general ambiguity in the literature, pointing out the pitfalls and also
a potential resolution to this problem.

\section{Summary of results}

The results of this thesis are presented in detail in \refchap{chap:rankN}
and \refchap{chap:comb}. I consider systems described by the time-dependent
Hamiltonian
\begin{equation*}
H(t) = H_0 + \lambda W \sum_{n=0}^\infty \delta(t-nT)
\end{equation*}
where $H_0$ is the time-independent Hamiltonian for some simple system
(e.g.,\ the harmonic oscillator), $W$ is an operator describing the
perturbation and $\lambda$ is a strength parameter. The spectral properties
of the time-evolution operator over one kick period, $T$, are investigated.

In \refchap{chap:rankN} I show that, for arbitrarily strong perturbation
strengths $\lambda$, the spectrum of the time-evolution operator remains
pure point if the perturbation operator $W$ satisfies the condition
\begin{equation*}
\sum_n \left| W^{1/2}\phi_n \right| < \infty\text{.}
\end{equation*}
The $\phi_n$ are the basis states of the $H_0$ system.
Essentially, we need $W^{1/2}\phi_n$, when written in terms of the basis
states of $H_0$, to be $l_1$ convergent.\footnote{To be a well defined state
in the Hilbert space, it must be $l_2$ convergent. That it is also $l_1$
convergent is a further restriction which a state in the Hilbert space may
or may not satisfy.} This is a non-perturbative result
and indicates that the system described by $H(t)$ is stable---the dynamics
does not change in a fundamental way due to the perturbation.

The condition on $W$ is relaxed in \refchap{chap:comb} and the possibility
for fundamentally different dynamical behaviour is shown to arise in the
case where $H_0$ is the harmonic oscillator, with eigenvalues (ignoring the
$1/2\hbar\omega$ term) given by $n\hbar\omega$. When the
ratio between the kicking period, $T$, and the natural frequency, $\omega$,
of the unperturbed system is irrational the Floquet operator is shown to
obtain a continuous spectral component. This result is proved for rank-$N$
perturbations (i.e.,\ $W$ is a rank-$N$ operator built from
$N$ projections), extending the previous rank-$1$ work of Combescure
\cite{Combescure90}.

The analysis is then extended to the case where $H_0$ is not the harmonic
oscillator, but some general pure point system with eigenvalues described
by an arbitrary order polynomial in $n$ (rather than a first order
polynomial as is the case for the harmonic oscillator). I show that the
question of whether a continuous Floquet spectrum will exist is equivalent
to a number-theoretic conjecture presented by Vinogradov \cite{Vinogradov}
over fifty years ago. The conjecture concerns the estimation of Weyl
sums---finite sums of the exponential of arbitrary order polynomials. The
optimal estimation of such sums for the case where the coefficients of the
polynomial are rational is known. For the case where the coefficients are
irrational, the current best estimates are not believed to be optimal.
Vinogradov's conjecture is that the optimal estimation for such sums is
that of the rational coefficient case.

If this conjecture were shown to be true, that is, if such Weyl sums were
shown to have a bound given by the rational coefficient case, then the
strongest conditions for the existence of a continuous spectrum for the
Floquet operator could be established.

The final piece of work presented in \refchap{chap:comb} concerns the
further classification of the continuous spectrum of the Floquet operator.
Extending the results of Milek and Seba \cite{Milek90.1}, I show that the
continuous Floquet spectrum is in fact singularly continuous. 
By the arguments presented in \refchap{chap:spectrum}, it is clear that the
existence of a singularly continuous component of the Floquet operator is a
necessary condition for the time evolution of such systems to show chaotic
structure.

\clearemptydoublepage
\chapter[Chaos in classical and quantum mechanics]
{\label{chap:chaos}Chaos in Classical and Quantum Mechanics}
\minitoc



The classical study of chaos is now a well established and flourishing
field of research in mathematics and mathematical physics. Chaotic behaviour
seems to pervade a large number of dynamical systems, and an appreciation
of it is essential in many contexts, both applied and theoretical. The
classic example of the earth's weather patterns always springs to mind when
chaos is mentioned.

The microscopic world, however, is not governed by the laws of classical
dynamics. In the realm of small quantum numbers the dynamics of a system is
governed by the Schr\"odinger equation. In such systems, the simple and
elegant definitions of chaos such as positive Lyapunov exponent, which hold
for classical systems, are not applicable. In fact, there is no universally
accepted definition for what constitutes chaos in quantum mechanics.  The
realisation that the concept of a phase space is not easily translated to 
quantum mechanics (the uncertainty relation makes it impossible to assign
both a position $\vec{x}$ and a momentum $\vec{p}$ simultaneously to a
quantum object) immediately precludes a simple link between classical
characterisations of chaos and quantum characterisations of chaos. The
route past this difficulty is not obvious, and much of the work so far has
been centred around finding appropriate tools. As yet, no consensus has been
reached on an appropriate definition of \emph{quantum chaos}.

That said, from the early 1980s onwards many papers have been published
that adapt classical concepts and bring them into the quantum world. There
are also papers that introduce completely new concepts not based on
equivalent classical ideas, in an attempt to make sense of quantum
dynamics. I will survey some of these ideas in this chapter.

\pagebreak

Three broad questions motivate the continued investigation of quantum
chaos.

\begin{enumerate}
\item Are there quantum systems that display chaotic behaviour at
the quantum level?
\item What properties of a quantum mechanical system determine whether or
not the corresponding classical system will display
chaotic behaviour? \label{page:ch_qb}
\item Do we come to the same conclusions irrespective of the choice of
definitional criteria for quantum chaos?
\end{enumerate}

The first question is very difficult to answer, in part because there is no
clear meaning to the expression ``chaotic behaviour'' when dealing with
microscopic phenomena. I will have more to say about this later.

The second question is intimately linked to the \emph{Correspondence
Principle} and theories of quantum measurement. It is a generally held
belief of physicists (although some do not agree, see \cite{Kruger}) that
the classical world is in fact quantum mechanical---how we obtain the
classical world
from quantum mechanics is a topic of extensive investigation, and needless to
say, this area of fundamental physics is infamous for its interpretational
difficulties and seemingly inconsistent behaviour. Mixing chaos into the
discussion can only make matters more interesting.

The final question is not independent of the previous two, but important to consider when
examining the literature. Many attempts to look at particular aspects of
the dynamics of quantum systems have been made. They are
based around certain toy systems and models. Some papers,
as a consequence of their definition of quantum chaos, come to the
conclusion that there is no such thing \cite{Partovi}. That is, they
conclude that quantum systems do not display chaotic behaviour. Other
papers, simply as a result of a different starting point, come to the
opposite conclusion. One can
certainly say that the pursuits of researchers has uncovered rich
variations in the dynamics of simple quantum systems, some of which are now
accessible to experiment. New experiments should provide great insight into
the numerous models and ideas currently being discussed in the literature.
The final word on this exciting field of research is a long way off.

I will now proceed to give an overview of the techniques used in classical
and quantum chaos research, and discuss model systems that receive the most
attention from the research community. The discussion leads naturally to
the new work presented in this thesis on the spectral analysis of unitary
operators.


\section{\label{sec:ch_classical}Chaos in classical systems}

The theory of chaos in classical mechanics is well established and understood.
The most accessible definition of chaos is obtained from the phase
space approach and the growth of the largest Lyapunov exponent for a
dynamical system. The path drawn out in phase space $(\vec{x},\vec{p})$ for
a particle is determined by solving the equations of motion. If initially
similar phase space points diverge exponentially as time progresses, a
Lyapunov exponent for the system is positive, and we say that the system is
chaotic. The system displays \emph{sensitivity to initial conditions}.
Conversely, systems which only experience power law type separation of
trajectories do not have any positive Lyapunov exponents and are deemed
non-chaotic.

The Lyapunov exponent definition of chaos is only one of the available
characterisations. The famous paper \emph{Period Three Implies Chaos} by
Tien-Yien~Li and James~A.~Yorke \cite{Li} shows that if a system described
by a continuous map\footnote{While Li and Yorke's original paper was
concerned with interval maps, it has been extended to the general frame of
topological dynamics. See \cite{Blanchard}.} has a period of
three\footnote{i.e.,\ After three iterations of the dynamics, we return to
the exact initial point} then it will have periods of all integers and,
importantly, also have an uncountable number of initial states which never
come even close to being periodic. That is, the evolution of those states
never returns to a point arbitrarily close to the initial state. 

A third way to characterise chaos in classical systems is to quantify the
information
required to describe the dynamics of a system. Essentially, if a dynamical
system requires an exponentially increasing number of bits to accurately
simulate its evolution for a linearly increasing time, then the information
content is high, and we say the system is chaotic. If the increase in bits
required follows only a power law as the desired time of simulation
increases, the system is non-chaotic. Information theory approaches to
chaos also include considerations of entropy production in systems.
For a beautiful discussion of the links between \emph{unpredictability},
\emph{information} and \emph{chaos}, see \cite{Caves}.

The different characterisations of chaos presented here are believed to be
essentially equivalent; attempting to establish this equivalence is an
active field of research. As an example, Blanchard \etal\ \cite{Blanchard}
have recently shown that the positive entropy characterisation of chaos
implies Li--Yorke chaos. For a review of the different approaches to chaos
see, for example, the review article by Kolyada and L'.~Snoha \cite{Kolyada}.

In classical mechanics, the simplest model systems in which we observe chaos
are iterative, rather than continuous. The logistic map and the
Henon-Heiles systems display chaotic behaviour for
certain parameter values. Of the continuous systems, the double pendulum
is arguably the most elegant example of a chaotic system. The equations
of motion are easily solvable, yet the resulting dynamics shows the
complexities of chaotic behaviour. Importantly, note that the system is
conservative---the total energy is constant over time.


\section{\label{sec:ch_kick}Kicked systems as a model testing ground for
chaos}

The field of non-conservative dynamics is a rich one, and chaotic behaviour
is a common feature. In particular, it turns out that systems perturbed by
an external sharp pulse (a ``kick'') have interesting dynamical properties.
For example, consider a spinning top---a childhood toy. If left to its own
devices, it will behave quite predictably, its main axis of rotation
precessing slowly, smoothly and \emph{predictably}. There is certainly no
chaotic motion. If, however, the spinning top is periodically kicked by a
short sharp pulse, the motion turns out to be quite unpredictable; the
system is chaotic \cite{Haake87}.

This observation has motivated many researchers in the field of quantum
chaos to consider quantum systems which are periodically perturbed. They
present an attractive combination of physical realisability and
mathematical tractability. Featuring strongly in the literature are the
kicked rotator (work by Izrailev and Shepelyanskii \cite{Izrailev80},
Grempel and Prange \cite{Grempel},
Casati \etal\ \cite{Casati86}, and Dittrich and Graham \cite{Dittrich90})
and the kicked
harmonic oscillator (see Combescure \cite{Combescure90}, Graham and
H\"ubner \cite{Graham}, and Daly and Hefernan \cite{Daly96.2}). The
quantum kicked top is also
examined in great detail by Haake \etal\ \cite{Haake87,Haake01}. A
whole array of
other kicked systems have been considered in relation to the study of
quantum chaos---see \cite{Milonni,Artuso92.1,Artuso92.2,Oelschlagel} for
just some examples. Many of the papers referred to in the next section are
also concerned with the analysis of kicked systems.


\section{\label{sec:ch_quantum}Characterisations of quantum chaos}

As already mentioned, there is no simple way to take the classical
characterisations of chaos and use them in the quantum context. As a
consequence, a great deal of research has been done attempting to find
appropriate characterisations of quantum systems which provide a clear link
to the classical concepts of chaos. In this section, I will review a number
of these attempts, highlighting the diversity in the field of quantum
chaos.  While I separate the discussion into broad sections, this is
somewhat artificial. There are significant overlaps between some of the
fields. Many of the papers referred to bring together many aspects of
quantum chaos.

The paper by Caves and Schack \cite{Caves}, concerning information-theoretic
approaches to dynamics, highlights a number of the issues that
make the study of chaos so much more difficult in quantum systems, as
opposed to the classical systems.

\subsection{Gutzwiller, periodic orbits and random matrix theory}

The field of quantum chaos was arguably born with the work of Gutzwiller
\cite{Gutzwiller71}. His semi-classical work on calculating the energy
eigenvalues for Helium via classical periodic orbits identified the
importance of the chaotic nature of the classical dynamics. For more
details on his work, see \cite{Gutzwiller82,Gutzwiller90}. An excellent
review was conducted by Heller and Tomsovic \cite{Heller}. A few examples
of the
influence of Gutzwiller's work in the literature is seen in the papers of
Eckhardt, Wintgen \etal, Tomsovic and Heller, and Keating
\cite{Eckhardt,Wintgen,Tomsovic,Keating}.

The ideas of Gutzwiller are still of central importance.
In an attempt to get to the essence of his work, the study of \emph{quantum
billiards} has been developed. Classically, the motion of a particle in a
two-dimensional bounded space (such as a billiard table) may be
chaotic. The motion is intimately linked to the geometry of the box. By
considering a quantum particle constrained to such a geometry, and using
Gutzwiller's ideas relating the classical periodic orbits to the quantum
energy levels, one can gain great insight into the dynamics of the quantum
system. Importantly, these investigations have also lead to actual
experimental investigations in microwave cavities, such as those by
Sridhar and Lu, presented in \cite{Sridhar} and references therein.

These considerations are intimately related to another technique of great
importance in the field of quantum chaos---\emph{random matrix theory}.
The eigenvalue statistics of non-chaotic and chaotic systems are, in a
sense, \emph{generic}, by which I mean that the energy eigenvalues of chaotic
systems typically have the same statistical structure. They are accurately
described by random matrix theory---the field of mathematics concerned with
the statistics of eigenvalues of classes of random matrices. Billiard systems
are an essential tool in work on random matrix theory \cite{Bohigas}. 
See the work of Smilansky \cite{Smilansky}, Kettemann \etal\ \cite{Kettemann}
and the proceedings \cite{p_Cohen86} along with references therein.

\subsection{de~Broglie Bohm theory}

Since the initial conception of quantum mechanics, there have been
persistent attempts to find a ``classical'' interpretation of the theory.
The dominant \emph{Copenhagen interpretation} of quantum mechanics is most
certainly not such an interpretation. The \emph{de~Broglie Bohm}
interpretation of quantum mechanics \cite{Holland}, largely ignored in the
physics community, attempts to provide a strong conceptual link between
quantum and classical dynamics; it is based around the Hamilton-Jacobi
equations of motion. A quantum particle is just like a classical particle,
except that the classical potential it exists in is supplemented by a
quantum potential.

The quantum potential is given through the solution of the Schr\"odinger
equation, and thus, results in de~Broglie Bohm theory are equivalent to
those in standard quantum mechanics.\footnote{For over seventy years, this
view of the equivalence has prevailed. This is in fact one of the reasons the
theory is usually dismissed---it is often seen as nothing more than a
rewrite of the normal quantum theory. However, in recent years there have
been some, such as Ghose \cite{Ghose01.2}, who claim there are
incompatibilities between the theories. Experimentally realisable tests
have even been suggested by Golshani and Akhavan \cite{Golshani02}. The
work is, however, highly controversial and disputed.  Struyve and De~Baere
\cite{Struyve} and
other authors referenced therein provide arguments against this suggested
incompatibility.} Given that a quantum particle now has a trajectory, a
phase space picture of the dynamics is realisable. Thus, both a quantum
Lyapunov exponent and hence quantum chaos are definable just as in
classical mechanics.  The quantum potential blurs the trajectory somewhat,
but one can think of a ``flux tube'' in phase space, describing the
evolution of the system.

Papers employing the de~Broglie Bohm theory to examine quantum chaos
include those by de~Alcantara Bonfim \etal\ \cite{AlcantaraBonfim},
de~Polavieja \cite{Polavieja96}, Schwengelbeck and Faisal
\cite{Schwengelbeck,Faisal}, and Wu and Sprung \cite{Wu}. The
de~Broglie Bohm theory also plays a central role in a number of the other
ideas listed in the following sections.

\subsection{Open quantum systems}

The vast literature on quantum mechanics in open systems also provides a
strong link to classical chaos. Tiny environmental interactions which have
no effect on the classical systems turn out to be strong enough (easily!)
to completely change the quantum dynamics. The coherent interference
effects responsible for the ``quantum suppression of chaos''\footnote{I
will detail the meaning of this expression in \refsec{sec:ch_results}.} are
destroyed and the quantum systems show rather classical behaviour,
including chaotic behaviour. Fritz Haake's book \emph{Quantum signatures of
Chaos} \cite{Haake01} and the many references therein provide an in depth
review and excellent discussion of these phenomena. The ideas of Zurek,
further discussed by Paz, have also been influential
\cite{Zurek95,Zurek91,Zurek02,Paz}. There has also been experimental work
in this area, including the early work of Bl\"umel \etal\
\cite{Blumel89,Blumel91}. On the theoretical side, of interest is much of
the work by Cohen, Dittrich \etal, Grobe \etal, and Kohler \etal\
\cite{Cohen91,Dittrich90,Dittrich93,Grobe,Kohler}. The links back to random
matrix theory and level statistics are strong. This work also provides
significant insight into Question (b) on page~\pageref{page:ch_qb}.
 
\subsection{Other interesting techniques}

Here I briefly list a number of other ideas which have been put forward in the
literature along with a number of references. Again, many of the
ideas draw upon the basic ideas of random matrix theory, de~Broglie
Bohm theory and open quantum systems.

\subsubsection{The Loschmidt echo}

A very interesting approach to unifying the concept of chaos in classical
and quantum mechanics has been presented by Jalabert and Pastawski
\cite{Jalabert} and
further reviewed by Cucchietti \etal\ \cite{Cucchietti}. In classical
mechanics,
small perturbations in the equations of motion lead to the exponential
divergence of trajectories in phase space. It turns out that in quantum
mechanics, while small changes to initial conditions do not lead to
significantly different dynamical behaviour, small changes in the
Hamiltonian can lead to significant variations in the time evolution. The
overlap of initially identical wave functions is measured and a Lyapunov
exponent is extracted.  This recent idea appears promising as it provides
an opportunity to directly compare chaotic structure in classical and
quantum systems.

\subsubsection{The Quantum action}

The quantum action was introduced by Jirari \etal\
\cite{Jirari01.1,Jirari01.3}
and utilised by Caron \etal\ \cite{Caron01.1,Caron01.2,Caron02}. The aim is to
unify the characterisation of chaos in classical and quantum mechanics by
introducing a \emph{quantum action} analogous to the classical action.
Jirari \etal\ \emph{conjecture} that \cite{Jirari01.3} (quote)
\begin{quote}
For a given classical action $S$ with a local
interaction $V(x)$ there is a renormalized/quantum action
\begin{equation*}
\tilde{S} = \int dt \frac{\tilde{m}}{2} \dot{x}^2 - \tilde{V}(x)\text{,}
\end{equation*}
such that the transition amplitude is given by
\begin{equation*}
G(x_\text{final}, t_\text{final}; x_\text{init}, t_\text{init})
 = \tilde{Z}\exp\left[ \frac{i}{\hbar}
 \tilde{S}\left[\tilde{x}_\text{class}\right]
 \bigg|^{x_\text{final},t_\text{final}}_{x_\text{init},t_\text{init}}
 \right]
\end{equation*}
where $\tilde{x}_\text{class}$ denotes the classical path corresponding to
the action $\tilde{S}$.
\end{quote}
Note that the mass term, $\tilde{m}$, and the potential, $\tilde{V}$, in the
action are quantum parameters. The quantum action takes into account
quantum corrections to the classical motion. Once obtained, the tools of
classical mechanics may be applied as the mathematical form corresponds
exactly with the classical action. The integral is taken only over the
classical path.

This direct link allows the definition of chaos in terms
of the classical action to be taken over to the quantum dynamics directly.

It must be noted, as acknowledged by Jirari \etal\ \cite{Jirari01.1}, that
there
is no proof of the conjecture. Numerical evidence is presented by Jirari
\etal\ and Caron \etal\ that indicates that it seems to be reasonable in
a range of cases.

\subsubsection{Entropy approaches and information theory}

As stated by S{\l}omczy\'nski and \.Zyczkowski in \cite{Slomczynski},
``the approach linking chaos with the unpredictability of the measurement
outcomes is the right one in the quantum case''. To measure this
unpredictability, they introduce a generalised notion of entropy. The
inclusion of the measurement process links this approach to some of the
open systems work already mentioned.  Other papers to recently use an entropy
approach include the work by Lahiri \cite{Lahiri03}.

The links between classical chaos and information theory are well known.
The ideas can be applied to quantum mechanics too.  Some recent work in
this field has been done by Inoue \etal\ \cite{Inoue01.2,Inoue04}. See the
references therein for historical details.

\subsubsection{Stochastic webs (kicked systems)}

The pioneering work of Zaslavski\u{\i} (see \cite{Zaslavsky85,Chernikov} for
results and background) investigating the effect of classical phase space
structures on quantum dynamics for kicked systems has been fruitful. The
ideas of the quantum suppression of chaos are clearly seen in many of these
works through an analysis of the diffusive behaviour. See the work by
Berman \etal, Borgonovi and Rebuzzini, Chernikov \etal, Daly and Hefernan,
Dana, Frasca, Korsch \etal, Sikri and Narchal,
Torres-Vega \etal, and Zaslavski\u{\i} \etal\
\cite{Berman,Borgonovi,Chernikov,Daly96.1,Daly96.2,Dana94,Frasca,Korsch,
Sikri93.1,Sikri93.2,TorresVega,Zaslavsky86.1,Zaslavsky86.2} for just a few
of the results obtained using these ideas.

\subsection{Spectral analysis of operators}

Finally, as alluded to throughout this review, the analysis of the
spectrum of certain operators can be related to the dynamical properties
of a quantum mechanical system. I mentioned this when discussing quantum
billiards and random matrix theory but it also plays a central role in
the examination of kicked systems.

The spectral analysis of operators is a rich mathematical field in its own
right and makes numerous claims relevant to the dynamics
of quantum systems. The mathematical link between the spectral properties
of operators and quantum dynamics is the motivation for the work in this
thesis and, accordingly, \refchap{chap:spectrum} is devoted to a fuller
exploration of this link.


\section{\label{sec:ch_results}Quantum chaos---some established results}

Having reviewed some of the broad research areas in quantum chaos, I now 
briefly present a number of established results from analytic examinations,
numerical simulation and experiment that have so far been established. They
cut across view points in the field of quantum chaos.

A common property of the time evolution of quantum systems is that
for a short time the classical and quantum evolutions correspond. This
correspondence is measured by, for example, the energy as a function of
time. This is a generic property, and can be attributed to the time it
takes for the quantum system to ``become aware'' of the finite
dimensionality of its phase space. After such a time, the
correspondence is lost and, while the classical system's energy continues
to increase (either in a diffusive way, corresponding to chaotic
motion, or in a ballistic fashion, corresponding to resonant energy growth),
the
quantum system shows recurrences. A great amount of work has been done
both analytically and numerically in identifying these timescales. See for
example \cite{Karkuszewski01.1,Berman,Haake01,Iomin}.

Related to this is the heuristic link between quantum recurrence and the
problem of conduction of electrons in a random lattice---the phenomena of
\emph{Anderson localisation}. Again, after a certain time, the classical
and quantum evolutions diverge. See
\cite{Fishman82,Grempel,Casati86,Fishman93} for an explanation of the
relation 
between quantum recurrence and Anderson localisation and also the
contrast between classical and quantum time evolutions.

The quantum recurrence results are so pervasive that they have lead to a new
concept, already mentioned, the \emph{quantum suppression of chaos}
\cite{Blumel84}. The concept encompasses all these results and reflects the
fact that the quantum equivalent of many classically chaotic systems seem
to be more well behaved and thus non-chaotic. In a general sense, this
behaviour is attributed to the interference effects in the quantum
evolution conspiring to suppress dynamical spreading in the wave packets.
This take on the results is beautifully explained by Haake \cite{Haake01}.
He also shows how the introduction of tiny environmental interactions
destroys these interferences effects, leading to rather chaotic-like
behaviour for the open quantum systems.


\section{\label{sec:ch_summary}Summary}

I have discussed characterisations of chaos in both classical and quantum
theory. While many of the simple ideas from classical mechanics cannot be
directly translated into the language of quantum mechanics, there are ways
around these problems, e.g.,\ both the de~Broglie Bohm theory and the
Loschmidt echo approach allow concepts from classical chaos to
be brought to the quantum theory.

The ``kicked'' systems were seen to be of fundamental interest in
investigations into quantum chaos, as was the broad field of spectral
analysis. \refchap{chap:floquet} now introduces the basic tools required to
consider the quantum evolution of kicked systems.

\clearemptydoublepage
\chapter[The Floquet operator]
{\label{chap:floquet}The Floquet Operator}
\minitoc



The kicked systems are an excellent ``testing ground'' for chaos in both
classical and quantum mechanics. In the quantum case, the time evolution of
the system, as governed by the Schr\"odinger equation, has a particularly
elegant form, allowing a stroboscopic analysis of the system to be made.
The mathematical tool for this is called the \emph{Floquet operator} and is
simply the time-evolution operator over a single kick period. It also goes
by the name of \emph{quasi-energy operator} or \emph{monodromy operator} in
the literature. The central role that the Floquet operator plays in the
work presented in this thesis warrants a detailed introduction, the topic
of this chapter. For more information and some simple examples of the
Floquet operator, see \cite{Haake01}.


\section{\label{sec:f_t_evol}Time evolution in quantum mechanics}

The dynamical evolution of a non-relativistic closed quantum system
$|\psi(t)\rangle$ is governed by the famous Schr\"odinger equation,
\begin{equation}
\label{eq:f_schr}
-\frac{i}{\hbar} \frac{\partial}{\partial t} |\psi(t)\rangle
  = H(t)|\psi(t)\rangle\text{.}
\end{equation}
With this notation, we have a rather simple looking partial differential
equation. Of course, this is only an illusion. The vast complexities of
quantum mechanics remain hidden.

The formal solution to \refeq{eq:f_schr} is given by
\begin{equation*}
\begin{split}
\label{eq:f_schro_sol}
|\psi(t)\rangle &= \left[\exp\left(-\frac{i}{\hbar}\int_0^t dt'\,H(t')\right)
  \right]_+ |\psi(0)\rangle \\
  &\equiv U(t,0)|\psi(0)\rangle
\end{split}
\end{equation*}
where the ``$+$'' subscript indicates that we must ensure that the
time ordering is done correctly. With correct time ordering, it becomes
evident that the relation
\begin{equation}
\label{eq:f_addU}
U(t,0) = U(t,s)U(s,0)
\end{equation}
holds and, due to the unitary nature of the time evolution,
\begin{equation*}
U(t,s) = U(t,0)U^\dagger(s,0).
\end{equation*}
Apart from for the simplest of systems such as the harmonic oscillator,
square well or the hydrogen atom (and minor variants on them), the
Schr\"odinger equation remains immune to analytic solution. Numerical
studies get us some way further, but even then, all but the most simple of
experimental situations remain intractable to detailed analysis.

The Hamiltonian $H(t)$ can, in general, be explicitly time dependant, as
indicated here. Solutions are just that bit more difficult. In some
idealised cases however, there is an effective way to make progress and it
turns out to be very useful in the study of potentially chaotic systems.
As discussed in \refsec{sec:ch_classical}, the study of classical systems
which experience a periodic (in time) perturbation is of great interest as
the motion is often chaotic. If the Hamiltonian is given by some easily
solvable system, say a harmonic oscillator or a quantum spinning top or
rotor which is then periodically perturbed, the exact solution to the
Schr\"odinger equation can be written down. If
\begin{equation}
\label{eq:f_hamil}
H(t) = H_0 + \lambda W\sum_{n=0}^\infty \delta(t-nT)
\end{equation}
where $H_0$ is the time-independent Hamiltonian of the simple system, $W$ is
an operator describing the perturbation and $\lambda$ is a strength parameter,
then the solution to the Schr\"odinger equation at times $nT$ is given by
\begin{equation*}
|\psi(nT)\rangle = V^n |\psi(0)\rangle\text{,}
\end{equation*}
where I have introduced the so called \emph{Floquet operator} $V$. $V$ is
simply the time-evolution operator for one period $T$,
\begin{equation*}
V \equiv U(T)\text{.}
\end{equation*}
Using \refeq{eq:f_addU}, the time evolution over one period is
\begin{equation*}
V = \exp\left[-\frac{i}{\hbar} \int_{\epsilon}^{T-\epsilon} H(t)\,dt \right]
 \exp\left[-\frac{i}{\hbar} \int_{T-\epsilon}^{T+\epsilon} H(t)\,dt \right]
 \text{.}
\end{equation*}
The first factor, for time $\epsilon \leq t < T-\epsilon$, is trivial. 
The delta function in \refeq{eq:f_hamil} is not acting, $H(t)=H_0$ is
independent of time and the system evolves freely via the time-evolution
operator
\begin{equation*}
U_0 = e^{-iH_0T/\hbar}\text{.}
\end{equation*}
The second factor is over an infinitesimal time period $2\epsilon$ when the 
delta function kick is acting. Over this infinitesimally short period of time
the influence from $H_0$ is zero. The system instantaneously evolves
via the operator
\begin{equation*}
e^{-i\lambda W/\hbar}\text{.}
\end{equation*}
Recombining these two parts of the evolution, the operator
describing the evolution of the system from just before one kick to just
before the next is seen to be
\begin{equation*}
V = e^{-iH_0T/\hbar}e^{-i\lambda W/\hbar}\text{.}
\end{equation*}
At times in this work, I will have need to consider the time-evolution
operator taking us from just after one kick to just after the next. By a
similar argument, this is given by
\begin{equation*}
V' = e^{-i\lambda W/\hbar}e^{-iH_0T/\hbar}\text{.}
\end{equation*}
At any rate, they are essentially equivalent---in a given context, one
may be more convenient than another. When used, it will always be made
clear how the Floquet operator has been defined.

Once the Floquet operator has been obtained, numerical investigations of
the system become far more tractable. A stroboscopic picture of the
evolution of complex systems can be obtained and investigated. It is also
hoped that the idealisation of the delta function pulse is useful when it
comes to predicting the behaviour of experimental situations where, for
example, a system may be perturbed by a periodic stream of short powerful
laser pulses. While each laser pulse clearly interacts with the system
over some finite time, the effect on the dynamics should be modelled well
by the Floquet operator type ``kicks'' discussed here.

The Floquet operator also proves useful in analytic work. Specifically, the
spectrum of the Floquet operator is of great use in discussing the dynamics
of a given quantum mechanical system. Characterisations of the spectrum for
particular classes of Hamiltonian systems may be of great value when it
comes to predicting how systems behave.


\section{\label{sec:f_kr}An example: the kicked top}

Here I introduce the simple example of the quantum kicked top to
demonstrate the usefulness of the Floquet operator. The analysis is covered
in significantly greater detail in the book by Haake \cite{Haake01}.

Both the unperturbed Hamiltonian, $H_0$, and perturbation operator, $W$,
from \refeq{eq:f_hamil} are polynomial functions of the total spin of the
top, $\vec{J}$. Conservation of $\vec{J}^2 = j(j+1)$, where I have set
$\hbar=1$ for convenience, implies that the kicked top exists in a
finite-dimensional Hilbert space. Choosing, somewhat arbitrarily (but for
convenience, see \cite{Haake01}), $H_0 \propto J_x$ and $W \propto J_z^2$
the Floquet operator is simply ($c_1$ and $c_2$ are the proportionality
constants)
\begin{equation*}
V = e^{-ic_1J_xT}e^{-i\lambda c_2 J_z^2/2j}\text{,}
\end{equation*}
where the $1/2j$ factor in the instantaneous evolution term is required for
the $j\rightarrow\infty$ classical limit to make sense. Again, see
\cite{Haake01} for a discussion of this.

To proceed, we must specialise to a particular spin system. The spin-$1/2$
system is trivial, so here I briefly describe the spin-$1$ system. For
spin-$1$ systems, states in the Hilbert space are represented by
$3$-component spinors (column vectors), and operators by $3\times3$
matrices which can be decomposed in terms of the Gell--Mann matrices
$\lambda_{0,\ldots,8}$. We have $(J_i)_{lm} = -i\hbar\epsilon_{ilm}$
\cite{Messiah2}, so
\begin{align*}
J_z &=
\begin{pmatrix} 0 & -i & 0 \\
		i & 0 & 0 \\
		0 & 0 & 0 \end{pmatrix} \\
   &= \lambda_2 \\
\intertext{and}
J_x &= 
\begin{pmatrix} 0 & 0 & 0 \\
		0 & 0 & -i \\
		0 & i & 0 \end{pmatrix} \\
    &= \lambda_7\text{.}
\end{align*}
Noting that
\begin{equation*}
\lambda_2^2 = 2/3I + \sqrt{3}/3 \lambda_8
\end{equation*}
the Floquet operator is
\begin{equation*}
\begin{split}
V &= e^{-ic_1\lambda_7T}e^{-i\lambda c_2(2/3I + \sqrt{3}/3\lambda_8)/2} \\
  &\equiv e^{-ic_1\lambda_7T}e^{-ic_3}e^{-ic_4\lambda_8}\text{.}
\end{split}
\end{equation*}
where the new constants $c_{3,4}$ have absorbed the various numerical
factors and the kicking strength $\lambda$. In a numerical study using this
operator, the strength is varied by adjusting $c_{4}$ appropriately. Note
that the term $e^{-ic_3}$ is simply a global phase and hence has no effect
on the dynamics.

The final step to be taken is to rewrite the Floquet operator directly as a
$3\times3$ unitary operator. This can be done either analytically or
numerically.

Now that we have an appropriate expression for the Floquet operator, the
dynamics of a quantum kicked top can now be examined. An initial state,
$|\psi_0\rangle$ is chosen
\begin{equation*}
|\psi_0\rangle = 
  \begin{pmatrix} 
  z_1 \\ z_2 \\ z_3
  \end{pmatrix}
\end{equation*}
and evolved using the matrix representation of the Floquet operator.
The resulting states, $|\psi(nT)\rangle$ are given by
\begin{equation*}
|\psi(nT)\rangle = V^n|\psi_0\rangle
\end{equation*}
and can be trivially generated numerically for further analysis.

As $V$ is a $3 \times 3$ unitary matrix in this case, its spectrum is
simply a point spectrum of three eigenvalues.


\section{\label{sec:f_summary}Summary}

The unitary Floquet operator just introduced provides one with a
stroboscopic view of the time evolution of periodically kicked systems.
This was demonstrated through the simple example of the spin-$1$
quantum kicked top.
The Floquet operator is the key tool in the work that follows, but in order
to understand the work a number of mathematical concepts must first be
introduced. These concepts are the topic of \refchap{chap:maths}.

\clearemptydoublepage
\chapter[Mathematical preliminaries]
{\label{chap:maths}Mathematical Preliminaries}
\minitoc



As mentioned in \refchap{chap:floquet}, and to be detailed in
\refchap{chap:spectrum}, knowledge of the spectrum of the Floquet operator
turns out to be important in understanding the dynamics of periodically
kicked Hamiltonian systems. To make analytic progress in this
field, we will need mathematical tools and concepts that are beyond
those usually employed by physicists. In this chapter, I will provide a
type of ``tutorial'' or, if you like, an overview of the fields of
measure theory, functional and spectral analysis, picking out those
concepts that will be
necessary when we come to the work contained in \refchap{chap:rankN}. I
begin with a review of integration and measure theory and then proceed to
discuss how the concepts of singular, absolute and point measures (to be
defined) apply in the context of operators on the physicists' Hilbert
space.


\section{\label{sec:m_meas}Measure theory}

The typical high school introduction to ``integration'' begins with the
idea that the integral of a function gives us the ``area under the graph''.
A suitably well behaved function is split up into smaller and smaller
intervals, and each interval approximated by a rectangle, whose height is
given by the function evaluated at the mid-point of the interval. The sum
of the areas of the rectangles gives an estimation of the area
under the function. By taking the limiting case of an infinite number of
infinitesimal intervals, the integral of the function is obtained.

This idea of integration is abstracted, formalised and extended to become
the Lebesgue integral of modern mathematics \cite{Riesz,RudinFA,Reed1,Saks}.
To go beyond this point, the concept of \emph{abstract measure theory} is
introduced.

When considering the integral of a function $f(x)$, one usually
thinks simply of
\begin{equation*}
\int f(x)\,dx
\end{equation*}
where $dx$ is the measure or ``size'' of an interval. Each interval along
the $x$-axis is given the same weight. An abstraction of this technique is
to allow the weight or ``size'' given to each interval to be determined
by a measure function $\alpha(x)$. Any positive, non-decreasing function,
$\alpha(x)$ will do. $\alpha(x)$ need not even be continuous. The measure
\begin{equation*}
\mu_\alpha([a,b]) = \alpha(b) - \alpha(a)
\end{equation*}
is formed and one obtains the Lebesgue--Steiltjes integral
\begin{equation*}
\int_a^b f\,d\mu_\alpha \equiv \int_a^b f\,d\alpha\text{.}
\end{equation*}
If $\alpha(x) = x$, we trivially obtain
\begin{equation*}
\mu_\alpha([a,b]) = \alpha(b) - \alpha(a) = b-a
\end{equation*}
and recover the usual Lebesgue integral
\begin{equation*}
\int_a^b f\,d\alpha = \int_a^b f\,dx\text{.}
\end{equation*}
In an intuitive sense, when integrating with respect to a measure
function $\alpha(x)$, the contribution to the integral from an interval is
proportional to the derivative $d\alpha(x)/dx$. For a given function
$\alpha(x)$, if the slope of $\alpha(x)$ is zero for a particular interval,
then that interval will contribute nothing, irrespective of the functional
value $f(x)$ on that interval.

\subsection{Point measures}

If the measure function is
\begin{equation*}
\alpha(x) = \sum_{x'} \theta(x-x')
\end{equation*}
where $\theta(x-x')$ is the usual step-function defined as
\begin{equation*}
\theta(x-x') = \begin{cases}
  0 \text{ if } x < x'\text{,} \\
  1 \text{ if } x \geq x'
\end{cases}
\end{equation*}
then, in rather loose, but intuitive notation
\begin{equation*}
d\alpha(x) = \sum_{x'}\delta(x-x')dx\text{.}
\end{equation*}
As we move along the $x$-axis, we obtain a contribution to the integral
$\int f\,d\alpha$ only at the points $x'$. Each contributing
``interval'' has a Lebesgue measure of zero---reflected in the fact that in
``ordinary integration'' the contribution to an integral from a single
point is zero, and thus, the occasional infinitely thin but high spike in a
function does not matter.

Our integral with the given measure above is now
\begin{equation*}
\int f\,d\alpha = \sum_{x'} f(x')
\end{equation*}
and we have converted the integral into a sum.

Measures of this type are called ``point measures'' for obvious reasons.
Compared to integrating a function with respect to the Lebesgue measure, a
point measure gives contributions to the integral from single discrete
points, exactly where the Lebesgue measure does not contribute. We say that
the Lebesgue measure and point measure are \emph{mutually singular}.

\subsection{Absolutely continuous measures}

The absolutely continuous measure is perhaps the simplest of measures apart
from the standard Lebesgue measure. If $\alpha(x)$ is a smooth, continuous,
everywhere differentiable function of $x$ then we can write
\begin{equation*}
\frac{d}{dx} \alpha(x) = g(x)
\end{equation*}
with $g(x)$ continuous and well-behaved. The integral $\int f\,d\alpha$ is
then
\begin{equation*}
\int f\,d\alpha = \int f(x)g(x)\,dx
\end{equation*}
and we essentially recover the standard Lebesgue integral, but now of the
function $f(x)g(x)$ rather than $f(x)$. The absolutely continuous measure
satisfies
\begin{equation*}
\mu_\alpha([a,b]) = 0 \text{ if and only if } b-a=0\text{.}
\end{equation*}
A simple example is to consider the function $\alpha(x) = (1/2)x^2$. Then
$g(x)=x$ and each interval in the integral, instead of having a constant
weight, is now weighted by its position on the $x$-axis.

The absolutely continuous measures are often referred to simply as the
continuous measures but I will not do so here, for reasons that will become
clear in the following sections. 

\subsection{Singular continuous measures}

The point and absolutely continuous measures discussed above are both fairly
straight-forward. They quantify concepts that we are already familiar with.
However, as is usually the case in mathematics, we can take things further.

It turns out that one can construct measures, $\mu_\alpha$, that contribute
to integrals exactly where the Lebesgue measure contributes zero (rather
like a point measure), but which are nevertheless continuous measures. This
seems contradictory but an example should clarify the idea. We start by
defining a particular set, the Cantor set.

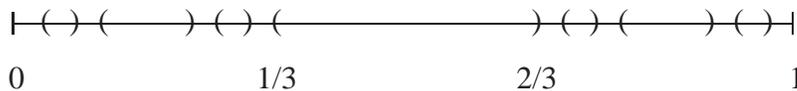
\begin{figure}[H]
\begin{center}
\begingroup
\setlength{\unitlength}{0.0200bp}%
\begin{picture}(18000,3000)(0,0)%
\put(2475,1100){\makebox(0,0){\strut{} 0}}%
\put(7375,1100){\makebox(0,0){\strut{} 1/3}}%
\put(12275,1100){\makebox(0,0){\strut{} 2/3}}%
\put(17175,1100){\makebox(0,0){\strut{} 1}}%
\thinlines
\put(2475,2100){\line(1,0){14700}}%
\put(2475,2300){\line(0,-1){400}}%
\put(17175,2300){\line(0,-1){400}}%
\put(3019,2170){\makebox(0,0){\strut{} (}}%
\put(3564,2170){\makebox(0,0){\strut{} )}}%
\put(4108,2170){\makebox(0,0){\strut{} (}}%
\put(5742,2170){\makebox(0,0){\strut{} )}}%
\put(6286,2170){\makebox(0,0){\strut{} (}}%
\put(6831,2170){\makebox(0,0){\strut{} )}}%
\put(7375,2170){\makebox(0,0){\strut{} (}}%
\put(12275,2170){\makebox(0,0){\strut{} )}}%
\put(12819,2170){\makebox(0,0){\strut{} (}}%
\put(13364,2170){\makebox(0,0){\strut{} )}}%
\put(13908,2170){\makebox(0,0){\strut{} (}}%
\put(15542,2170){\makebox(0,0){\strut{} )}}%
\put(16086,2170){\makebox(0,0){\strut{} (}}%
\put(16631,2170){\makebox(0,0){\strut{} )}}%
\end{picture}%
\endgroup
 
  \caption{The Cantor set}
  \label{fig:m_cantor_set}
\end{center}
\end{figure}

Consider the subset $S$ of $[0,1]$ given by
\begin{equation*}
S = \left(\frac{1}{3},\frac{2}{3}\right) \cup
    \left(\frac{1}{9},\frac{2}{9}\right) \cup
    \left(\frac{7}{9},\frac{8}{9}\right) \cup
    \left(\frac{1}{27},\frac{2}{27}\right) \cup \ldots
\end{equation*}
The Lebesgue measure of $S$ is
\begin{equation*}
\frac{1}{3} + 2\left(\frac{1}{9}\right) +
4\left(\frac{1}{27}\right) + \ldots = 1\text{.}
\end{equation*}
The complement of this set, $C=[0,1]\setminus S$, has Lebesgue
measure zero and is known as the Cantor set. It contains an infinite number
of points but has a size of zero---it is an uncountable set of (Lebesgue)
measure 0. See \reffig{fig:m_cantor_set}.
The Cantor function, $\alpha(x)$, is defined by setting
\begin{align*}
\alpha(x) &= 1/2 \text{ on }
\left(1/3, 2/3 \right)\text{,} \\
\alpha(x) &= 1/4 \text{ on }
\left(1/9, 2/9\right)\text{,} \\
\alpha(x) &= 3/4 \text{ on }
\left(7/9, 8/9\right) \ldots
\end{align*}
$\alpha(x)$ becomes a continuous function by continuing this idea all the
way to $\alpha(x)$ on $[0,1]$. The Cantor function $\alpha(x)$ is a
non-constant continuous function on $[0,1]$ whose derivative exists almost
everywhere (with respect to the Lebesgue measure) and is zero almost
everywhere!

\begin{figure}[H]
\begin{center}
  \input{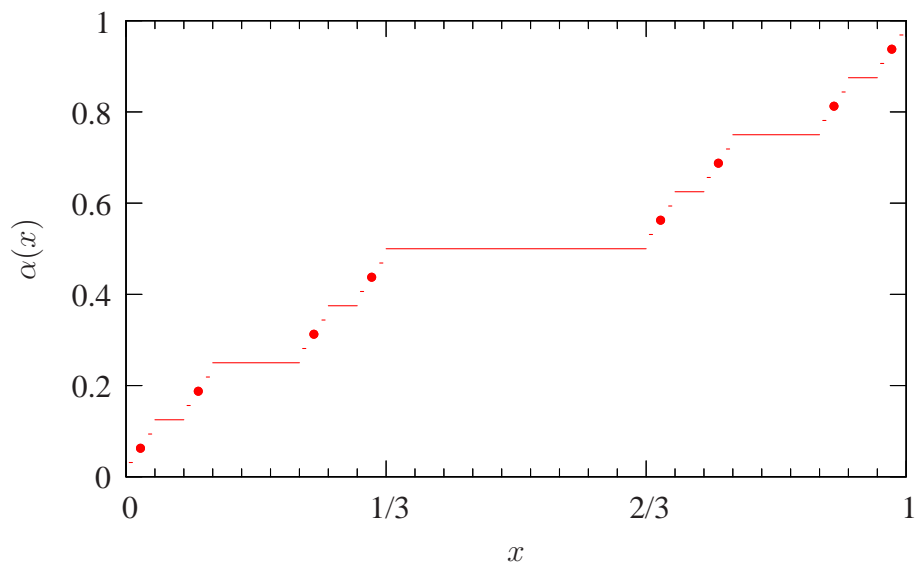}
  \caption{The Cantor function}
  \label{fig:m_cantor_func}
\end{center}
\end{figure}

The Cantor function, aptly coined the ``Devil's staircase'', is shown in
\reffig{fig:m_cantor_func}. The function has zero slope almost everywhere,
but still manages to rise from $0$ to $1$ across the finite interval $[0,1]$
without ever jumping by a finite amount.

The measure, $\mu_\alpha$, formed using the Cantor function is continuous.
That is, $\mu_\alpha(\{p\})=0$ for any set $\{p\}$ with only a single
point. $\mu_\alpha$ contributes to the integral only on the set $C$,
exactly the set on which the normal Lebesgue measure contributes nothing.
When integrating with this measure, almost every interval in $[0,1]$
contributes nothing to the integral as the slope of $\alpha(x)$ is zero
almost everywhere.

The Cantor measure is an example of a continuous measure, but one which is
mutually singular to the Lebesgue measure. It satisfies
\begin{equation*}
\mu_\alpha([a,b]) = 0 \text{ if and only if } b-a \neq 0\text{.}
\end{equation*}
Such measures are known as singularly continuous measures.

\subsection{Summary}

I have introduced three types of measure---the point, absolutely
continuous and singularly continuous measures. It turns out that an
arbitrary measure can always be decomposed into three parts,
\begin{equation}
\label{eq:m_meas_decomp}
\mu = \mu_\pp + \mu_\ac + \mu_\cs\text{.}
\end{equation}
This decomposition is unique and, importantly, the three pieces are
mutually singular. It should be noted that there are other ways to
decompose the spectrum into mutually singular parts. While important, in
this work I will not have need for such characterisations. The interested
reader is referred to \cite{Reed1}.


\section{\label{sec:m_func_anal}The tools of functional analysis}

Having introduced the notion of a generalised measure of a set on the
real line, I now provide an overview of the field of functional analysis,
through the introduction of the Hilbert space and operators on the Hilbert
space. The abstract measure theory just presented is brought into the
picture when discussing the spectral properties of operators on the
Hilbert space.  It is via this route that the power of functional and
spectral analysis enters into mathematical physics.  Classifications of the
spectrum of operators (the eigenvalues) corresponds to finding the relevant
measure to describe the eigenvalues.

Here I only provide a brief overview, defining the pertinent concepts for
my work and introducing notation.  For a mathematically rigorous, step by
step derivation of the concepts discussed, Chapters~1 and 2 in \cite{Reed1}
provide an excellent introduction. Much of the chosen notation is inherited
from the work of Howland \cite{Howland87}. When, as will sometimes be the
case, I need to depart from this notation, I will so indicate.

\subsection{Hilbert spaces}

A vector space which is complete, i.e.,\ one in which all Cauchy sequences
converge to an element of the space, is called a \emph{Hilbert space}. The
elements of the space are the vectors which physicists use to represent
states of a quantum mechanical system. Throughout this work, I will always
denote Hilbert spaces by $\mathcal{H}$ or $\mathcal{K}$. Subspaces of a
given Hilbert space will often be referred to as $\mathcal{S}$. A Hilbert
space is said to be \emph{separable} if and only if it has a countable
orthonormal basis. I will only ever consider separable Hilbert spaces in
this work. Elements of the space will generally be referenced by either
$|\psi\rangle$, $|\phi\rangle$, or in a mathematical, rather than physical
context, may often be denoted by $x$ or $y$.

A subset $\mathcal{S}$ of $\mathcal{H}$ may or may not be complete. The
closure of $\mathcal{S}$, $\overline{\mathcal{S}}$ is obtained by adding
to $\mathcal{S}$ all the limit points of sequences of elements of
$\mathcal{S}$. By closing $\mathcal{S}$, a complete subspace is obtained.
A set $\mathcal{S}$ is said to be \emph{dense} in $\mathcal{H}$ if
$\overline{\mathcal{S}}=\mathcal{H}$.

The inner product of two vectors $x,y\in\mathcal{H}$ is
$\langle x,y \rangle$, and
the norm of a vector $x$ is $\lpar x \rpar \;=\langle x,x \rangle^{1/2}$.

\subsection{Operators}

An operator $A:\mathcal{H}\rightarrow\mathcal{K}$ acts on elements of the
Hilbert space $\mathcal{H}$ and returns elements in the Hilbert space
$\mathcal{K}$. It is often the case that $\mathcal{K}$ is either a subspace
of, or in fact is, $\mathcal{H}$. Operators on a Hilbert space will always be
referenced by uppercase Latin characters.

For an operator $A:\mathcal{H}\rightarrow\mathcal{K}$ we define
\begin{itemize}
\item the domain $D(A)$ ; the vectors $x\in\mathcal{H}$ for which $Ax$ is
defined,
\item the range $R(A) = \left\{y\in\mathcal{K} : y=Ax \text{ for some
$x\in\mathcal{H}$} \right\}$,
\item the kernel $\ker(A) = \left\{ x\in\mathcal{H} : Ax = 0 \right\}$, and
\item the operator norm $\lpar A \rpar \;=
  \sup_{x\in D(A)\;:\; \parallel x\parallel=1}\left\{\lpar Ax \rpar \right\}$.
\end{itemize}
An operator is said to be \emph{densely defined} if $R(A)$ is dense in
$\mathcal{H}$.

In this work I will have need to consider the family of all operators that
can act on a Hilbert space $\mathcal{H}$ and produce an element of
$\mathcal{K}$. This space, denoted $\mathcal{L}(\mathcal{H}, \mathcal{K})$
turns out to be a Banach space (a complete normed linear space). The Banach
space is a generalisation of a Hilbert space, and thus some of the
properties of Hilbert spaces discussed here do not apply to the space of
operators.  However, many do. Banach spaces will often be denoted by $X$
and $Y$ but context will avoid confusion with operators. See Chapter~3 of
\cite{Reed1} for an introduction to Banach spaces.

\subsection{Invariance and reducing operators}

A subspace, $\mathcal{S}$, of $\mathcal{H}$ is called \emph{invariant} for
an operator $A$ if, for all $x\in\mathcal{S}$, $Ax\in\mathcal{S}$. That is,
action with $A$ on $\mathcal{S}$ does not take us out of $\mathcal{S}$.
Further to this definition, a set $\mathcal{S}\in\mathcal{H}$ is said to
\emph{reduce} an operator $A$ if both $\mathcal{S}$ and its ortho-complement
$\mathcal{H}\ominus\mathcal{S}$ are invariant subspaces for $A$.

\subsection{Cyclicity}

A vector $\phi$ is cyclic for an operator $A$ if and only if finite linear
combinations of elements of $\{A^n\phi\}_{n=0}^\infty$ are dense in
$\mathcal{H}$. This motivates the definition that a set $\mathcal{S}$ is
cyclic for $\mathcal{H}$ if and only if the smallest closed reducing
subspace of $\mathcal{H}$ containing $\mathcal{S}$ is $\mathcal{H}$. This
deserves some explanation. The existence of a cyclic vector means that by
acting with the operator $A$ and taking linear combinations of the results,
the whole Hilbert space $\mathcal{H}$ can be explored. Now consider the
reducing subspace $\mathcal{S}$. Action on elements of $\mathcal{S}$ with
$A$ always leaves us within $\mathcal{S}$. If repeated operations of this
fashion end up reaching all vectors in the full Hilbert space $\mathcal{H}$
(i.e.,\ there are cyclic vectors in $\mathcal{S}$) then the smallest reducing
subspace of $\mathcal{H}$ containing $\mathcal{S}$ is $\mathcal{H}$ itself.
Hence, the definition above for a set $\mathcal{S}$ to be cyclic corresponds
to the existence of at least one cyclic vector in $\mathcal{S}$.

\subsection{Limits}

If $A_n$ is a sequence of operators, s-lim~$A_n$ (also
$A_n\overset{s}{\rightarrow}A$) denotes the strong limit, defined by $\lpar
(A_n - A)g \rpar \rightarrow 0$ for all $g\in\mathcal{H}$.  w-lim~$A_n$
(also $A_n\overset{w}{\rightarrow}A$) denotes the weak limit, defined by $|
\langle A_ng,f \rangle - \langle Ag,f \rangle | \rightarrow 0$ for all $g,f
\in\mathcal{H}$. By the Schwartz inequality, the weak limit exists if the
condition is satisfied for $f=g$. I will also have need for the norm limit
of an operator, $A_n\overset{n}{\rightarrow}A$, defined by $\lpar A_n - A
\rpar \rightarrow 0$.

As the names suggest,
\begin{equation*}
A_n\overset{n}{\rightarrow}A \Rightarrow
A_n\overset{s}{\rightarrow}A \Rightarrow
A_n\overset{w}{\rightarrow}A
\end{equation*}
but the converse need not be true. A trivial application of the Schwartz
inequality demonstrates this.


\section{\label{sec:m_spec_thrm}The spectral theorem}

I refer the reader to Chapters~4, 6 and 7 in \cite{Reed1} for a full
discussion of the following results. Here I will skip over much of the
mathematical detail, only highlighting the pertinent definitions and
theorems.

The topic of spectral analysis is essentially an extension of the familiar
linear algebra results on matrices of complex numbers to the action of
operators on a Hilbert space---the spectrum is constructed
in the same way that eigenvalues of a matrix are constructed.
Many familiar mathematical expressions and definitions carry over, but
great care must be taken to not incorrectly infer results based on those
from finite dimensional linear algebra. The change from elements of
$\mathbb{C}^n$ to arbitrary vectors in a (possibly infinite dimensional)
Banach (or Hilbert) space allows for much more complex behaviour.

I will start by defining the spectrum for an operator $A: X \rightarrow X$,
mapping elements of the Banach space $X$ into the same space $X$. The space
of such operators $\mathcal{L}(X,X)$ is itself a Banach space. Note that I
generalise to consider the operator $A$ acting on a Banach space, rather
than a Hilbert space. It is normal practice in the literature to write
$\mathcal{L}(X)$ for $\mathcal{L}(X,X)$ when the target space for $A$ is
also the initial space. With the spectrum defined, I will then jump ahead
to the spectral theorem and the decomposition of the Hilbert space into
point, absolutely continuous and singularly continuous reducing subspaces
for a particular operator.

\subsection{The spectrum}

Consider $A\in\mathcal{L}(X)$. For $\alpha\in\mathbb{C}$, the operator
\begin{equation*}
(\alpha I - A)^{-1}
\end{equation*}
is called the \emph{resolvent} of $A$ at $\alpha$.

The resolvent set, $\rho(A)$, is defined by
\begin{equation*}
\rho(A)
  = \{\alpha \in \mathbb{C}: (\alpha I - A) \text{ is a bijection with a
  bounded inverse}\}\text{.}
\end{equation*}
If $\alpha\notin\rho(A)$ then $\alpha$ is in the spectrum, $\sigma(A)$, of
$A$. Essentially, the spectrum of $A$ is the set of $\alpha$s for which
the resolvent is not invertible. We now consider an element $x$ of the
Banach space $X$. If there exists an $\alpha\in\mathbb{C}$ such that
\begin{equation*}
Ax = \alpha x
\end{equation*}
then $x$ is an \emph{eigenvector} of $A$ and $\alpha$ is the corresponding
\emph{eigenvalue}. For now, note that it is possible to have
$\alpha\in\sigma(A)$ without $\alpha$ being an eigenvector. See
(p.~188, \cite{Reed1}).

\subsection{Spectral measures}

I now specialise to operations on Hilbert spaces. As I will be concerned
in this work with unitary operators, I state the following theorems in
terms of unitary operators. The literature, however, generally introduces
these concepts in terms of self-adjoint operators. As such, the references
in this section refer to results on self-adjoint operators, but the results
all pass through to unitary operators without much change.

Fix $A$ to be some unitary operator (typically it will be a
time-evolution operator) and choose $\psi \in \mathcal{H}$. As discussed
in Section~VII.2 in \cite{Reed1}, one can show that there exists a unique
measure, $\mu_\psi$, on the compact set $\sigma(A)$ with
\begin{equation}
\label{eq:m_fcalc}
\langle \psi, f(A)\psi \rangle
  = \int_{\sigma(A)} f(\alpha)\,d\mu_\psi\text{.}
\end{equation}
The left-hand side of the above equation is a typical inner product that a
physicist needs to calculate. On the right-hand side is an integral
in the complex plane, which we (hopefully!) have the tools to evaluate.
The measure $\mu_\psi$ is called the \emph{spectral measure} for the vector
$\psi$. We now begin to see a link between physics and the mathematics
presented earlier in \refsec{sec:m_meas}. The discussions have lead to a
rigorous (although I have not really shown the details) derivation
of the common physics practice of evaluating inner products via the
insertion of a ``complete set of states''. The flip-side is that beyond the
standard discrete (point) and continuous states found in most physics books,
we can now deal with arbitrary measures---point, absolutely continuous and
singularly continuous measures.

\subsection{The spectral theorem}

First, I briefly return to the notion of cyclicity introduced
earlier. Given an operator $A$ on a separable Hilbert space $\mathcal{H}$,
one can always find a decomposition of $\mathcal{H}$ into reducing 
subspaces
\begin{equation*}
\mathcal{H} = \bigoplus_{n=1}^N \mathcal{H}_n
\end{equation*}
such that for each subspace $\mathcal{H}_n$ there exists a
$\phi_n\in\mathcal{H}_n$ that is cyclic for $A$ restricted to operating
on $\mathcal{H}_n$. Note that in the above, $N$ can be finite or countably
infinite. See (Lemma~2, p.~226, \cite{Reed1}).

Defining $\mathbb{T}$ to be the unit circle in the complex plane, the
spectral theorem states that for a unitary operator, $A$, on
$\mathcal{H}$, there exist measures $\{\mu_n\}_{n=1}^N$
($N=1,2,\ldots$ or $\infty$) on $\sigma(A)$ and a unitary operator
\begin{equation*}
U: \mathcal{H} \rightarrow \bigoplus_{n=1}^N L^2(\mathbb{T}, d\mu_n)
\end{equation*}
such that
\begin{equation*}
\left(UAU^{-1}\psi\right)_n(\alpha) = \alpha\psi_n(\alpha)
\end{equation*}
where $\psi$ is a $N$ component vector
$(\psi_1(\alpha),\ldots,\psi_N(\alpha))
\in\bigoplus_{n=1}^N L^2(\mathbb{T}, d\mu_n)$. Essentially, there exists a
unitary operator $U$ that converts operation with $A$ on $\mathcal{H}$ into
a linear combination of multiplications by $\alpha$ on elements of the
complex unit circle.

After introducing a few more concepts, I will return to another form of the
spectral theorem which I will actually use for the rest of this work. It
turns out that it is also the most familiar to physicists and provides a
strong conceptual link between the mathematics here and the usual treatment
of quantum mechanics.

As discussed in \refsec{sec:m_meas}, $\mu_{\pp,\ac,\cs}$ are mutually
singular, and thus
\begin{equation*}
L^2(\mathbb{R}, d\mu) = L^2(\mathbb{R},d\mu_\pp)
  \oplus L^2(\mathbb{R},d\mu_\ac)
  \oplus L^2(\mathbb{R},d\mu_\cs)
\end{equation*}
where I have introduced the space $L^2(\mathbb{R})$, the completion of the
space $C(\mathbb{R})$ of real continuous functions.\footnote{The completion
implies the existence of a metric (or ``distance'' function). It is simply
$\lpar x - y \rpar$ for $x,y\in\mathcal{H}$. It is normal
practice to drop the measure, $d\mu$, when there can be no confusion, or
when it is not necessary for an understanding of the concept being
discussed.}
The spectral theorem says that operation with an operator $A$ on
$\mathcal{H}$ is always
equivalent to multiplication operations on the space $L^2(\sigma(A))$ for
some appropriate measure $\mu$. Thus, the decomposition of
$L^2(\mathbb{R}, d\mu)$ into point, absolutely continuous and singularly
continuous parts allows one to conclude that for an operator $A$ acting on
$\mathcal{H}$, one can always write
\begin{equation*}
\mathcal{H} = \mathcal{H}_\pp \oplus \mathcal{H}_\ac
 \oplus \mathcal{H}_\cs
\end{equation*}
where
\begin{align*}
\mathcal{H}_\pp &= \{\psi | \mu_\psi \text{ is pure point}\}
  = \overline{\{\alpha_n: \alpha_n \text{ is an eigenvalue of } A\}}
\text{,} \\
\mathcal{H}_\ac &= \{\psi | \mu_\psi
\text{ is absolutely continuous}\}\text{,} \\
\mathcal{H}_\cs &= \{\psi | \mu_\psi
\text{ is singularly continuous}\}\text{.}
\end{align*}
Each subspace reduces the operator $A$. Introducing the notation
$A \upharpoonright \mathcal{H}_\text{x}$ for the restriction of $A$
acting only on the subspace $\mathcal{H}_\text{x}$, we may conclude that $A
\upharpoonright \mathcal{H}_\pp$ has a complete set of eigenvectors while
$A \upharpoonright \mathcal{H}_\ac$ and $A \upharpoonright \mathcal{H}_\cs$
have only absolutely continuous and singularly continuous spectral measures
respectively.

The pure point, absolutely continuous and
singularly continuous spectrum for an operator are now defined by
\begin{equation*}
\sigma_\text{x}(A) = \sigma(A \upharpoonright \mathcal{H}_\text{x})\text{.}
\end{equation*}
It is possible to combine the sets in various ways to produce
the singular (point plus singular continuous) spectrum and the
continuous (absolutely continuous plus singularly continuous) spectrum 
\begin{align*}
\sigma_\sing(A) &= \sigma_\pp(A) \cup \sigma_\cs(A)\text{,} \\
\sigma_\cont(A) &= \sigma_\ac(A) \cup \sigma_\cs(A)\text{.}
\end{align*}
As already alluded to, the form of the spectral theorem above is still not
quite what is required in this work. By introducing the concept of
\emph{spectral projections} I will now rewrite the spectral
theorem in a way that is very familiar to the physicist.

\subsection{Spectral projections}

Section~VII.3 in \cite{Reed1} covers the following rigorously as does
Chapter~VII in \cite{Riesz}.

Spectral projections are, intuitively speaking, the building blocks for
``operator measures''. Just as we considered the integral
\begin{equation*}
\int f(x)\,d\alpha(x)
\end{equation*}
and characterised (decomposed) the measure function $\alpha(x)$ in terms
of point and continuous measures, here I wish to decompose an operator $A$
on the Hilbert space and write it as
\begin{equation*}
A = \int_{\sigma(A)} x\,dE(x)
\end{equation*}
where $E(x)$ is an ``operator measure''.

The \emph{characteristic function} $\chi_{\text{\tiny\itshape S}}(x)$ for
$S \subset \mathbb{R}$ and $x\in\mathbb{R}$ is defined by
\begin{equation*}
\chi_{\text{\tiny\itshape S}}(x) = \begin{cases}
  1 \text{ if } x \in S\text{,} \\
  0 \text{ if } x \notin S\text{.}
  \end{cases}
\end{equation*}
The characteristic function provides the foundation for the step-function
already introduced in \refsec{sec:m_meas} and is essential in a rigorous
development of integration.

The notion of the characteristic function is now extended to operators.
To extend the definition, we must first understand how to form functions of
operators.

A function of an operator is, in general, defined through a
limiting process of polynomials. Just as for functions on the real line,
a general function of an operator can be considered as the limit of an
appropriate sum of powers of the operator. The characteristic function is
just one example. See \cite{Riesz,Reed1} for the details and the
justification for why this formalism is self consistent, i.e.,\ why it
makes sense to talk of functions of an operator in this way. Note that this
fact is essentially what allowed us to write down \refeq{eq:m_fcalc} earlier.

For a unitary operator $A$ and a Borel (think ``measurable'') set
$\Omega\in\mathbb{T}$, the \emph{spectral projection},
$E_\Omega: \mathcal{H}\rightarrow\mathcal{H}$ is defined by
\begin{equation*}
E_\Omega \equiv \chi_\Omega(A)\text{.}
\end{equation*}
The spectral projections $E_\Omega$ are orthogonal projections as
$E_\Omega^2 = E_\Omega$. A family of spectral projections, $\{E_\Omega\}$,
has a number of important and rather intuitive properties that mirror the
properties of ordinary measures. For a family of sets $\Omega\in\mathbb{T}$,
the orthogonal projections have the following properties:
\begin{itemize}
\item Each $E_\Omega$ is an orthogonal projection,
\item $E_\emptyset = 0$; $E_{\mathbb{T}} = I$,
\item If $\Omega = \cup_{n=1}^\infty \Omega_n$ with $\Omega_n \cap \Omega_m
= \emptyset$ for all $n \neq m$, then
\begin{equation*}
E_\Omega = \text{s-}\lim_{N\rightarrow\infty}
  \left(\sum_{n=1}^\infty E_{\Omega_n}\right)\text{,}
\end{equation*}
\item $E_{\Omega_n}E_{\Omega_m} = E_{\Omega_n \cap \Omega_m}\text{.}$
\end{itemize}
These properties are exactly those you would expect for a measure---a
function that returns the ``size'' of a set. Any family of projections
satisfying the above properties is called a \emph{projection-valued
measure} and is, rather confusingly, also denoted by $E_\Omega$, where
it is now understood that this is a family of spectral projections, rather
than a single spectral projection. As expected, it turns out that for any
$\phi\in\mathcal{H}$,
\begin{equation*}
\langle \phi, E_\Omega \phi \rangle
\end{equation*}
is just an ordinary measure which one can integrate with respect to.
The projections, $E_\Omega$, are self-adjoint\footnote{Actually, they are
\emph{symmetric}. An operator $A$ is symmetric if, for
$f,g\in\mathcal{H}$, $(Af,g) = (f,Ag)$. For bounded operators, as
considered here, self-adjoint and symmetric are equivalent concepts and
hence the use of ``self-adjoint'' is perfectly acceptable.} and thus
$\langle \phi, E_\Omega \phi \rangle$ is real.

Armed with the spectral projections, we can now write an operator
(introducing the notation used for the rest of this work), $A$, simply
as\footnote{The subtle change of notation where I have replaced
$dE(\alpha)$ with $E(d\alpha)$ is a relic of the development of my work
and the conflicting inherited notation from \cite{Reed1} and
\cite{Howland87}. In the rest of this work, I generally use the notation
$E(d\alpha)$.}
\begin{equation*}
A = \int_{\sigma(A)} \alpha\,dE(\alpha)
  \equiv \int_{\sigma(A)} \alpha E(d\alpha)\text{.}
\end{equation*}
If $A$ is unitary, then the eigenvalues are on the unit circle in the
complex plane, and
\begin{equation*}
A = \int e^{-i\theta} E(d\theta)\text{.}
\end{equation*}
If $A$ is self-adjoint, then
\begin{equation*}
A = \int_{-\infty}^\infty \alpha E(d\alpha)\text{.}
\end{equation*}
In the physicist's language, the $E(\alpha)$ are the projections
$|\phi_n\rangle\langle\phi_n|$ formed from the eigenvectors
$|\phi_n\rangle$ for the operator $A$. If the eigenvectors are
discrete, $A$ is expressed in terms of a sum. If the eigenvectors are
continuous\footnote{We now know that it always eventuated that we had a
discrete or absolutely continuous spectrum in the standard physics
examples.} then $A$ is written in terms of a standard integral (with measure
$\alpha$ the Lebesgue measure).


\section{\label{sec:m_discussion}Some results and discussion}

Having presented most of the concepts and definitions required for the rest
of the work, I now summarise a few simple results that will be of use, and
discuss some consequences of the theory so far presented.

For any operator, $A$, the corresponding family $E(\alpha)$ form a general
resolution of the identity
\begin{equation*}
I = \int E(d\alpha)\text{.}
\end{equation*}
Any function of an operator can always be written as
\begin{equation*}
f(A) = \int_{\sigma(A)} f(\alpha) E(d\alpha)\text{.}
\end{equation*}
As the $E(d\alpha)$ are orthogonal projection operators, i.e.,\ $E^2 = E$,
we have the very useful result that
\begin{equation*}
\int \left|f(\alpha)\right|^2E(d\alpha)
 = \left|\int f(\alpha)E(d\alpha)\right|^2\text{.}
\end{equation*}
The $E(\alpha)$, being operator measures, can be decomposed into their
point, absolutely continuous and singularly continuous components. Thus, we
can decompose an operator $A$ into three components: $A_\pp$, $A_\ac$ and
$A_\cs$. We also form the singular and continuous parts of the operator
$A$,
\begin{align*}
A_\sing &= A_\pp + A_\cs\text{,} \\
A_\cont &= A_\ac + A_\cs\text{.}
\end{align*}
Another minor change in notation due to the way in which this work developed
is now introduced. As mentioned earlier in \refsec{sec:m_func_anal}, in a
mathematical context I usually use $x$ and $y$ to refer to elements of the
Hilbert space $\mathcal{H}$. The ordinary measure formed from a
projection-valued measure $E(\alpha)$ for the operator $A$ is now written
as $m_x$, rather than $\mu_x$. I define, for a vector $x\in\mathcal{H}$,
the measure
\begin{equation*}
m_x(S) = \langle E(S)x,x \rangle
\end{equation*}
where $S$ is a Borel set in $\sigma(A)$. Note that as $E(S)$ is self-adjoint
one is free to move it to the other side of the inner product.

Having obtained a full characterisation of operators in terms of their
decomposition into point, absolutely continuous and singularly continuous
parts, a final definition is  now introduced. The operator $A$ is
\emph{pure point} if and only if the eigenvectors of $A$ form a basis of
$\mathcal{H}$. That is, $A$ is pure point if and only if
$\mathcal{H}_\cont = \emptyset$ for the operator $A$.


\section{\label{sec:m_example}Two examples---hydrogen and the
harmonic oscillator}

To clearly link the preceding discussion back to familiar physics,
consider the operator $A$ to be the Hamiltonian, $H$, for hydrogen and
consider the spectrum, $\sigma{H}$, of $H$.

The bound states of hydrogen are a countable number of isolated, discrete
energies. Each energy corresponds to an eigenvalue of the system and the
set of these points makes up the point energy spectrum. The positive energy
scattering states form the continuous energy spectrum. Thus, the energy
spectrum for the hydrogen system consists of two disjoint parts: the
negative energy discrete (or ``point'') spectrum, and the positive energy
continuous spectrum. For hydrogen,
$\sigma_p(H) = \{\alpha_n ; \alpha_n \approx -13.6/n^2
\text{ for } n\in \mathbb{N}\}$, and $\sigma_{cont}(H) = (0,\infty)$.

The Hilbert space splits into two subspaces $\mathcal{H}_\pp$ and
$\mathcal{H}_\cont =  \mathcal{H}_\ac$. There is no singularly continuous
component to the Hamiltonian for hydrogen. This is typical of most physics
Hamiltonians. The singularly continuous component of an operator is rather
abstract and not commonly considered. In certain contexts however, we will
see that it becomes an essential tool in obtaining a better understanding
of the dynamics of those systems.

As another simple example, the harmonic oscillator quantum system has only
discrete energy levels, and thus is said to be ``pure point''. That is, the
eigenvectors of the harmonic oscillator form a basis of the Hilbert Space.

The Hamiltonian of a typical quantum mechanical system does not posses a
singularly continuous spectral component and thus, singular continuity is
not usually mentioned in texts on quantum mechanics.  The physical
interpretation of a singularly continuous component to the energy spectrum
is murky to say the least.


\section{\label{sec:m_summary}Summary}

I am now ready to move on and discuss why the spectrum of the Floquet
operator is an interesting object to study in the context of quantum chaos.
Results on the characterisation of the Floquet operator spectrum in terms
of properties of the perturbation to the base Hamiltonian can then be
presented.

\clearemptydoublepage
\chapter[Quantum chaos: the spectrum of the Floquet operator]
{\label{chap:spectrum}Quantum Chaos: The Spectrum of the Floquet Operator}
\minitoc



Having introduced the concept of the Floquet operator in
\refchap{chap:floquet} and the mathematical ideas of measure and
spectrum in \refchap{chap:maths}, I can now bring them together to discuss
the dynamics of periodically perturbed quantum systems. One
can argue that knowledge of the spectrum of the Floquet operator provides
great insight into the dynamics of the system of interest.

The links between spectral decomposition and dynamics is still an active
field of research. Model cases provide concrete examples of the links, but
the interplay between spectra and dynamics is far from fully understood.
Here, I outline the main results and provide references to the current
research efforts.

I first consider the case of time-independent systems where the spectrum of
the Hamiltonian is important. The RAGE theorem is the starting point for
the discussion. Time-dependent systems are then considered---in particular,
time-periodic systems whence the Floquet operator's spectrum is of
importance when considering the dynamics. The work of Yajima and Kitada
\cite{Yajima} on RAGE-like theorems is discussed, as well as the work of
Milek and Seba
\cite{Milek90.1}. The important contributions of Antoniou and Suchanecki
\cite{Antoniou02,Antoniou03} are also noted.  A good overview of the field
is provided by Combes \cite{Combes} and also Last \cite{Last}. The
introduction in the paper by Enss and Veseli\'{c} \cite{Enss} provides the
most physically intuitive discussion of the RAGE theorem and is essential
reading.

Finally, I also comment on a point of ambiguity seen in a number of papers.
I detail the problem in \refsec{sec:s_clarification} and suggest a possible
reason for why the confusion has survived for nearly twenty years.


\section{\label{sec:s_t_indep}Time-independent systems}

For time-independent systems, characterisations of the dynamics is linked
to the spectrum of the Hamiltonian through, in part, the RAGE theorem
\cite{Reed3}. A large literature considers these systems and the link
between dynamics and the spectrum of $H$. See \cite{Last} and the
references therein.

A quantum system is characterised by its energy eigenstates, requiring
knowledge of the spectrum of $H$. The time evolution of these states is
then examined. Thus, one considers terms like $e^{-iHt}\psi$, where
$\psi$ is spectrally decomposed with respect to the operator $H$.

The RAGE theorem simply states that states in the point subspace,
$\mathcal{H}_\pp(H)$, survive as time evolves, while states in the continuous
subspace, $\mathcal{H}_\cont(H)$, decay. Put another way, for systems with a
potential, states in the point subspace are \emph{bound states} and
essentially remain in a bounded region of space as time evolves. States in
the continuous subspace spread in space as time evolves---they are the
\emph{scattering states}.

The RAGE theorem is as follows \cite{Reed3}. Let $H$ be a self-adjoint
operator and $C$ a compact operator. Then for all $\psi \in
\mathcal{H}_\cont(H)$,
\begin{equation*}
\lim_{T\rightarrow\pm\infty}
  \frac{1}{T} \int_0^T \lpar C \exp (-itH) \psi \rpar^2\,dt = 0\text{.}
\end{equation*}
The square on the norm can be dropped by technical considerations
\cite{Reed3}.

To better understand the RAGE theorem, consider the case where $C$ is a
projection, namely \cite{Enss},
\begin{equation*}
C = F(|x| < R)\text{.}
\end{equation*}
$F(|x|<R)$ is the multiplication operator in $x$-space with the
characteristic function of the ball of radius $R$. The RAGE theorem then
says that for $\psi\in\mathcal{H}_\cont(H)$ and a wide class of potentials,
\begin{equation*}
\lim_{T\rightarrow\pm\infty}
  \frac{1}{T} \int_0^T \lpar F(|x|<R) \exp (-itH) \psi \rpar\,dt = 0
\end{equation*}
for any $R < \infty$. That is, as time evolves, the wave-function $\psi$
has negligible components within an arbitrarily sized ball in space. It
spreads to arbitrarily large spatial distances from its initial
location.

Conversely, for $\psi \in \mathcal{H}_\point(H)$, for all $\epsilon>0$,
there exists an $R(\epsilon)$ such that
\begin{equation*}
\sup_{t\in\mathbb{R}} \lpar F(|x|>R) \exp (-itH) \psi \rpar
  < \epsilon\text{.}
\end{equation*}
That is, there is some ball of radius $R$ such that the wave-function has
negligible components outside that ball for all time. The wave-function
remains \emph{localised} in space when $\psi\in\mathcal{H}_\point(H)$.

The RAGE theorem provides a course grained overview of the possible
dynamics. For $\psi\in\mathcal{H}_\cont(H)$ the rate of decay is further
dependant upon the characterisation of the continuous spectrum. Singularly
continuous states are typically expected to provide the ``chaotic''
behaviour, sitting somewhere between the ballistic (continuously
accelerated) absolutely continuous states and the recurrent point states.


\section{\label{sec:s_t_dep}Time-dependent systems}

First and foremost, I refer here to the work by Yajima and Kitada
\cite{Yajima}. For systems with a time-periodic Hamiltonian
(e.g.,\ kicked systems) a RAGE-like theorem exists, but now the spectrum of
interest is that of the Floquet operator, rather than the Hamiltonian.

The numerous discussions on the link between dynamics and the spectrum of
$H$ for time-independent systems then apply.

Yajima and Kitada show that for time-periodic systems the Floquet operator
spectrum
determines the dynamics in the same way that the Hamiltonian spectrum
determines the dynamics for time-independent systems.

For time-dependent systems as described in \refchap{chap:floquet}, the
following two results \cite{Yajima} apply. The decomposition of the
Hilbert space, $\mathcal{H}$, is with respect to the Floquet operator, $V$.
Just as in the work by Enss and Veseli\'{c} \cite{Enss}, Yajima and Kitada's
work \cite{Yajima} is
presented in the case where $C$ is the projection operator
\begin{equation*}
C = F(|x| < R)\text{.}
\end{equation*}
For $\psi \in \mathcal{H}_\cont(V)$,
\begin{equation*}
\lim_{N\rightarrow\pm\infty}
  \frac{1}{N} \sum_{n=0}^N \lpar C V^n \psi \rpar^2 = 0\text{.}
\end{equation*}
For $\psi \in \mathcal{H}_\point(V)$, for all $0<\epsilon<1$, there exists
an $R(\epsilon)$ such that
\begin{equation*}
\inf_n \lpar C V^n \psi \rpar \geq (1-\epsilon)\lpar \psi \rpar\text{.}
\end{equation*}

Simply put, for $\psi \in \mathcal{H}_\point(V)$, the norm does not decay.
After some number of kicks, $n$, the change in norm is arbitrarily small
and the state, initially localised in space, remains localised. The state
$\psi$ is a bound state of the system. If $\psi \in \mathcal{H}_\cont(V)$
the norm does decay and the state $\psi$ is a scattering state of the
system.

The work of Hogg and Huberman \cite{Hogg82,Hogg83} takes the point spectrum
part of the result one step further. They show that the full wave-function
``reassembles'' itself infinitely often when the system has a discrete
Floquet operator spectrum. The energy also shows recurrent behaviour and
the system is seen to be quasi-periodic.

The recurrence of the energy means that for a system to be chaotic,
characterised by a diffusive growth in energy, it must have a continuous
Floquet spectral component. The exact requirements are a topic of ongoing
research. An intuitive discussion is provided by Milek and Seba
\cite{Milek90.1} but it does have its flaws. I discuss this in the next
section.

\subsection{Milek and Seba's work}

The singular continuous spectrum of the Floquet operator can, and
does, exist in physical systems
of interest. With an understanding that Milek and Seba meant to refer to
Yajima and Kitada's RAGE-like theorem in \cite{Yajima} rather than the
RAGE theorem itself,
the argument presented in Section~II of their paper \cite{Milek90.1} shows
that if a system possesses a singularly continuous quasi-energy spectrum
then its energy growth over time \emph{may} be characteristic of a classically
chaotic system.  Thus, establishing the existence or otherwise of singular
continuous spectra of the Floquet operator is seen to be of central
importance to the question of whether or not a quantum mechanical system is
chaotic. It must be noted that the arguments presented by Milek and Seba
are acknowledged to be anything but rigorous---a point clearly established
by Antoniou and Suchanecki \cite{Antoniou02,Antoniou03} who split the
singularly continuous spectrum (and of course Hilbert space) into two
parts,
\begin{equation*}
\mathcal{H}_\cs = \mathcal{H}_\cs^{\text{Decaying}} \oplus
\mathcal{H}_\cs^{\text{Non-decaying}}\text{.}
\end{equation*}
The Hilbert space decomposition is reducing. Vectors in the decaying
subspace do not survive as $t\rightarrow\infty$
while those in the non-decaying subspace do survive. The singular
continuous part can act like either a point spectrum or an
absolutely continuous spectrum. On reflection, this is to be
expected. It is true, as argued by Milek and Seba, that parts of
the singularly continuous spectrum will display chaotic behaviour. But
there are also parts of the singularly continuous
spectrum which do not. The game is wide open in terms of the
details and subtleties, but there is no doubt that singularly
continuous states can manifest as ``chaotic like'' behaviour.

With the caveat that the described behaviour is not guaranteed, I now detail
the argument of Milek and Seba. It has been inherited from the argument in
\cite{Casati86}.

Consider the probability $p_{k,l}(n)$ of exciting the $k^{th}$ state
of $H_0$ after $n$ cycles/kicks to the $l^{th}$ state,
\begin{equation*}
\begin{split}
p_{k,l} &= \bigl|\langle k | V^n | l \rangle \bigr|^2 \\
 &\equiv \bigl| P_{k,l}(n)\bigr|^2\text{.}
\end{split}
\end{equation*}
To obtain the probability amplitude $P_{k,l}(n)$ I decompose
the Floquet operator $V$ into its point, singularly continuous and
absolutely continuous parts.
As $V$ is unitary we can write, for $\omega$ an eigenvalue,
\begin{equation*}
V | \omega \rangle = e^{i\omega}|\omega \rangle\text{.}
\end{equation*}
We have
\begin{equation*}
p_{k,l} = \left| \int_{\sigma(V)} e^{i\omega n}
  \langle k | E(d\omega) | l \rangle \right|^2\text{.}
\end{equation*}
Using \refeq{eq:m_meas_decomp} gives three parts,
\begin{equation*}
\begin{split}
P_{k,l} &= \int e^{i\omega n}\,d\mu_{k,l}(\omega) \\
  &= \sum_{\sigma_\pp(V)} e^{i\omega n} \langle k | \omega \rangle
  \langle \omega | l \rangle
  + \int_{\sigma_\ac(V)} e^{i\omega n} f_{k,l}(\omega)\,d\omega
  + \int_{\sigma_\cs(V)} e^{i\omega n}\,d\mu_{\cs(k,l)}(\omega)\text{.}
\end{split}
\end{equation*}
The transition $k \rightarrow l$ will only occur if a state, labelled by
$\omega$, links them. That is, for the sum part above, there must be at
least one state $|\omega\rangle$ such that
$\langle k | \omega \rangle \langle \omega | l \rangle \neq 0$.
This demonstrates the importance of the spectral nature of $V$.
If $V$'s spectrum is discrete (only the sum above remains) then the
$|\omega\rangle$ states are localised (like $\delta$-function spikes) and
hence they can connect only a few states of $H_0$. In this case one would
expect \emph{recurrent behaviour}. As already discussed, this is the
essence of Hogg and Huberman's work \cite{Hogg82,Hogg83}.


If the spectrum is absolutely continuous, the Riemann--Lebesgue lemma
(Theorem~IX.7, p.~10, \cite{Reed2}) shows that
\begin{equation*}
\lim_{n\rightarrow\infty} P_{k,l}(n) = 0 \quad \forall k,l\text{.}
\end{equation*}
The system is continuously accelerated and we obtain resonant energy
growth. This is not chaotic behaviour as the system's future evolution
will remain predictable.

If the spectrum of $V$ has a singularly continuous component the
possibilities for the dynamics are more varied. The RAGE theorem, as
presented in Reed and Simon's book (Theorem~XI.115, p.~341, \cite{Reed3}),
allows for a very slow diffusive growth in the energy in
the presence of a singularly continuous spectral component
(p.~343 and Problem~149 on p.~403, \cite{Reed3}).\footnote{Note that I
am referencing the time-independent theory, but as shown by Yajima and
Kitada the results flow through to the time-dependent theory which is of
interest here.} As pointed out by Antoniou and Suchanecki \cite{Antoniou02},
this behaviour is not guaranteed---a point which Milek
and Seba \cite{Milek90.1} seem to have missed or at least glossed over.
Anyway, it is certainly possible that
\begin{align*}
\lim_{M\rightarrow\infty} \sum_{n=0}^M p_{k,l}(n) &= \infty \text{,} \\
\lim_{M\rightarrow\infty} \frac{1}{M} \sum_{n=0}^M p_{k,l}(n) &= 0
\end{align*}
There is a very slow increase in the energy of the occupied state
over time. As the first equation diverges we see that the system spends an
infinite amount of time in the ``lower'' states.  That is, the system
energy does not grow resonantly, but continues to explore the lower energy
states as time progresses. The second equation shows, however, that on
average the system will eventually escape any fixed chosen state---the
system is not bounded in energy.

This slow energy growth (diffusive) behaviour is typical of that seen
in classically chaotic systems as mentioned in \refchap{chap:chaos}.

It should be noted that if there is no singularly continuous spectrum, then
the diffusive growth described is certainly \emph{not} possible. The
existence of a singularly continuous spectrum is a \emph{necessary}
condition for chaotic behaviour. It is not a \emph{sufficient}
condition.

\subsection{The quasi-energy self-adjoint operator, $K$}

I have shown in the previous chapters that time-periodic systems are
characterised by the Floquet operator---the unitary time-evolution
operator over one kick period.

There is an alternative way to access information on the spectral properties
and dynamics of such systems. Developed in papers by Howland
\cite{Howland74,Howland79,Howland89.1}, the self-adjoint quasi-energy
operator, or \emph{Floquet Hamiltonian},
\begin{equation*}
K = -id/dt + H(t)
\end{equation*}
turns out to provide a different way to access similar information to what
I am seeking from the unitary Floquet operator $V$. The spectrum of $K$
is directly related to the spectrum of $V$, as clearly shown in 
(p.~808, \cite{Bunimovich}).

$K$ also plays a central role in the work, already discussed, of Yajima and
Kitada \cite{Yajima}. To obtain the time-periodic equivalent of the RAGE
theorem, Yajima and Kitada first introduce $K$, apply the original RAGE
theorem results, and then convert them into the result on the spectrum of
the unitary Floquet operator.

A significant amount of the work on time-dependent systems utilises the
Floquet Hamiltonian, $K$, because, in some sense, working with $V$ proved
difficult.  The large body of knowledge on self-adjoint operators provides
a mature basis for proving theorems about $K$. As discussed in
\cite{Bunimovich}, the spectrum of $K$ is easily related algebraically to
that of $V$, so results on the spectrum for $K$ and $V$ are equivalent.

The trade off is that, especially for a physicist, $V$ is a far more
intuitive operator than $K$. The abstraction involved in working with $K$
must be balanced against the loss of a simple physical picture.


\section{\label{sec:s_clarification}A point of clarification}
\longpage
It must be noted that the concept of spectrum is associated with a
particular operator.  Typically, physicists talk of the energy spectrum,
associated with the Hamiltonian.  However, all operators (e.g.,\
Hamiltonian, Floquet etc.)\ have a spectrum. A failure to realise, or at the
very least to explain, this has lead to a number of potentially
misleading papers (see for example \cite{Milek90.1}, but
they inherited their argument from \cite{Casati86}) which used results on
the spectrum of the Hamiltonian in a discussion of the spectrum of the
Floquet operator.  While it is probable that the authors
are aware of the jump they have made, referring to the work of
\cite{Yajima}, rather than the original RAGE theorem, would be of
significant benefit.

Similarly, papers such as that by Guarneri \cite{Guarneri89} do not make it
clear which operator they are referring to in discussions of the spectrum.
The issue is one of language. When physicists refer to ``the spectrum'' in
quantum mechanics, it is generally assumed they mean ``the spectrum of the
Hamiltonian''.  That is, the spectrum has come to mean the ``energy
spectrum'' in the language of quantum mechanics. This, however, is not the
mathematical definition. The spectrum is associated with a particular
operator, as clearly discussed in \refchap{chap:maths}. This
misunderstanding has lead to a number of the more ``physical'' papers in
the literature incorrectly drawing conclusions from the rigorous
mathematical literature. It does however turn out that most of the
conclusions arrived at are valid. The time-independent system theorems
(such as the RAGE theorem) all have equivalent theorems in the
time-dependent theory (e.g.,\ the work of Yajima and Kitada, Hogg
and Huberman). That
no obvious numerical inconsistencies have arisen means that the subtle
flaws and potential misunderstandings in the physical literature have gone
largely unnoticed for close to twenty years. 

I believe that it is of paramount importance that authors exercise great
caution when discussing the spectrum of an operator. It should be made
clear which operator is being investigated, especially when it is not
the Hamiltonian.

The work of Yajima and Kitada \cite{Yajima} is, in this field, the key link
due to the fundamental importance of the RAGE theorem for time-independent
systems.  Unfortunately, it is scantly referenced outside the mathematical
literature. Increasing awareness of this important work would be greatly
beneficial.


\section{\label{sec:s_summary}Summary}

It is with the application of the RAGE-like theorem in mind \cite{Yajima},
that I undertook the following work on the analysis of the quasi-energy
spectrum of the class of Hamiltonians as defined by
\refeq{eq:f_hamil}. The aforementioned work by Milek and Seba
\cite{Milek90.1}, utilising the rank-$1$ work of Combescure, has shown the
manifestation of singularly continuous spectra in numerical simulations of
rank-$1$ kicked rotor quantum systems. The work now presented in
\refchap{chap:rankN} and \refchap{chap:comb} extends these results and
provides a rigorous mathematical basis to numerical
calculations on the time evolution of higher rank kicked quantum systems.

\clearemptydoublepage
\chapter[Spectral analysis of rank-N perturbed Floquet operators]
{\label{chap:rankN}Spectral Analysis of Rank-N Perturbed Floquet Operators}
\minitoc



This chapter constitutes the first part of the research undertaken in my
PhD. The aim is to characterise the spectrum of the Floquet operator for
kicked systems as defined by \refeq{eq:f_hamil}. The
method used parallels the investigation into the spectrum of the
Hamiltonian itself undertaken by Howland \cite{Howland87} and relies on the
mathematical background presented in \refchap{chap:maths}. This work is an
extension of a result of Combescure \cite{Combescure90}. The motivation for
investigating the spectrum of the Floquet operator has been discussed
in detail in \refchap{chap:spectrum}.


\section{\label{sec:r_outline}Outline and summary of results}

I will derive conditions on the time-periodic perturbations to the base
Hamiltonian for the spectrum of the Floquet operator to remain pure point.
Equation \refeq{eq:f_hamil} is replaced with a more ``technical'', but
equivalent form,
\begin{equation}
\label{eq:r_hamiltonian}
  H(t) = H_0 + A^*WA \sum_{n=0}^\infty \delta(t-nT)\text{,}
\end{equation}
where $A$ is bounded, $W$ is self adjoint and $H_0$ has pure point
(discrete) spectrum. In terms of \refeq{eq:r_hamiltonian}, the Floquet
operator is
\begin{equation}
\label{eq:r_floqop}
V=e^{iA^*WA / \hbar}e^{-iH_0 T / \hbar}\text{.}
\end{equation}
If $A$ is a rank-$1$ perturbation,
\begin{equation*}
\begin{split}
A &= | \psi\rangle\langle\psi | \\
W &= \lambda I
\end{split}
\end{equation*}
then I reproduce the work of Combescure \cite{Combescure90}. The vector
$|\psi\rangle$ is a linear combination of the orthonormal basis states,
$|\phi_n\rangle$, of the unperturbed Hamiltonian $H_0$
\begin{equation}
|\psi\rangle = \sum_{n=0}^\infty a_n|\phi_n\rangle\text{.}
\end{equation}
Combescure showed that if $\psi \in l_1(H_0)$, that is if
\begin{equation}
\sum_{n=0}^\infty |a_n| < \infty
\end{equation}
then the quasi-energy spectrum remains pure point for almost every
perturbation strength $\lambda$. I will generalise this result to all
finite rank perturbations
\begin{equation}
\begin{split}
\label{eq:r_rankNpert}
A &= \sum_{k=1}^N A_k = \sum_{k=1}^N | \psi_k\rangle\langle\psi_k |\text{,} \\
W &= \sum_{k=1}^N \lambda_k | \psi_k\rangle\langle\psi_k |
\end{split}
\end{equation}
where $\lambda_k \in \mathbb{R}$ and each vector $|\psi_k\rangle$ is a
linear combination of the $H_0$ basis states, $|\phi_n\rangle$,
\begin{equation}
\label{eq:r_psi_k}
|\psi_k\rangle = \sum_{n=0}^\infty (a_k)_n | \phi_n\rangle\text{.}
\end{equation}
The states $| \psi_k\rangle$ are orthogonal,
\begin{equation}
\label{eq:r_orth_states}
\langle\psi_k | \psi_l\rangle = \delta_{kl}\text{.}
\end{equation}
The basic result is that if each $|\psi_k\rangle$ is in $l_1(H_0)$, the
spectrum of $V$ will remain pure point for almost every perturbation
strength.

The perturbation for which I prove that the quasi-energy spectrum remains
pure point is, in fact, more general than the finite rank perturbation
presented above. The finite rank result is, however, the motivation for
undertaking this work.

Howland \cite{Howland87} showed that the Hamiltonian \refeq{eq:r_hamiltonian}
has a pure point spectrum if the $\psi_k$s are in $l_1(H_0)$. Here I
follow a similar argument, showing that the continuous part of the spectrum
of $V$ is empty, allowing one to conclude that the spectrum of $V$ must be
pure point.

Before proceeding, it should be mentioned that there are alternative routes
to results similar to those I present. As mentioned in \refsec{sec:s_t_dep},
associated with the unitary Floquet operator $V$ is the self-adjoint
Floquet Hamiltonian $K$ \cite{Howland74}. Utilising $K$ allows the self-adjoint work of Howland
\cite{Howland87} to be used directly. This was done by Howland himself
\cite{Howland89.1}. As my work is a unitary equivalent to the work of
Howland \cite{Howland87} the results obtained correspond to those
determined in \cite{Howland89.1}. The relationship between my work and
Howland's work \cite{Howland87,Howland89.1} is similar to the
relationship between the self-adjoint rank-$1$ work of Simon and Wolff
\cite{Simon86} and the unitary rank-$1$ work of Combescure
\cite{Combescure90}.

The techniques developed in this chapter provide new, general theorems
applicable to unitary operators and show that it is possible to develop
the theory of the spectrum of time-evolution operators directly, without
need for the techniques of \cite{Howland74} briefly mentioned earlier in
\refsec{sec:s_t_dep}.

In \refsec{sec:r_spectra} I will present the main theorems of the
chapter, concerned with establishing when systems of the form given by
\refeq{eq:r_hamiltonian} maintain a pure point quasi-energy
spectrum. Parallelling Howland's paper \cite{Howland87} on self-adjoint
perturbations of pure point Hamiltonians, the key ideas are those of
$U$-finiteness and the absolute continuity of the multiplication
operator $\mathbb{V}$. To establish the second of these concepts for the
unitary case (remember that we are concerned with the spectral properties
of the unitary time-evolution operator and not with the spectral properties
of the self adjoint Hamiltonian), I will require a modified version of
the Putnam--Kato theorem \cite{Reed4}. This, and associated theorems are the
topic of \refsec{sec:r_pk}. \refsec{sec:r_finite} uses the results of
\refsec{sec:r_spectra} and \refsec{sec:r_pk} to give the final results,
which are then discussed in \refsec{sec:r_discussion}.


\section{\label{sec:r_spectra}Spectral properties of the Floquet operator}

Let $U$ be a unitary operator on $\mathcal{H}$ and let $\mathcal{K}$ be an
auxiliary Hilbert space. Define the closed operator
$A:\mathcal{H}\rightarrow\mathcal{K}$, with dense domain $D(A)$. For our
purposes, $A$ bounded on $\mathcal{H}$ is adequate. I work with a
modification (multiplication by $e^{i\theta}$) of the resolvent of $U$,
\begin{equation}
\label{eq:r_Fdef}
F(\theta;U) = \left(1-Ue^{i\theta}\right)^{-1}
\end{equation}
and define for $\theta \in [0,2\pi)$ and $\epsilon > 0$ the
function $G_\epsilon:\mathcal{K}\rightarrow\mathcal{K}$,
\begin{equation}
\label{eq:r_Geps}
G_\epsilon(\theta;U,A)
  = AF^*(\theta_+;U)F(\theta_+;U)A^*\text{,}
\end{equation}
where $\theta_\pm = \theta \pm i \epsilon$. Let $J$ be a subset of $[0,2\pi)$. 


\begin{defn}[U-finite]
\tcaption{\textbf{Definition:} The operator $A$ is $U$-finite}
\label{defn:r_1}
The operator $A$ is $U$-finite if and only if the operator
$G_\epsilon(\theta;U,A)$ has a bounded extension to $\mathcal{K}$, and
\begin{equation}
\label{eq:r_strongG}
G(\theta;U,A) = \text{\textup{s-}}\underset{\epsilon\downarrow 0}{\lim}
  \;G_\epsilon(\theta;U,A)
\end{equation}
exists for a.e.\ $\theta \in J$.
\end{defn}

We define the function
\begin{equation}
\begin{split}
\label{eq:r_delta_epsilon}
\delta_\epsilon(t) &= \frac{1}{2\pi}\left(
    \sum_{n=0}^{\infty}e^{in(t+i\epsilon)} +
    \sum_{n=-\infty}^0 e^{in(t-i\epsilon)} - 1\right) \\
  &= \frac{1}{2\pi}\frac{1-e^{-2\epsilon}}
    {1-2e^{-\epsilon}\cos(t)+e^{-2\epsilon}}\text{.}
\end{split}
\end{equation}
The limit as $\epsilon\rightarrow 0$ of $\delta_\epsilon(t)$ is a series
representation of the $\delta$-function. The proof is based on showing that
\begin{equation*}
\underset{\epsilon\downarrow 0}{\lim}
  \int_{-\pi}^\pi g(t)\delta_\epsilon(t)\,dt = 0
\end{equation*}
where $g(t) = f(t) - f(0)$ and $f(t)$ is bounded in $(-\pi,\pi)$. We split
the integral into three parts,
$\int_{-\pi}^{-\xi} + \int_{-\xi}^\xi + \int_\xi^\pi$. We must assume that
$f(t)$ is continuous at $t=0$ (otherwise $\int f(t)\delta(t)\,dt$ is not well
defined) so that
\begin{equation*}
\forall\eta, \exists \xi > 0 \text{ s.t. }
\forall t, |t|<\xi \text{ we have } |f(t)-f(0)|<\eta\text{.}
\end{equation*}
We have
\begin{equation}
\label{eq:r_delta_split}
\int_{-\pi}^{\pi}g(t)\delta_\epsilon(t)\,dt
  = \int_{-\pi}^{-\xi}g(t)\delta_\epsilon(t)\,dt
  + \int_{-\xi}^{\xi}g(t)\delta_\epsilon(t)\,dt
  + \int_{\xi}^{\pi}g(t)\delta_\epsilon(t)\,dt\text{.}
\end{equation}
Consider the third term in \refeq{eq:r_delta_split}. For $\xi\leq t\leq\pi$,
$\cos t < \cos \xi$, so
\begin{equation*}
1+e^{-2\epsilon}-2e^{-\epsilon}\cos t
  \geq 1+e^{-2\epsilon}-2e^{-\epsilon}\cos \xi\text{.}
\end{equation*}
Therefore
\begin{equation*}
\left|\int_\xi^\pi g(t)\delta_\epsilon(t)\,dt\right|
  \leq \int_\xi^\pi \left|g(t)\right| \frac{1}{2\pi}
  \frac{1-e^{-2\epsilon}}{1-2e^{-\epsilon}\cos\xi +
  e^{-2\epsilon}}\,dt\text{.}
\end{equation*}
Since $g(t)$ is also bounded in $(-\pi,\pi)$, we have
\begin{equation*}
\left|g(t)\right| \leq K \text{ for } t\in(-\pi,\pi)
\end{equation*}
for some $K\in\mathbb{R}$. Thus,
\begin{equation*}
\begin{split}
\left|\int_\xi^\pi g(t)\delta_\epsilon(t)\,dt\right|
  &\leq \frac{K}{2\pi}
    \frac{1-e^{-2\epsilon}}{1-2e^{-\epsilon}\cos\xi + e^{-2\epsilon}}
    \int_\xi^\pi dt \\
  &\leq \frac{K\pi}{2\pi}
    \frac{1-e^{-2\epsilon}}{1-2e^{-\epsilon}\cos\xi +
    e^{-2\epsilon}}\text{.}
\end{split}
\end{equation*}
Taking $\epsilon\rightarrow 0$, the denominator
\begin{equation*}
1-2e^{-\epsilon}\cos\xi + e^{-2\epsilon} \rightarrow 2(1-\cos\xi) \neq 0
\end{equation*}
as $\xi>0$. The numerator $1-e^{-2\epsilon}\rightarrow 0$. Therefore,
\begin{equation*}
\left|\int_\xi^\pi g(t)\delta_\epsilon(t)\,dt\right|\rightarrow 0
  \text{ as } \epsilon\rightarrow 0\text{.}
\end{equation*}
Similarly, the first term in \refeq{eq:r_delta_split},
\begin{equation*}
\int_{-\pi}^{-\xi} g(t)\delta_\epsilon(t)\,dt\rightarrow 0
  \text{ as } \epsilon\rightarrow 0
\end{equation*}
tends to zero. Now consider the second term in \refeq{eq:r_delta_split}. We
have
\begin{equation*}
\left| \int_{-\xi}^\xi g(t)\delta_\epsilon(t)\,dt \right|
  \leq \int_{-\xi}^\xi \left| g(t) \right| \delta_\epsilon(t)\,dt
\end{equation*}
as $\delta_\epsilon(t)$ is a positive function. As $f(t)$ is continuous at
$t=0$, $\left| g(t) \right| \leq \eta$ for all $|t| < \xi$, so
\begin{equation*}
\begin{split}
\int_{-\xi}^\xi \left| g(t) \right| \delta_\epsilon(t)\,dt
  &\leq \eta \int_{-\xi}^\xi \delta_\epsilon(t)\,dt \\
  &\leq \eta \int_{\pi}^\pi \delta_\epsilon(t)\,dt\text{,}
\end{split}
\end{equation*}
where we have again used the positivity of $\delta_\epsilon(t)$ to extend
the limits of the integral. This integral is equal to one by
((3.792.1), p.~435, \cite{Gradshteyn}) so we conclude that
\begin{equation*}
\left|\int_{-\xi}^\xi g(t)\delta_\epsilon(t)\,dt\right| \leq \eta\text{,}
\end{equation*}
which can be made as small as desired and thus has the limit zero.
We have shown that
\begin{equation*}
\underset{\epsilon\downarrow 0}{\lim}
  \int_{-\pi}^\pi \left[f(t)-f(0)\right]\delta_\epsilon(t)\,dt = 0\text{.}
\end{equation*}
Therefore,
\begin{equation*}
\begin{split}
\underset{\epsilon\downarrow 0}{\lim}
  \int_{-\pi}^\pi f(t)\delta_\epsilon(t)\,dt
  &= \underset{\epsilon\downarrow 0}{\lim}
  \int_{-\pi}^\pi f(0)\delta_\epsilon(t)\,dt
  + \int_{-\pi}^\pi \left[f(t)-f(0)\right]\delta_\epsilon(t)\,dt \\
  &= \underset{\epsilon\downarrow 0}{\lim}
  \int_{-\pi}^\pi f(0)\delta_\epsilon(t)\,dt \\
  &= f(0)
\end{split}
\end{equation*}
and we have proved that
$\lim_{\epsilon\downarrow 0}\delta_\epsilon(t)$ as defined in
\refeq{eq:r_delta_epsilon} is an appropriate series representation of
the $\delta$-function.

Given \refeq{eq:r_delta_epsilon} and the spectral decomposition of $U$, we
may write
\begin{equation}
\begin{split}
\label{eq:r_delta_resolvent}
\delta_\epsilon\left(1-Ue^{i\theta}\right)
  &= \int\delta_\epsilon\left(1-e^{i(\theta-\theta')}\right)E(d\theta') \\
  &= \int\delta_\epsilon(\theta-\theta')E(d\theta') \\
  &= \frac{1}{2\pi}\int E(d\theta')
    \left[\sum_{n=0}^\infty e^{in(\theta_+-\theta')}
    + \sum_{n=0}^\infty e^{-in(\theta_--\theta')} -1 \right] \\
  &= \frac{1}{2\pi} \left[ \sum_{n=0}^\infty e^{in\theta_+}U^n
    + \sum_{n=0}^\infty e^{-in\theta_-}\left(U^*\right)^n - 1 \right] \\
  &= \frac{1}{2\pi}\bigl[F(\theta_+;U) + F^*(\theta_+;U) -1 \bigr] \\
  &= \frac{1}{2\pi}\left(1-e^{-2\epsilon}\right)
    F^*(\theta_+;U)F(\theta_+;U)\text{.}
\end{split}
\end{equation}
The existence of a non-trivial $U$-finite operator will have important
consequences for the spectrum of the Floquet operator $V$. We introduce the
set
\begin{equation*}
N(U,A,J) = \{\theta\in J : 
  \text{s-}\underset{\epsilon\downarrow 0}{\lim}
  G_\epsilon(\theta;U,A) \text{ does not exist} \}
\end{equation*}
of measure zero, which enters the theorem. I will often refer to
this set simply as $N$ during proofs.


\begin{thrm}
\tcaption{\textbf{Theorem:} The unitary time-evolution operator,
$U=e^{-iH_0 T}$ has $\sigma_\ac(U) = \emptyset$ and $\sigma_\sing(U)$ is
supported on the set $N$}
\label{thrm:r_2}
If $A$ is $U$-finite on $J$ and $R(A^*)$ is cyclic for $U$, then
\begin{enumerate}
\item $U$ has no absolutely continuous spectrum in $J$, and
  \label{thrm:r_2a}
\item the singular spectrum of $U$ in $J$ is supported by $N(U,A,J)$.
  \label{thrm:r_2b}
\end{enumerate}
\end{thrm}


\emph{Proof.} (\ref{thrm:r_2a}) Following Howland, note that the
absolutely continuous spectral measure, $m_y^{ac}(J)$, is the
$\epsilon\rightarrow 0$ limit of
$\left\langle\delta_\epsilon\left(1-Ue^{i\theta}\right)y,y\right\rangle$
for $\theta \in J$.
If $y\in\mathcal{H}$ is in $R(A^*)$, allowing one to write $y=A^*x$ for some
$x\in\mathcal{K}$, then
\begin{equation*}
\begin{split}
\underset{\epsilon\downarrow 0}{\lim}
  &\left\langle\delta_\epsilon\left(1-Ue^{i\theta}\right)y,y\right\rangle \\
  &= \underset{\epsilon\downarrow 0}{\lim}
    \left\langle\delta_\epsilon\left(1-Ue^{i\theta}\right)A^*x,A^*x
    \right\rangle \\
  &= \underset{\epsilon\downarrow 0}{\lim}\frac{\epsilon(1-\epsilon)}{\pi}
    \left\langle A F^*(\theta_+;U)F(\theta_+;U) A^*x,x\right\rangle \\
  &= \underset{\epsilon\downarrow 0}{\lim}\frac{\epsilon(1-\epsilon)}{\pi}
    \left\langle G_\epsilon(\theta;U,A)x,x\right\rangle = 0
\end{split}
\end{equation*}
for a.e.\ $\theta\in J$. The set $\mathcal{Y}$ of vectors $y$ for which
$m_y^{ac}(J)=0$ is a closed reducing subspace of $\mathcal{H}$, and by
construction contains the cyclic set $R(A^*)$ as a subset. Because
$\mathcal{Y}$ is invariant, finite linear combinations of action with $U^n$
leaves us in $\mathcal{Y}$. Due to the cyclicity, these same linear
combinations allow us to reach any $y\in\mathcal{H}$. Thus, the set
$\mathcal{Y}$ of vectors $y$ with $m_y^{ac}(J)=0$ must be the whole Hilbert
space $\mathcal{H}$. So there is no absolutely continuous spectrum of $U$
in $J$.

\vspace{\baselineskip}


(\ref{thrm:r_2b}) A theorem of de~la~Vall\'{e}e Pousin
((9.6), p.~127, \cite{Saks}) states that the singular part of the
spectrum of a function is
supported on the set where the derivative is infinite. In our case, this
corresponds to finding where $m_y(d\theta)\rightarrow\infty$. We calculate
\begin{equation*}
\begin{split}
\underset{\epsilon\downarrow 0}{\lim}
  \langle \delta_\epsilon\left(1-Ue^{i\theta}\right)y,y\rangle
  &= \int\delta(\theta - \theta') \langle E(d\theta')y,y\rangle \\
  &= \int\delta(\theta - \theta') m_y(d\theta') \\
  &= m_y(d\theta)\text{.}
\end{split}
\end{equation*}

Thus, $m_y^s = m_y^{sc} + m_y^{pp}$ is supported on the set where
\begin{equation}
\label{eq:r_vallee}
\underset{\epsilon\downarrow 0}{\lim}
  \left\langle\delta_\epsilon\left(1-Ue^{i\theta}\right)y,y\right\rangle
  = \infty\text{.}
\end{equation}
From the proof to part (\ref{thrm:r_2a}), if $y=A^*x$ then the limit
\refeq{eq:r_vallee} is zero for $\theta \in J$, $\theta \notin N$, so $m_y^s$
in $J$ must be supported by $N$. The set of vectors $y$ with
$m_y^s(J \cap N^c) \equiv m_y^s (J \sim N) = 0$ is closed, invariant and
contains $R(A^*)$, so must be $\mathcal{H}$ by the argument above. Thus,
the singular spectrum of $U$ is supported on the set $N$.\proofend


Now define a new operator, $Q(z):\mathcal{K}\rightarrow\mathcal{K}$,
\begin{equation*}
Q(z) = A(1-Uz)^{-1}A^*\text{.}
\end{equation*}
Note that
\begin{equation}
\label{eq:r_QfromF}
Q\left(e^{i\theta_\pm}\right) = AF(\theta_\pm;U)A^*\text{.}
\end{equation}
$Q(z)$ is clearly well defined for $|z|\neq 1$. Proposition~\ref{prop:r_3}
shows that the definition can be extended to $|z|=1$.


\begin{prop}
\tcaption{\textbf{Proposition:} The self-adjoint operator $Q$ is bounded and
$Q_\epsilon \rightarrow Q$ strongly as $\epsilon\rightarrow\infty$}
\label{prop:r_3}
Let $A$ be bounded. If $\theta \in J$, but $\theta \notin N(U,A,J)$, then
\begin{enumerate}
\item the operator $Q\left(e^{i\theta}\right)
  = A\left(1-Ue^{i\theta}\right)^{-1}A^*$
  is bounded on $\mathcal{K}$, and
  \label{prop:r_3a}
\item one has $\text{\textup{s-}}\underset{\epsilon\downarrow 0}{\lim}
  Q\left(e^{\pm i(\theta\pm i\epsilon)}\right)
  = Q\left(e^{\pm i\theta}\right)$.
  \label{prop:r_3b}
\end{enumerate}
\end{prop}


\emph{Proof.} (\ref{prop:r_3a}) Without loss of generality, take
$\theta = 0$ ($z=1$). By Theorem~\ref{thrm:r_2},
$e^{-i0} \notin \sigma_p(U)$, so $(1-Ue^{i0})^{-1}$ exists as a
densely defined operator. As $A$ is a bounded operator, it suffices to show
that $\left(1-Ue^{i0}\right)^{-1}A^*$ is bounded. We have
\begin{equation}
\begin{split}
\label{eq:r_Qpbounded}
  \lpar \left(1-Ue^{i0_+}\right)^{-1}A^*x \rpar^2
  &= \langle F(0_+;U)A^*x, F(0_+;U)A^*x \rangle \\
  &= \langle AF^*(0_+;U)F(0_+;U)A^*x,x \rangle \\
  &= \langle G_\epsilon(0;U,A)x,x \rangle \leq C|x|^2
    \text{ (as $\theta \notin N$)}
\end{split}
\end{equation}
for some real constant $C$.
If $y=A^*x$, noting $U = \int e^{-i\theta}E(d\theta)$, we also have
\begin{equation*}
  \lpar \left(1-Ue^{i0_+}\right)^{-1}A^*x \rpar^2 =
    \int\left(\frac{1}{1-e^{-i\theta}e^{-\epsilon}}\right)
    \left(\frac{1}{1-e^{i\theta}e^{-\epsilon}}\right)
    \langle E(d\theta)y,y\rangle\text{.}
\end{equation*}
In light of \refeq{eq:r_Qpbounded}, $\epsilon$ may safely be taken to zero
to obtain
\begin{equation}
\label{eq:r_Q_bounded}
\int\left(\frac{1}{1-e^{-i\theta}}\right)\left(\frac{1}{1-e^{i\theta}}\right)
  \langle E(d\theta)y,y \rangle \leq C\lpar x \rpar^2 < \infty\text{.}
\end{equation}
From \refeq{eq:r_Q_bounded}, we have
\begin{multline}
\int\left(\frac{1}{1-e^{-i\theta}}\right)\left(\frac{1}{1-e^{i\theta}}\right)
  \langle E(d\theta)y,y \rangle \\
  = \left\langle [1-U]^{-1}y,[1-U]^{-1}y\right\rangle \leq C\lpar x \rpar^2
    < \infty
\end{multline}
so $y\in D\left[(1-U)^{-1}\right]$. Thus, $Q(1) = A(1-U)^{-1}A^*$
is defined on all $\mathcal{K}$ and bounded.
\vspace{\baselineskip} 


(\ref{prop:r_3b}) For $y \in D\left((1-U)^{-1}\right)$, we show that the
difference between $Q\left(e^{\pm i(0 \pm i\epsilon)}\right)$ and
$Q\left(e^{\pm i0}\right)$ tends to zero as $\epsilon\rightarrow 0$. Again,
due to the boundedness of $A$, we need only show that
\begin{equation*}
\lpar \left( \left(1-Ue^{i0_+}\right)^{-1} - \left(1-U\right)^{-1} \right)
  A^*x \rpar
\end{equation*}
tends to zero. Consider
\begin{equation}
\begin{split}
\label{eq:r_lim_Q_exists}
\bigl| &\bigl(1-Ue^{-\epsilon}\bigr)^{-1}y - (1-U)^{-1}y \bigr|^2 \\
  &=\left| \int \left(1-e^{-i\theta}e^{-\epsilon}\right)^{-1} E(d\theta)y -
    \int \left(1-e^{-i\theta}\right)^{-1} E(d\theta)y \right|^2 \\
  &=\left| \int\left(\frac{1}{1-e^{-i\theta}e^{-\epsilon}} -
    \frac{1}{1-e^{-i\theta}}\right) E(d\theta)y \right|^2 \\
  &=\int \left|\frac{1}{1-e^{-i\theta}e^{-\epsilon}}-
    \frac{1}{1-e^{-i\theta}} \right|^2 \langle E(d\theta)y,y
    \rangle\text{.}
\end{split}
\end{equation}
To show that this has a limit of zero, write the numerical factor in
\refeq{eq:r_lim_Q_exists} as
\begin{equation*}
\begin{split}
\left|\frac{1}{1-e^{-i\theta}e^{-\epsilon}}-\frac{1}{1-e^{-i\theta}} \right|^2
  &= \left|\frac{e^{-i\theta}\left(1-e^{-\epsilon}\right)}
    {\left(1-e^{-i\theta}e^{-\epsilon}\right)\left(1-e^{-i\theta}\right)}
    \right|^2 \\
  &= \frac{1}{\left(1-e^{i\theta}\right)\left(1-e^{-i\theta}\right)}
    \left(\frac{\left(1-e^{-\epsilon}\right)^2}
    {1-2e^{-\epsilon}\cos\theta+e^{-2\epsilon}}\right)\text{.}
\end{split}
\end{equation*}
Equation \refeq{eq:r_lim_Q_exists} now equals
\begin{equation}
\label{eq:r_Q_diff_zero}
\int \left(
  \frac{\left(1-e^{-\epsilon}\right)^2}
  {1-2e^{-\epsilon}\cos\theta+e^{-2\epsilon}}\right)
  \frac{\langle E(d\theta)y,y \rangle}
  {\left(1-e^{-i\theta}\right)
  \left(1-e^{i\theta}\right)}\text{.}
\end{equation}
The first factor is bounded and tends to zero for $\theta \neq 0$.
The second factor is the measure from \refeq{eq:r_Q_bounded}. Clearly, away
from the origin, the integral tends to zero. About
the origin, some care must be taken to show that there is no contribution to
the integral.

Using \refeq{eq:r_delta_epsilon}, we have
\begin{equation*}
\frac{\left(1-e^{-\epsilon}\right)^2}
  {1-2e^{-\epsilon}\cos\theta+e^{-2\epsilon}}
  = \frac{\left(1-e^{-\epsilon}\right)^2}{1-e^{-2\epsilon}}
    2\pi\delta_\epsilon(\theta)\text{.}
\end{equation*}
On substitution into \refeq{eq:r_Q_diff_zero}, we obtain
\begin{equation*}
\frac{\left(1-e^{-\epsilon}\right)^2}{1-e^{-2\epsilon}} 2\pi
  \int_{-\alpha}^\alpha \delta_\epsilon(\theta)
    \frac{m_y(d\theta)}{2(1-\cos\theta)}
  = \frac{\left(1-e^{-\epsilon}\right)^2}{1-e^{-2\epsilon}} 2\pi
       \int_{-\alpha}^\alpha \frac{d\Theta_\epsilon}{d\theta}
       \frac{g_y(\theta)}{2(1-\cos\theta)}\,d\theta\text{.}
\end{equation*}
The function
$\Theta_\epsilon(\theta) = \int \delta_\epsilon(\theta')\,d\theta'$ is the
step function in the
$\epsilon\rightarrow 0$ limit. For non--zero $\epsilon$ it is positive,
monotonic, increasing and bounded by unity. As $\theta\notin N$ I have
also written $m_y(d\theta) = g_y(\theta)d\theta$ for some well behaved
positive  function $g_y(\theta)$. By integration by parts (see
p.~32, \cite{jeffreys} for existence conditions, which are satisfied) we
obtain
\begin{equation*}
\frac{\left(1-e^{-\epsilon}\right)^2}{1-e^{-2\epsilon}} 2\pi \left\{
  \left[ \Theta_\epsilon(\theta)
    \frac{g_y(\theta)}{2(1-\cos\theta)}\right]^\alpha_{-\alpha} -
  \int_{-\alpha}^\alpha \Theta_\epsilon(\theta)
    \frac{d}{d\theta}\frac{g_y(\theta)}{2(1-\cos\theta)}\,d\theta
    \right\}\text{.}
\end{equation*}
The first term within the curly braces is clearly some finite value. The
second term is less than
\begin{equation*}
\int_{-\alpha}^\alpha
  \frac{d}{d\theta} \frac{g_y(\theta)}{2(1-\cos\theta)}\,d\theta
 = \left[ \frac{g_y(\theta)}{2(1-\cos\theta)}\right]^\alpha_{-\alpha}
\end{equation*}
from the properties of the $\Theta_\epsilon$ function mentioned above. As
with the first term, it is clearly some finite value. Noting that
\begin{equation*}
\underset{\epsilon\downarrow 0}{\lim}
  \frac{\left(1-e^{-\epsilon}\right)^2}{1-e^{-2\epsilon}} = 0\text{,}
\end{equation*}
part (\ref{prop:r_3b}) follows.\proofend


\begin{thrm}
\tcaption{\textbf{Theorem:} The Floquet operator $V$ has
$\sigma_\ac(V) = \emptyset$ and $\sigma_\cs(V)$ is supported on the
set $N\cup M$}
\label{thrm:r_4}
Let A be bounded and $U$-finite on $J$, with $R(A^*)$ cyclic for $U$. Let
$W$ be bounded and self-adjoint on $\mathcal{K}$, and define the Floquet
operator,
\begin{equation*}
V = e^{iA^*WA/\hbar}U\text{.}
\end{equation*}
Assume that for $|z|  \neq 1$, $Q(z)$ is
compact, and that $Q\left(e^{\pm i(\theta \pm i\epsilon)}\right)$ converges
to $Q\left(e^{\pm i\theta}\right)$ in operator norm as $\epsilon\to 0$ for
a.e.\ $\theta$ in $J$. Define the set
\begin{equation*}
M(U,A,J) = \{\theta\in J : Q\left(e^{\pm i(\theta \pm i0)}\right)
 \text{ does not exist in norm} \}\text{.}
\end{equation*}
Then
\begin{enumerate}
\item $V$ has no absolutely continuous spectrum in $J$, and
  \label{thrm:r_4a}
\item the singular continuous part of the spectrum of $V$ in $J$ is
supported by the set $N(U,A,J) \cup M(U,A,J)$.
  \label{thrm:r_4b}
\end{enumerate}
\end{thrm}


\emph{Proof.} (\ref{thrm:r_4a}) 
For convenience, write the Floquet operator as
\begin{equation*}
V = (1+A^*ZA)U\text{,}
\end{equation*}
where $Z$ is defined appropriately by requiring\footnote{For the
rank-N perturbation case where
$W=\sum_{k=1}^N \lambda_k|\psi_k\rangle\langle\psi_k|$ and
$A = \sum_{k=1}^N |\psi_k\rangle\langle\psi_k|$, we have
$Z= \sum_{k=1}^N (\exp(i\lambda_k / \hbar)-1)|\psi_k\rangle\langle\psi_k|$.}
$\exp(iA^*WA / \hbar) = 1+A^*ZA$.
Noting \refeq{eq:r_Fdef} and \refeq{eq:r_QfromF} allows one to define
\begin{equation*}
\begin{split}
Q_1\left(e^{i\theta}\right) &= AF(\theta;V)A^* \\
  &= A\left(1-Ve^{i\theta}\right)^{-1}A^*\text{.}
\end{split}
\end{equation*}
Consider some vector $y' \in \mathcal{H}$. 
$Ay'= x \in \mathcal{K}$ is defined for such $y'$. $A^*x = y''$ is some
vector in $\mathcal{H}$. The cyclicity of $R(A^*)$ means that action with
linear combinations of powers of $U$ on $y''$ allows one to obtain any
$y \in \mathcal{H}$, the original $y'$ being one of them. Thus, we have a
construction of $A^{-1}$, namely, operation with $A^{*}$ followed by the
linear combination of powers of $U$. As $y'$ was arbitrary, $A^{-1}$ exists
for all $y \in \mathcal{H}$. This allows one to introduce $I=A^{-1}A$ in what
follows.\footnote{The particular choice of $A$ as a projection in
\refeq{eq:r_rankNpert} does not have an inverse, but I will show in
\refsec{sec:r_finite} that one can define a subspace of $\mathcal{H}$ on which
$R(A^*)$ is cyclic, and apply this theorem.}

We now proceed by use of the resolvent equation,
\begin{equation}
\begin{split}
\label{eq:r_q1-q}
Q_1 &- Q \\
  &=
  A\left\{\frac{1}{1-Ve^{i\theta}} - \frac{1}{1-Ue^{i\theta}}\right\}A^* \\
  &= A\left\{\frac{1}{1-Ve^{i\theta}}(V-U)e^{i\theta}\frac{1}{1-Ue^{i\theta}}
    \right\}A^* \\
  &= A\left\{\frac{1}{1-Ve^{i\theta}}\left([1+A^*ZA-1]Ue^{i\theta}\right)
    \frac{1}{1-Ue^{i\theta}}\right\}A^* \\
  &= A\left\{\frac{1}{1-Ve^{i\theta}}\left(A^*ZAUe^{i\theta}\right)
    \frac{1}{1-Ue^{i\theta}}\right\}A^* \\
  &= Q_1\left(e^{i\theta}\right)ZAUA^{-1}
    e^{i\theta}Q\left(e^{i\theta}\right)\text{.}
\end{split}
\end{equation}
Thus, briefly using $L = ZAUA^{-1}e^{i\theta}$ for clarity, we have
\begin{alignat}{2}
\label{eq:r_q1_inverse_q}
&&LQ_1 - LQ &= LQ_1LQ \notag \\
&\Rightarrow&\qquad 1 - LQ + LQ_1 - LQ_1LQ &= 1 \notag \\
&\Rightarrow& (1+LQ_1)(1-LQ) &= 1 \notag \\
&\Rightarrow& 1+e^{i\theta}ZAUA^{-1}Q_1\left(e^{i\theta}\right)
  &= \notag \\
&& \bigl[1-e^{i\theta}ZAU&A^{-1}Q\bigl(e^{i\theta}\bigr)\bigr]^{-1}\text{.}
\end{alignat}
Denote by $N$ and $M$ the sets $N(U,A,J)$ and $M(U,A,J)$. If
$\theta \in (J \sim N) \sim M$, i.e.,\ $\theta \in J \cap N^c \cap M^c$,
and $1-e^{i\theta}ZAUA^{-1}Q\left(e^{i\theta}\right)$ is not invertible,
then the compactness of $-LQ\left(e^{i\theta}\right)$ (which follows from the
compactness of $Q\left(e^{i(\theta+i\epsilon)}\right)$, the norm
convergence of $Q\left(e^{i(\theta+i\epsilon)}\right)$ and
(Theorem~VI.12, \cite{Reed1})) allows one to use the Fredholm Alternative
(Theorem~VI.14, p.~201, \cite{Reed1}) to assert that
\begin{equation*}
\exists x \in \mathcal{K},\text{ s.t. }
\left[1-e^{i\theta}ZAUA^{-1}Q\left(e^{i\theta}\right)\right]x = 0\text{.}
\end{equation*}
That is, there is some vector $x\in\mathcal{K}$ which satisfies the
equation
\begin{equation}
\label{eq:r_x_exists}
x - e^{i\theta}ZAUA^{-1}A\left(1-Ue^{i\theta}\right)^{-1}A^*x = 0\text{.}
\end{equation}
As $\theta \in J \sim N$, by Proposition~\ref{prop:r_3}
$y=A^*x \in D\left[\left(1-Ue^{i\theta}\right)^{-1}\right]$ so define $\phi$ as
\begin{equation}
\label{eq:r_phi}
\phi = \left(1-Ue^{i\theta}\right)^{-1}A^*x\text{.}
\end{equation}
$\phi$ is a well defined vector on $\mathcal{H}$ and we have
\begin{equation*}
x - e^{i\theta}ZAU\phi = 0\text{,}
\end{equation*}
which implies that
\begin{equation*}
x = e^{i\theta}ZAU\phi\text{.}
\end{equation*}
By \refeq{eq:r_phi}, $x \neq 0$ implies $\phi \neq 0$, so we have
\begin{alignat}{2}
&&\qquad\left(1-Ue^{i\theta}\right)\phi = A^*x &=
 e^{i\theta}A^*ZAU\phi\text{,} \notag \\
&\text{whence}&\qquad \left(1+A^*ZA\right)U\phi &= e^{-i\theta}\phi\text{,}
\notag \\
&\text{or}&\qquad V\phi &=e^{-i\theta}\phi\text{.} \label{eq:r_f_e_value}
\end{alignat}
We conclude that $e^{-i\theta}\in\sigma_p(V)$.

The multiplicity of the eigenvalue is given by the dimension of the kernel
of $1-e^{i\theta}ZAUA^{-1}Q$, which is finite by the
compactness of $Q$ and (Theorem~4.25, \cite{RudinFA}).

Therefore, if $\theta \in J \sim (N \cup M \cup \sigma_p(V))$, which is a
set of full Lebesgue measure,\footnote{That the set $M$ has
measure zero is a consequence of Lemma~\ref{lemma:r_5} on page
\pageref{lemma:r_5}.} then the vector
\begin{equation}
\begin{split}
\label{eq:r_xeps}
x(\epsilon)
  &= \left[1+e^{i(\theta+i\epsilon)}ZAUA^{-1}Q_1\left(e^{i(\theta+i\epsilon)}
    \right)\right]x \\
  &\equiv \left[1+L_+Q_1\left(e^{i\theta_+}\right)\right]x
\end{split}
\end{equation}
must be bounded in norm as $\epsilon\to 0$ because we have just seen
that if it is unbounded we have an eigenvalue of the operator $V$. For
$y=A^*x \in R(A^*)$, the absolutely continuous spectrum, $m_y^{ac}$, of $V$
is the limit of
\begin{equation*}
\left\langle\delta_\epsilon\left(1-Ve^{i\theta}\right)y,y\right\rangle
  = \left\langle A \delta_\epsilon\left(1-Ve^{i\theta}\right)A^*x,x
  \right\rangle\text{.}
\end{equation*}
The aim is to show that this is zero for all $y\in\mathcal{H}$. Define
\begin{align}
F_1(\theta) &= \left(1-Ve^{i\theta}\right)^{-1}\text{,} \\
F(\theta) &= \left(1-Ue^{i\theta}\right)^{-1}
\end{align}
and in a similar fashion to \refeq{eq:r_q1-q} and \refeq{eq:r_q1_inverse_q},
we obtain
\begin{equation}
\label{eq:r_F1resolvent}
F_1(\theta) = F(\theta)\left[1+ (V-U)e^{i\theta}F_1(\theta)\right]
\end{equation}
and
\begin{equation*}
\left(1+(V-U)e^{i\theta}F_1(\theta)\right)
  = \left(1-(V-U)e^{i\theta}F(\theta)\right)^{-1}\text{.}
\end{equation*}
Writing $X=V-U$, on substituting \refeq{eq:r_F1resolvent} into the expression
for the $\delta$-function \refeq{eq:r_delta_resolvent} we obtain
\begin{equation*}
\begin{split}
2\pi \delta_\epsilon\left(1-Ve^{i\theta}\right)
  &= \left(1-e^{-2\epsilon}\right)
    F_1^*(\theta_+)F_1(\theta_+)\\	
  &= \left[1+e^{i\theta_+}XF_1(\theta_+)\right]^*
    2\pi \delta_\epsilon\left(1-Ue^{i\theta}\right)
    \left[1+e^{i\theta_+}XF_1(\theta_+)\right]\text{.}
\end{split}
\end{equation*}
Substitution of \refeq{eq:r_QfromF} and noting that
\begin{equation*}
X = V - U = (1+A^*ZA)U - U = A^*ZAU
\end{equation*}
gives
\begin{equation*}
\begin{split}
A&\delta_\epsilon\left(1-Ve^{i\theta}\right) A^* \\
  &= A\left[1+Xe^{i\theta_+}F_1(\theta_+)\right]^*
    \delta_\epsilon\left(1-Ue^{i\theta}\right)
    \left[1+Xe^{i\theta_+}F_1(\theta_+)\right]A^* \\
  &= A\left[1+e^{-i\theta_-}F_1^*(\theta_+)U^*A^*Z^*A\right]
    \delta_\epsilon\left(1-Ue^{i\theta}\right)
    \left[1+e^{i\theta_+}A^*ZAUF_1(\theta_+)\right]A^* \\
  &= \left[A + e^{-i\theta_-}AF_1^*(\theta_+)A^*\left(A^*\right)^{-1}
    U^*A^*Z^*A\right]
    \delta_\epsilon\left(1-Ue^{i\theta}\right) \\
    &\qquad\qquad\qquad\qquad\qquad\qquad
    \times \left[A^* + e^{i\theta_+}A^*ZAUA^{-1}AF_1(\theta_+)A^*\right] \\
  &= \left[1+e^{-i\theta_-}Q_1^*(\theta_+)\left(A^*\right)^{-1}
    U^*A^*Z^*\right]
    A\delta_\epsilon\left(1-Ue^{i\theta}\right)A^* \\
    &\qquad\qquad\qquad\qquad\qquad\qquad
    \times \left[1 + e^{i\theta_+}ZAUA^{-1}Q_1(\theta_+)\right] \\
  &= \left[1 + L_+Q_1(\theta_+)\right]^*
    A\delta_\epsilon\left(1-Ue^{i\theta}\right)A^*
    \left[1 + L_+Q_1(\theta_+)\right]\text{.}
\end{split}
\end{equation*}
The absolutely continuous spectrum, $m_y^{ac}$ of $V$ is the
$\epsilon\rightarrow 0$ limit of
\begin{equation}
\begin{split}
\label{eq:r_V_ac}
\bigl\langle &A\delta_\epsilon\left(1-Ve^{i\theta}\right)A^*x,x
  \bigr\rangle \\
  &= \left\langle\left[1+L_+Q_1(\theta_+)\right]^*
    A\delta_\epsilon\left(1-Ue^{i\theta}\right)A^*
    \left[1+L_+Q_1(\theta_+)\right]x,x\right\rangle \\
  &= \left\langle A\delta_\epsilon\left(1-Ue^{i\theta}\right)A^*
    x(\epsilon),x(\epsilon)\right\rangle \\
  &= \frac{\epsilon(1-\epsilon)}{\pi}
    \left\langle G_\epsilon(\theta;U,A)x(\epsilon),x(\epsilon)\right\rangle
\end{split}
\end{equation}
which tends to zero as $\epsilon\to 0$ if both $G_\epsilon(\theta;U,A)$
and $x(\epsilon)$ are bounded.
$G_\epsilon(\theta;U,A)$ is bounded as $\theta \in J \sim N$ and
$x(\epsilon)$ is bounded by \refeq{eq:r_xeps}.

Part (\ref{thrm:r_4a}) follows since $R(A^*)$ cyclic for $U$ implies that
$R(A^*)$ is cyclic for $V$.
\vspace{\baselineskip}  


(\ref{thrm:r_4b}) Let $N_1=N(V,A,J)$. We have just shown that
$\theta \in J \sim (N \cup M \cup \sigma_p(V))$ implies that
\begin{equation}
\label{eq:r_g_epsilon_v}
\frac{\epsilon}{\pi}\langle G_\epsilon(\theta;V,A)
  x(\epsilon),x(\epsilon)\rangle \to 0
\end{equation}
and therefore
\begin{equation}
\label{eq:r_delta_v}
\langle \delta_\epsilon\left(1-Ve^{i\theta}\right)y,y\rangle \to 0\text{.}
\end{equation}
If we can infer the strong limit from this weak limit then we
have established that $\theta \notin N_1$. We use the result that if
$x_n\overset{w}{\rightarrow} x$ and
$\lpar x_n \rpar \rightarrow \lpar x \rpar$, then
$x_n\overset{s}{\rightarrow} x$
(\cite{Bachman}, p 244). Writing $G_\epsilon$ and $G$ for
$G_\epsilon(\theta;V,A)$ and $G(\theta;V,A)$, and $F_\epsilon$ and $F$ for
$F(\theta_+;V)$ and $F(\theta;V)$, consider
\begin{equation*}
\begin{split}
\bigl| \lpar G_\epsilon &x \rpar^2 - \lpar G x \rpar^2 \bigr| \\
  &= \left| \langle (G_\epsilon^2 - G^2)x, x \rangle \right| \\
  &= \left| \langle A\left\{\left(F^*_\epsilon F_\epsilon - F^*F\right)
     A^*AF^*_\epsilon F_\epsilon + F^*FA^*A
     \left(F^*_\epsilon F_\epsilon - F^*F\right)
     \right\}A^* x, x \rangle \right|\text{.}
\end{split}
\end{equation*}
If $A$, $F_\epsilon$ and $F$ are bounded operators, then if
$F^*_\epsilon F_\epsilon - F^*F$ tends to zero as $\epsilon\rightarrow 0$
we can conclude that the strong limit exists. A short calculation
shows that
\begin{equation*}
F^*_\epsilon F_\epsilon - F^*F
  = \left[\left(1-e^{-2\epsilon}\right)
    - \left(1-e^{-\epsilon}\right)
    \left(Ue^{i\theta} + U^*e^{-i\theta}\right) \right]
    F^*_\epsilon F_\epsilon F^* F
\end{equation*}
which trivially tends to zero as $\epsilon\rightarrow 0$ given the
boundedness of $F_\epsilon$ and $F$. Finally, $A$ is bounded by assumption
and \refeq{eq:r_xeps} shows that $Q_1(\theta_+)$ is a bounded operator as
$\epsilon\rightarrow 0$ and thus both $F_\epsilon$ and $F$ are bounded.

Moving on from \refeq{eq:r_delta_v}, we have now established that
$N_1 \subset N \cup M \cup \sigma_p(V)$ so $N_1$ must have measure zero,
again remembering that we need Lemma~\ref{lemma:r_5} below to prove that $M$
has measure zero. By Theorem~\ref{thrm:r_2}, $N_1$ supports the singular
spectrum of $V$. That is,
\begin{equation*}
m^s\left(N_1^\text{c}\right) = 0
\end{equation*}
where the set $N_1^\text{c}$ is the complement of $N_1$.
As the measure is positive and $m^s = m^{sc} + m^p$, we know that
\begin{equation*}
m^{sc}\left(N_1^\text{c}\right) = 0\text{.}
\end{equation*}
Trivially, $(N \cup M)\sim \sigma_p(V)$ contains $N_1 \sim \sigma_p(V)$.
Thus
\begin{equation*}
\begin{split}
m^{sc}\left(\left[N_1 \cap \sigma_p(V)^\text{c}\right]^\text{c}\right)
  &= m^{sc}\left(N_1^\text{c} \cup \sigma_p(V)\right) \\
  &= m^{sc}\left(N_1^\text{c}\right) + m^{sc}\left(\sigma_p(V)\right) \\
  &= 0 + 0 \\
  &= 0
\end{split}
\end{equation*}
as the (continuous) measure of single points is zero.

The set $N \cup M \cap \sigma_p(V)^\text{c}$ must support
$m^{sc}$ as $N_1 \cap \sigma_p(V)^\text{c}$ is a subset. Therefore
\begin{equation*}
m^{sc}\left(\left[N \cup M \cap \sigma_p(V)^\text{c}\right]^\text{c}\right)
  = 0\text{.}
\end{equation*}
This equals
\begin{equation*}
\begin{split}
 m^{sc}\left(\left[N \cup M\right]^\text{c} \cup \sigma_p(V)\right)
  &= m^{sc}\left(\left[N \cup M\right]^\text{c}\right) +
      m^{sc}\left(\sigma_p(V)\right) \\
 &= m^{sc}\left(\left[N \cup M\right]^\text{c}\right)
\end{split}
\end{equation*}
so we conclude that the set $N \cup M$ supports the singular continuous
part of the spectrum.\proofend


Theorem~\ref{thrm:r_4} has shown us that $V$ has an empty absolutely
continuous component, and that the singular continuous component is
supported by the set $N \cup M$, which is independent of $\lambda$.
We know that $N$ has measure zero, and
Lemma~\ref{lemma:r_5} below shows us that $M$ also has measure zero. This
will allow us to apply Theorem~\ref{thrm:r_6} to show that the singular
continuous spectrum of $V$ is also empty. Thus, with both the absolutely
continuous and singularly continuous spectra empty, we can conclude that
$V$ must have pure point spectrum.


\begin{lemma}
\tcaption{\textbf{Lemma:} The self-adjoint operator
$Q\left(e^{i(\theta + i\epsilon)}\right)$ converges in Hilbert Schmidt
norm as $\epsilon\rightarrow 0$}
\label{lemma:r_5}
Let $Q(z)$ be a trace class valued analytic function inside the complex
unit circle, with $|z|<1$. Then for a.e.\ $\theta$
\begin{equation*}
\underset{\epsilon\downarrow 0}{\lim} Q\left(e^{i(\theta +i\epsilon)}\right)
  \equiv
  Q\left(e^{i(\theta+i0)}\right)
\end{equation*}
exists in Hilbert Schmidt norm.
\end{lemma}


\emph{Proof.}
%
We parallel the proof of de~Branges theorem (see \cite{deBranges} and
p.~149--150, \cite{Kato71}). Consider
\begin{equation*}
\begin{split}
Q\left(e^{i(\theta+i\epsilon)}\right)
  &+ Q^*\left(e^{i(\theta+i\epsilon)}\right) \\
 &= \int A^*\left\{\frac{1}{1-e^{-i(\theta'-\theta)}e^{-\epsilon}}
   + \frac{1}{1-e^{i(\theta'-\theta)}e^{-\epsilon}} \right\} AE(d\theta') \\
 &= \int A^*\left\{\frac{2\left(1-e^{-\epsilon}\cos(\theta'-\theta)\right)}
   {1+e^{-2\epsilon}-2e^{-\epsilon}\cos(\theta'-\theta)}\right\}
   AE(d\theta')\text{.}
\end{split}
\end{equation*}
The factor within the curly braces is greater than zero for all
$\theta',\theta$ and thus we have
\begin{equation*}
Q\left(e^{i(\theta+i\epsilon)}\right) + Q^*\left(e^{i(\theta+i\epsilon)}\right)
 \geq 0 \phantom{a}\forall \epsilon \geq 0\text{.}
\end{equation*}
Therefore, following de~Branges,
\begin{equation*}
\begin{split}
\left| \det \left(1+Q\left(e^{i(\theta+i\epsilon)}\right)\right) \right|^2
 &\geq \det \left( 1+Q^*\left(e^{i(\theta+i\epsilon)}\right)
   Q\left(e^{i(\theta+i\epsilon)}\right) \right) \\
 &= \prod\left(1 + |\alpha_n|^2\right) \\
 &\geq
    \begin{cases}
      \sum|\alpha_n|^2 =
        \lpar Q\left(e^{i(\theta+i\epsilon)}\right) \rpar^2_{H.S.} \\
      1\text{.}
    \end{cases}
\end{split}
\end{equation*}
$\{\alpha_n\}$ are the eigenvalues of $Q\left(e^{i(\theta+i\epsilon)}\right)$.
From these two bounds we obtain
\begin{equation*}
\left|\left| \frac{Q\left(e^{i(\theta+i\epsilon)}\right)}
  {\det \left(1+Q\left(e^{i(\theta+i\epsilon)}\right)\right)}
  \right|\right|_{H.S.} \leq 1 \text{ and }
\left| \frac{1}{\det \left(1+Q\left(e^{i(\theta+i\epsilon)}\right)\right)}
  \right| \leq 1\text{.}
\end{equation*}
The definition of an analytic operator (p.~189, \cite{Reed1}) implies the
analyticity of the eigenvalues, and thus the operations of taking the
determinant and the Hilbert Schmidt norm are analytic. Hence, both functions
above are analytic and bounded within the complex unit circle
($\epsilon > 0$). Application of Fatou's theorem (p.~454, \cite{Dienes})
establishes the existence in the limit as $\epsilon\rightarrow 0$ and hence
both functions exist on the boundary almost everywhere. Taking the quotient
we establish the existence of $Q\left(e^{i(\theta+i0)}\right)$ in the
Hilbert Schmidt norm.\proofend


Let $(\Omega,\mu)$ be a separable measure space, and
\begin{equation*}
V(\lambda) = \int e^{-i\theta} E_\lambda(d\theta)
\end{equation*}
a measurable family of unitary operators on $\mathcal{H}$. We denote by
\begin{equation*}
\mathbb{V} = \int e^{-i\theta} \mathbb{E}(d\theta)
\end{equation*}
the multiplication operator
\begin{equation*}
(\mathbb{V}u)(\lambda) = V(\lambda)u(\lambda)
\end{equation*}
on $L^2(\Omega,\mu;\mathcal{H})$, where
$u(\lambda) \in L^2(\Omega,\mu;\mathcal{H})$.

A vector $u(\lambda)$ is an element of $L^2(\Omega,\mu;\mathcal{H})$ if,
for $u(\lambda)\in\mathcal{H}$,
\begin{equation*}
\int_{-\infty}^\infty \lpar u(\lambda) \rpar ^2\,d\mu < \infty\text{.}
\end{equation*}
It is important to note the difference between $V(\lambda)$ acting on
$\mathcal{H}$ and $\mathbb{V}$ acting on $L^2(\Omega,\mu;\mathcal{H})$. To
obtain our goal of showing that for a.e.\ $\lambda$, $V(\lambda)$ has a pure
point spectrum, we must show that $\mathbb{V}$ is absolutely continuous as
a function of $\lambda$ on the space $L^2(\Omega,\mu;\mathcal{H})$.

Theorem~\ref{thrm:r_6} is taken directly from \cite{Howland87}. The proof given
is, apart from some small notational changes,
identical to that in \cite{Howland87}. Due to a number of typographical
errors however, I have reproduced the proof here for reference and clarity.


\begin{thrm}
\tcaption{\textbf{Theorem:} The Floquet operator $V$ has
$\sigma_\cs(V) = \emptyset$}
\label{thrm:r_6}
Let $\mathbb{V}$ be absolutely continuous on $L^2(\Omega,\mu;\mathcal{H})$,
and assume that there is a fixed set $S$ of Lebesgue measure zero which
supports the singular continuous spectrum of $V(\lambda)$ in the interval $J$
for $\mu$-a.e.\ $\lambda$. Then $V(\lambda)$ has no singular continuous
spectrum in $J$ for $\mu$-a.e.\ $\lambda$.
\end{thrm}
 

\emph{Proof.} For fixed $x\in\mathcal{H}$, and any measurable subset
$\Gamma$ of $\Omega$, let
$u(\lambda) = \chi_{\text{\tiny\itshape $\Gamma$}}(\lambda)x$ be a vector
in $L^2(\Omega,\mu;\mathcal{H})$. Then
\begin{equation*}
\begin{split}
\int_\Gamma \left| E_\lambda^{sc}[J]x\right|^2\mu(d\lambda)
  &\leq \int_\Gamma \left| E_\lambda[S]x\right|^2\mu(d\lambda) \\
  &= \int \left| E_\lambda[S]u(\lambda)\right|^2\mu(d\lambda) \\
  &= \int \left| \mathbb{E}[S]u(\lambda) \right|^2\mu(d\lambda) \\
  &= \;\lpar \mathbb{E}[S]u(\lambda)\rpar^2 \;= 0\text{.}
\end{split}
\end{equation*}
$\int_\Gamma \left| E_\lambda^{sc}[J]x\right|^2\mu(d\lambda) = 0$ implies
that $\left|E_\lambda^{sc}[J]x\right|^2 = 0$ for $\mu$-a.e.\ $\lambda$. Thus
\begin{equation*}
E_\lambda^{sc}[J]x = 0
\end{equation*}
for every $x\in\mathcal{H}$.\proofend


The application of Theorem~\ref{thrm:r_6} relies on finding a fixed set $S$
of measure zero which supports the singularly continuous spectrum.
$S = N \cup M$ is sufficient.

I have now established all the basic requirements for $V$ to be pure
point, given $U$ pure point. They are now combined to produce the main
theorem of the chapter. There is still quite a lot of manipulation to satisfy
the condition $\mathbb{V}$ absolutely continuous on
$L^2(\mathbb{R};\mathcal{H})$ of Theorem~\ref{thrm:r_6}, and this will be
the focus for the remainder of \refsec{sec:r_spectra} and \refsec{sec:r_pk}.

\pagebreak


\begin{thrm}
\tcaption{\textbf{Theorem:} The Floquet operator $V$ is pure point if
$A$ is $U$-finite}
\label{thrm:r_7}
Let $U$ and $A$ satisfy the hypotheses of Theorem~\ref{thrm:r_4} and define
for $\lambda \in \mathbb{R}$
\begin{equation*}
V(\lambda) = e^{i\lambda A^*A / \hbar}U\text{.}
\end{equation*}
Then $V(\lambda)$ is pure point in $J$ for a.e.\ $\lambda$.
\end{thrm}


\emph{Proof.} By Theorem~\ref{thrm:r_4}, with $W=\lambda I$,
$V(\lambda)$ has no absolutely continuous spectrum in $J$, and its singularly
continuous spectrum is supported on the fixed set $S = N \cup M$. Application
of Lemma~\ref{lemma:r_5} shows that $S$ is of measure zero.
If we can show that $\mathbb{V}$ is absolutely continuous on
$L^2(\mathbb{R};\mathcal{H})$ then Theorem~\ref{thrm:r_6} applies and shows
that the singular continuous spectrum is empty. I prove the absolute
continuity of $\mathbb{V}$ in the following sections.

As I have shown that both the absolutely continuous and singular continuous
parts of the spectrum are empty, we conclude that $V(\lambda)$ is pure point
for a.e.\ $\lambda \in \mathbb{R}$.\proofend


To show that $\mathbb{V}$ is absolutely continuous, I apply a modified
version of the Putnam--Kato theorem which is proved in \refsec{sec:r_pk}.
The unitary Putnam--Kato theorem is:

\vspace{\baselineskip}
\noindent{\textbf\itshape Theorem~\ref{thrm:r_unitpk}}
{\itshape Let $V$ be unitary, and $D$ a self-adjoint bounded operator. If
$C=V[V^*,D] \geq 0$, then $V$ is absolutely continuous on $R(C^{1/2})$.
Hence, if $R(C^{1/2})$ is cyclic for $V$, then $V$ is absolutely
continuous on $\mathcal{H}$.}
\vspace{\baselineskip}  

I apply this theorem on the space $L^2(\mathbb{R};\mathcal{H})$. A naive
application to obtain the desired result is as follows. I slightly change
notation and explicitly include the $\lambda$ dependence of $W$ in the
definition of $V$. If we choose $\mathbb{D} = -i(d/d\lambda)$, with
$V = e^{i\lambda A^*WA}U$, then
\begin{equation*}
-i\frac{dV^*}{d\lambda} = -U^*A^*WA e^{-i\lambda A^*WA} = -V^*A^*WA\text{,}
\end{equation*}
so that for some $u \in L^2(\mathbb{R};\mathcal{H})$,
\begin{equation*}
\begin{split}
[\mathbb{V}^*,\mathbb{D}]u &= (\mathbb{V}^*\mathbb{D} -
  \mathbb{D}\mathbb{V}^*)u = -\mathbb{D}\mathbb{V}^*u \\
  &= i\frac{d}{d\lambda}(\mathbb{V}^*u) = \mathbb{V}^*A^*WAu\text{.}
\end{split}
\end{equation*}
Therefore,
\begin{equation*}
\mathbb{C} = \mathbb{V}[\mathbb{V}^*,\mathbb{D}] = A^*WA\text{.}
\end{equation*}
With $W=I$, we obtain $\mathbb{C} = A^*A \geq 0$ and thus
$R(\mathbb{C}^{1/2})=R(A^*)$ (see
the proof to (Theorem~VI.9, \cite{Reed1})) is cyclic for $V$. Hence,
$\mathbb{V}$ is absolutely continuous and all the requirements of
Theorem~\ref{thrm:r_7} are satisfied.

The problem here is that $\mathbb{D}$ is not bounded, and boundedness of
$\mathbb{D}$ is essential in the proof of the Putnam--Kato theorem. I use
a similar technique as Howland \cite{Howland87} to overcome this issue. 

As the norm of $A^*A$ may be scaled arbitrarily, we can rewrite $V$, for
real $t$, as
\begin{equation}
\label{eq:r_scaled_v}
V(t) = e^{ictA^*A}U
\end{equation}
for some real $c>0$.


\begin{prop}
\tcaption{\textbf{Proposition:} The operator $C=\mathbb{V}[\mathbb{V}^*,D]$
is positive definite with cyclic range for the multiplication operator
$\mathbb{V}$}
\label{prop:r_8}
On $L^2(\mathbb{R};\mathcal{H})$, consider the unitary multiplication operator
$\mathbb{V}$, defined by
\begin{equation*}
\mathbb{V}u(t) = V(t)u(t) = e^{ictA^*A}Uu(t)
\end{equation*}
and the bounded self-adjoint operator $\mathbb{D}=-\arctan(p/2)$,
where $p=-id/dt$.
Then $C=\mathbb{V}[\mathbb{V}^*,D]$ is positive definite, and
$R\left(C^{1/2}\right)$ is cyclic for $\mathbb{V}$. Hence, the requirements
of Theorem~\ref{thrm:r_7} are fully satisfied.
\end{prop}


\emph{Proof.} The operator $\mathbb{D}$ on $L^2(\mathbb{R};\mathcal{H})$
is convolution by the Fourier transform of $-\arctan(x/2)$ \cite{Howland87},
which is $i\pi t^{-1} e^{-2|t|}$ ((3), p.~87, \cite{Erdelyi}). This is a
singular (principal value) integral operator, because $\arctan(p/2)$ does
not vanish at infinity. Thus, for $u(t) \in L^2(\mathbb{R};\mathcal{H})$,
\begin{equation*}
\mathbb{D}u(t) 
  = i\pi P \int_{-\infty}^{\infty} \frac{e^{-2|t-y|}}{t-y}u(y)\,dy
\end{equation*}
and
\begin{equation*}
[\mathbb{V}^*,\mathbb{D}]u(t) = i\pi P \int_{-\infty}^{\infty}
  e^{-2|t-y|}\frac{V^*(t)-V^*(y)}{t-y}u(y)\,dy
\end{equation*}
so
\begin{equation}
\begin{split}
\label{eq:r_C}
\mathbb{C}u(t) &= \mathbb{V}[\mathbb{V}^*,\mathbb{D}]u(t) \\
  &= i\pi P \int_{-\infty}^{\infty}
    e^{-2|t-y|}\frac{1-V(t)V^*(y)}{t-y}u(y)\,dy\text{.}
\end{split}
\end{equation}
Inserting expression \refeq{eq:r_scaled_v} for $V(t)$, we obtain
\begin{equation}
\begin{split}
\label{eq:r_pk_int}
\mathbb{C}u(t) &=i\pi \int_{-\infty}^{\infty}
    e^{-2|t-y|}\frac{1-e^{ic(t-y)A^*A}}{t-y}u(y)\,dy \\
  &=i\pi \int_{-\infty}^{\infty}
    e^{-2|t-y|}\frac{1-\cos\left(A^*Ac(t-y)\right)
    -i\sin\left(A^*Ac(t-y)\right)}{t-y}u(y)\,dy\text{.}
\end{split}
\end{equation}
Note that this is no longer a singular integral. To show that $\mathbb{C}$
is positive, we must show that
\begin{equation*}
(u(t),\mathbb{C}u(t)) > 0
  \phantom{a}\forall u(t)\in L^2(\mathbb{R};\mathcal{H}).
\end{equation*}
Note that the inner product on $L^2(\mathbb{R};\mathcal{H})$ is given by
\begin{equation}
\label{eq:r_L_inner_prod}
(u(t),u'(t)) = \int_{-\infty}^\infty u^*(t)u'(t)\,dt\text{.}
\end{equation}
The operator $A$ is now written in terms of its spectral components. Note
that here $\lambda$ decomposes $A$ and bears no relation
to the strength parameter used at other stages in this chapter. When required
for clarity, I write $\int_\lambda$ to identify the integral over the
variable $\lambda$,
\begin{equation*}
A = \int \lambda E(d\lambda)\text{.}
\end{equation*}
A general vector $u(t)$ may be written
\begin{equation*}
u(t) = \int E(d\lambda)u(t)\text{.}
\end{equation*}
Then
\begin{equation*}
f(A)u(t) = \int f(\lambda) E(d\lambda)u(t)
\end{equation*}
which implies that we may rewrite \refeq{eq:r_pk_int} as
\begin{equation*}
\begin{split}
\mathbb{C}u(t)
  &= i\pi \int_{-\infty}^\infty dy\,\int_\lambda
    e^{-2|t-y|}\frac{1-e^{ic(t-y)|\lambda|^2}}{t-y}
    E(d\lambda)u(y) \\
  &= \int_{-\infty}^\infty dy\,\int_\lambda \phi_\lambda(t-y)
  E(d\lambda)u(y) \\
  &= \int_\lambda E(d\lambda)\mathcal{C}_\lambda(t)
\end{split}
\end{equation*}
where
\begin{equation*}
\mathcal{C}_\lambda(t)
  = \int_{-\infty}^\infty dy\,\phi_\lambda(t-y)u(y)
\end{equation*}
and we have defined the new function
\begin{equation*}
\phi_\lambda(t) = i\pi e^{-2|t|}t^{-1}\left(1-e^{ict|\lambda|^2}
  \right)\text{.}
\end{equation*}
By the convolution theorem, note that
\begin{equation*}
\tilde{\mathcal{C}}_\lambda(\omega)
  = \tilde{\phi}_\lambda(\omega)\tilde{u}(\omega)
\end{equation*}
where the ``$\tilde{\phantom{a}}$'' indicates Fourier transform.

Using this decomposition of $u(t)$ and Parseval's theorem, we can now easily
write down $(u(t),\mathbb{C}u(t))$. I use $(x,y)_\mathcal{H}$ to indicate
the inner product on the Hilbert Space $\mathcal{H}$, reserving $(x,y)$ for
the inner product on $L^2(\mathbb{R};\mathcal{H})$ as in
\refeq{eq:r_L_inner_prod}.
\begin{equation*}
\begin{split}
\left(u(t),\mathbb{C}u(t)\right)
  &= \int_{-\infty}^\infty 
  dt\,\left(u(t),\mathbb{C}u(t)\right)_{\mathcal{H}} \\
  &= \int_{-\infty}^\infty dt\,\left(u(t),
    \int_\lambda E(d\lambda)\mathcal{C}_\lambda(t)\right)_{\mathcal{H}} \\
  &= \int_{-\infty}^\infty dt\,\left(u(t), \int_\lambda E(d\lambda)
    \int_{-\infty}^\infty dy\,\phi_\lambda(t-y)u(y)\right)_{\mathcal{H}} \\
  &= \int_{-\infty}^\infty dt\,\left(u(t), \int_\lambda E(d\lambda)
    \int \frac{d\omega}{2\pi}\,e^{i\omega t}\tilde{\mathcal{C}}_\lambda(\omega)
    \right)_{\mathcal{H}} \\
  &= \int_\lambda E(d\lambda) \int \frac{d\omega}{2\pi}\,
    \left(\tilde{u}(\omega), \tilde{\phi}_\lambda(\omega)\tilde{u}(\omega)
    \right)_{\mathcal{H}} \\
  &= \int_\lambda E(d\lambda) \int \frac{d\omega}{2\pi}\,
    \left|\tilde{u}_\lambda(\omega)\right|^2 \tilde{\phi}_\lambda(\omega)
    \text{.}
\end{split}
\end{equation*}
It is clear that if $\tilde{\phi}_\lambda(\omega)$ is positive for all
$\lambda$ then $\mathbb{C}$ will be positive.

In the following calculation we will find the need to bound $c|\lambda|^2$.
The restriction $0 \leq c|\lambda|^2\leq 1$ will be employed. I argue that
as $A^*A$ is a positive self-adjoint bounded operator we can restrict the
integral over $\lambda$ to (\cite{Riesz}, p.~262, 273)
\begin{equation}
A^*A = \int_{-\infty}^\infty |\lambda|^2 E(d\lambda)
  = \int_{m-0}^M |\lambda|^2 E(d\lambda)
\end{equation}
where $M$ is the least upper bound and $m$ the greatest lower bound of
$A^*A$. The norm of $A^*A$ is given by $\max(|m|,|M|)$. Thus, by setting
\begin{equation*}
c = \frac{1}{\lpar A^*A \rpar}
\end{equation*}
then each $c|\lambda|^2$ is guaranteed to be less than unity.

Proceeding, the Fourier transform, $\tilde{\phi}_\lambda(\omega)$, of
\begin{equation}
\label{eq:r_phi_convolution}
\phi_\lambda(t) 
  = i\pi e^{-2|t|}t^{-1}\left[ 1-\cos ct|\lambda|^2 - i\sin ct|\lambda|^2
    \right]
\end{equation}
is now calculated. Split \refeq{eq:r_phi_convolution} into two parts,
\begin{align}
\label{eq:r_phi_1}
\phi_{\lambda1}(t) &= i\pi e^{-2|t|}t^{-1}\left[1-\cos ct|\lambda|^2\right]
\text{,} \\
\label{eq:r_phi_2}
\phi_{\lambda2}(t) &= \pi e^{-2|t|}t^{-1}\sin ct|\lambda|^2\text{.}
\end{align}
The Fourier transform of \refeq{eq:r_phi_1} is
\begin{equation*}
\begin{split}
\tilde{\phi}_{\lambda1}(\omega) &= i\pi\int_{-\infty}^{\infty}
  e^{-2|t|}t^{-1}(1-\cos ct|\lambda|^2)e^{-i\omega t}\,dt \\
  &= i\pi\Biggl[
    \int_0^{\infty} e^{-2t}t^{-1}(1-\cos ct|\lambda|^2)e^{-i\omega t}\,dt \\
  &\qquad  
    + \int_0^{\infty} e^{-2t}(-t^{-1})(1-\cos ct|\lambda|^2)e^{i\omega t}\,dt
    \Biggr]\text{.}
\end{split}
\end{equation*}
Using (\cite{Erdelyi}, p.~157, (59)), and setting
$S=c|\lambda|^2/(2+i\omega)$, we obtain
\begin{equation*}
\tilde{\phi}_{\lambda1}(\omega)
  = \frac{i\pi}{2}\log\left(\frac{1+S^2}{1+S^{*2}}\right)\text{.}
\end{equation*}
The logarithm of a complex number can in general be written as
\begin{equation*}
\log(z) = \log(|z|) + i\Arg z
\end{equation*}
so noting that $\left|(1+S^2)/(1+S^{*2})\right|=1$, we see that
\begin{equation*}
\begin{split}
\tilde{\phi}_{\lambda1}(\omega)
  &= -\frac{\pi}{2}\Arg\left(\frac{1+S^2}{1+S^{*2}}\right) \\
  &= -\pi\Arg\left(1+S^2\right)\text{.}
\end{split}
\end{equation*}
With $\kappa = c|\lambda|^2$, the real and imaginary parts of $1+S^2$ are
\begin{align*}
\Re\left(1+S^2\right)
  &= \frac{\left(4+\omega^2\right)^2 + \kappa^2\left(4-\omega^2\right)}
          {\left(4+\omega^2\right)^2}\text{,} \\
\Im\left(1+S^2\right)
  &= \frac{-4\kappa^2\omega}{\left(4+\omega^2\right)^2}\text{.}
\end{align*}
With the restriction that $0 \leq \kappa \leq 1$, the real part is positive
for all $\omega$ and thus $\Arg(z) = \arctan(\Im z / \Re z)$. Thus,
\begin{equation*}
\tilde{\phi}_{\lambda1}(\omega)
  = -\pi\arctan\left(\frac{\Im\left(1+S^2\right)}{\Re\left(1+S^2\right)}
    \right)\text{.}
\end{equation*}
$\arctan(z)$ is the principal part of $\Arctan(z)$, with range
$-\pi/2 < \arctan(z) < \pi/2$. The Fourier transform of \refeq{eq:r_phi_2} is
similarly calculated, using (\cite{Erdelyi}, p.~152, (16)), to be
\begin{equation*}
\begin{split}
\tilde{\phi}_{\lambda2}(\omega)
  &= \pi\left[\arctan S + \arctan S^* \right] \\
  &= \pi\left[\arctan\left(\frac{c|\lambda|^2}{2+i\omega}\right)
    + \arctan\left(\frac{c|\lambda|^2}{2-i\omega}\right)\right]\text{.}
\end{split}
\end{equation*}
Repeated application of the formula
$\arctan(z_1)+\arctan(z_2)=\arctan(z_1+z_2/1-z_1z_2)$, valid when
$z_1z_2 < 1$ (true for $0 \leq\kappa\leq 1$), yields\footnote{
This result is not valid for values of $\kappa$ larger than
around $2$, at which point the $\arctan$ addition formulas fail---this is
a moot point however, as we may trivially restrict $\kappa$ as already
explained.}
\begin{equation}
\begin{split}
\label{eq:r_Phi}
\tilde{\phi}_\lambda(\omega)
  &= \tilde{\phi}_{\lambda1}(\omega)+\tilde{\phi}_{\lambda2}(\omega) \\
  &= \pi\arctan\left(\frac{n(\omega,c|\lambda|^2)}
    {d(\omega,c|\lambda|^2)}\right)\text{,}
\end{split}
\end{equation}
where
\begin{align}
\label{eq:r_numerator}
n(\omega,\kappa) &= 4\kappa\bigl[
  \left(4+\omega^2\right)^2 + \kappa\omega\left(4+\omega^2\right)
  + \kappa^2\left(4-\omega^2\right) - \kappa^3\omega \bigr] \\
\intertext{and}
d(\omega,\kappa) &= \left(4+\omega^2\right)^3
   -2\kappa^2\omega^2\left(4+\omega^2\right)
   - 16\kappa^3\omega - \kappa^4\left(4-\omega^2\right)\text{.}
\end{align}
We can easily confirm that for $0 \leq \kappa \leq 1$,
$n(\omega,\kappa)/d(\omega,\kappa)$ and hence $\tilde{\phi}_\lambda(\omega)$
is strictly positive by noting that there are four distinct regions of
interest for $\omega$, in which terms in $n$ and $d$ do not change sign.
\reftab{table:fourier} shows these regions and the sign of each term in
the region. Note that the global (positive and hence irrelevant) $\kappa$
factor from \refeq{eq:r_numerator} is dropped from the numerator for the
following discussion.
\begin{table}[H]
\begin{tabular}{|c|c|c|c|c|}\hline
$n(\omega,\kappa)=$ & $\left(4+\omega^2\right)^2$ &
  $+\kappa\omega\left(4+\omega^2\right)^2$ &
  $+\kappa^2\left(4-\omega^2\right)$ & $-\kappa^3\omega$ \\ \hline

$\omega < -2$ & +ve & -ve & -ve & +ve \\
$-2 < \omega < 0$ & +ve & -ve & +ve & +ve \\
$0 < \omega < 2$ & +ve & +ve & +ve & -ve \\
$\omega > 2$ & +ve & +ve & -ve & -ve \\ \hline \hline

$d(\omega,\kappa)=$ & $\left(4+\omega^2\right)^3$ &
  $-2\kappa^2\omega^2\left(4+\omega^2\right)$ &
  $-16\kappa^3\omega$ & $-\kappa^4\left(4-\omega^2\right)$ \\ \hline

$\omega < -2$ & +ve & -ve & +ve & +ve \\
$-2 < \omega < 0$ & +ve & -ve & +ve & -ve \\
$0 < \omega < 2$ & +ve & -ve & -ve & -ve \\
$\omega > 2$ & +ve & -ve & -ve & +ve \\ \hline
\end{tabular}
\caption{Sign of each term in the numerator $n(\omega,\kappa)$ and the
denominator $d(\omega,\kappa)$ of \refeq{eq:r_Phi}.}
\label{table:fourier}
\end{table}
For each row in the table, we simply need to show that the terms add to
produce a strictly positive number. First note that the first column for
both the numerator and denominator is independent of $\kappa$. To show the
positivity of each row, set all positive $\kappa$-dependent terms to
zero and then take $\kappa=1$ for the negative terms to maximise their
contribution. Expanding out terms, it is then trivially seen in all cases
that the first column ($\left(4+\omega^2\right)^2$ for the numerator and
$\left(4+\omega^2\right)^3$ for the denominator) dominates. Thus, no row is
negative and we conclude that $\tilde{\phi}_\lambda$ is positive definite.

\begin{figure}[H]
\begin{center}
  \input{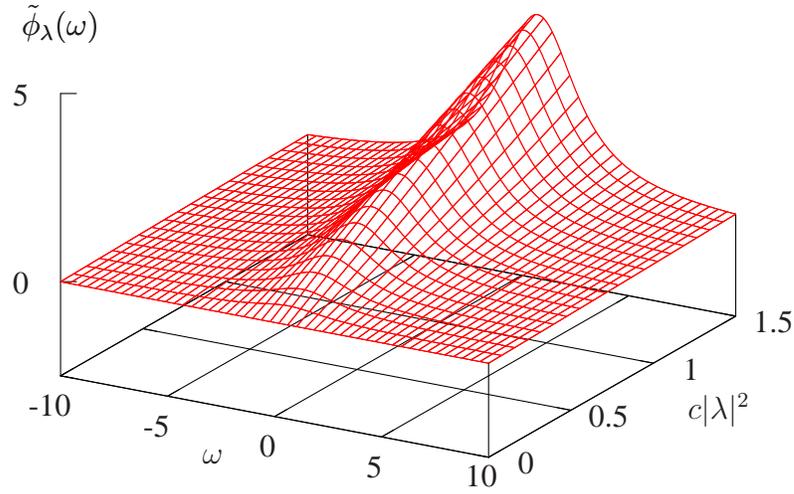}
  \caption{Plot of $\tilde{\phi}_\lambda(\omega)
  = \tilde{\phi}_{\lambda1}(\omega) + \tilde{\phi}_{\lambda2}(\omega)$, the
  Fourier transform of
  $\phi_\lambda(t) = i\pi e^{-2|t|}t^{-1}\left(1-\cos(ct|\lambda|^2)
  - i\sin(ct|\lambda|^2)\right)$. $\tilde{\phi}_\lambda(\omega)$ is strictly
  positive for all $\omega$ when $c|\lambda|^2 \leq 1$.}
  \label{fig:r_fourier}
\end{center}
\end{figure}

We have established that the Fourier transform of $\phi_\lambda$ is positive
definite for $c|\lambda|^2 \leq 1$. As a visual aid, \reffig{fig:r_fourier}
shows $\tilde{\phi}_\lambda(\omega)$. The positivity for
$c|\lambda|^2 \leq 1$ is clear.

Thus,
$\mathbb{C}$ is strictly positive and $\mathbb{V}$ is absolutely continuous
on $R(\mathbb{C}^{1/2})$. As $A^*A$ is a factor of $1-e^{ictA^*A}$
(i.e.,\ $A^*A$ is a factor of $\mathbb{C}$), $R(\mathbb{C}^{1/2}) = R(A^*)$.
Noting that $R(A^*)$ is cyclic for $U$ and hence cyclic for $V$, we conclude
that $R(\mathbb{C}^{1/2})$ is cyclic for $\mathbb{V}$. Thus, $\mathbb{V}$ is
absolutely continuous on $L^2(\mathbb{R};\mathcal{H})$.\proofend

I have now satisfied all the requirements of Theorem~\ref{thrm:r_7}.


\section{\label{sec:r_pk}The unitary Putnam--Kato theorem}

In this section, I will prove a modified version of the Putnam--Kato
theorem, as used in the preceding section. The theorems and proofs follow a
similar argument to that of Reed and Simon
(Theorem~XIII.28, p.~157, \cite{Reed4}) and are motivated by the
stroboscopic nature of the kicked Hamiltonian.


\begin{defn}[V-Smooth]
\tcaption{\textbf{Definition:} The operator $A$ is $V$-smooth}
\label{defn:r_vsmooth}
Let $V$ be a unitary operator. $A$ is $V$-smooth if and only if for all
$\phi \in \mathcal{H}$, $V(t)\phi\in D(A)$ for almost every
$t\in\mathbb{R}$ and for some constant $C$,
\begin{equation*}
\sum_n \lpar AV^n\phi \rpar^2 \;\leq C\lpar\phi\rpar^2\text{.}
\end{equation*}
\end{defn}


\begin{thrm}
\tcaption{\textbf{Theorem:} The closure of the range, $\overline{R(A^*)}$,
of the operator $A^*$ is a subset of $\mathcal{H}_{ac}(V)$}
\label{thrm:r_ranA_ac}
If $A$ is $V$-smooth, then
$\overline{R(A^*)} \subset \mathcal{H}_{ac}(V)$.
\end{thrm}


\emph{Proof.} Since $\mathcal{H}_{ac}(V)$ is closed, we need only show 
$R(A^*) \subset \mathcal{H}_{ac}(V)$. Let $\phi \in D(A^*)$,
$\psi = A^*\phi$, and let $d\mu_\psi$ be the spectral measure for $V$
associated with $\psi$. Define, for the period, $T$, in
\refeq{eq:r_hamiltonian},
\begin{equation}
\label{eq:r_f_n}
\mathcal{F}_n(T) = \frac{1}{\sqrt{2\pi}}\left(A^*\phi,[V(T)]^n\psi
  \right)\text{.}
\end{equation}
We calculate, dropping the $T$ for clarity,
\begin{equation*}
\begin{split}
|\mathcal{F}_n|
  &= \frac{1}{\sqrt{2\pi}}\left| \left(\phi, AV^n\psi\right) \right| \\
  &\leq \frac{1}{\sqrt{2\pi}}\lpar\phi\rpar\lpar AV^n\psi \rpar\text{.}
\end{split}
\end{equation*}
Because $A$ is $V$-smooth, we see that
\begin{equation*}
\begin{split}
\sum_n |\mathcal{F}_n|^2
  &\leq \frac{1}{2\pi} \lpar\phi\rpar^2
    \sum_n \lpar AV^n\psi \rpar^2 \\
  &\leq \frac{C}{2\pi} \lpar\phi\rpar^2 \lpar\psi\rpar^2 \\
  &< \infty\text{.}
\end{split}
\end{equation*}
Thus, $\mathcal{F}_n \in L^2(\mathbb{R})$. By the Riesz--Fischer theorem
(4.26 Fourier Series, p.~96--7, \cite{RudinRC}),
$\mathcal{F}(\theta) = \frac{1}{\sqrt{2\pi}}
  \sum_n \mathcal{F}_n e^{-in\theta} \in L^2$. 

The spectral resolution of $V[T]$ is
\begin{equation*}
V[T] = \int_0^{2\pi} e^{i\theta}\,dE_T(\theta)\text{,}
\end{equation*}
so
\begin{equation*}
(V[T])^n = \int_0^{2\pi} e^{in\theta}\,dE_T(\theta)\text{.}
\end{equation*}
Therefore, from \refeq{eq:r_f_n} we obtain
\begin{equation*}
\begin{split}
\mathcal{F}_n &= \frac{1}{\sqrt{2\pi}}\int_0^{2\pi}
  \left(A^*\phi, e^{in\theta}\,dE_T(\theta)\psi\right) \\
  &= \frac{1}{\sqrt{2\pi}}\int_0^{2\pi}
    e^{in\theta}\left(\psi,dE_T(\theta)\psi\right) \\
  &= \frac{1}{\sqrt{2\pi}}\int_0^{2\pi}
    e^{in\theta}\,d\mu_\psi(\theta)\text{.}
\end{split}
\end{equation*}
Using the inverse of the expression above for $\mathcal{F}(\theta)$ gives
\begin{equation*}
  \mathcal{F}_n = \frac{1}{\sqrt{2\pi}}\int_0^{2\pi}
    e^{in\theta} \mathcal{F}(\theta)\,d\theta\text{.}
\end{equation*}
As we have just shown that $\mathcal{F}(\theta) \in L^2$,
$d\mu_\psi(\theta) = \mathcal{F}(\theta)d\theta$ is absolutely continuous,
which implies that $\psi \in R(A^*)$ is in $\mathcal{H}_{ac}(V)$ and so
$\overline{R(A^*)} \subset \mathcal{H}_{ac}(V)$.\proofend


\begin{thrm}[Unitary Putnam--Kato theorem]
\tcaption{\textbf{Theorem:} The unitary equivalent of the Putnam--Kato
theorem. If $C=V[V^*,A] \geq 0$, then $V$ is absolutely continuous
on $R(C^{1/2})$}
\label{thrm:r_unitpk}
Let $V$ be a unitary operator, and $A$ a self-adjoint bounded operator. If
$C=V[V^*,A] \geq 0$, then $V$ is absolutely continuous on $R(C^{1/2})$.
Hence, if $R(C^{1/2})$ is cyclic for $V$, then $V$ is absolutely
continuous.
\end{thrm}


\emph{Proof.} The discrete time evolution of an operator $A$ is given by
\begin{equation*}
\mathcal{F}_n = V^{-n}AV^n\text{.}
\end{equation*}
Calculate
\begin{equation*}
\begin{split}
\mathcal{F}_n - \mathcal{F}_{n-1}
  &= V^{-n}AV^n - V^{-(n-1)}AV^{(n-1)} \\
  &= V^{-n}[AV - VA]V^{(n-1)} \\
  &= V^{-n}V[V^{-1}A - AV^{-1}]VV^{n-1} \\
  &= V^{-n}V[V^*,A]V^n \\
  &\equiv G_n\text{,}
\end{split}
\end{equation*}
so
\begin{equation*}
\begin{split}
\sum_{n=a}^{b}\left(\phi,G_n\phi\right)
  &= \sum_{n=a}^{b}\left(\phi,V^{-n}V[V^*,A]V^n\phi\right) \\
  &= \sum_{n=a}^{b}\left(V^n\phi,V[V^*,A]V^n\phi\right) \\
  &= \sum_{n=a}^{b}
    \left(C^{\frac{1}{2}}V^n\phi, C^{\frac{1}{2}}V^n\phi\right) \\
  &= \sum_{n=a}^{b}\lpar C^{\frac{1}{2}}V^n\phi\rpar^2\text{,}
\end{split}
\end{equation*}
where $C = V[V^*,A]$. We also have
\begin{equation*}
\sum_{n=a}^{b}\left(\phi,G_n\phi\right) = 
  \left(\phi,V^{-b}AV^b\phi \right) -
  \left(\phi,V^{-(a-1)}AV^{(a-1)}\phi \right)\text{.}
\end{equation*}
Taking the modulus and using the Schwartz inequality yields
\begin{equation*}
\begin{split}
\sum_{n=a}^{b}\lpar C^{\frac{1}{2}}V^n\phi\rpar^2
  &\leq 2 \left| \left(\phi, V^{-b}AV^b\phi\right) \right| \\
  &= 2 \left| \left(V^b\phi, AV^b\phi\right) \right| \\
  &\leq 2\lpar A \rpar \lpar V^b\phi \rpar^2 \\
  &= 2 \lpar A \rpar \lpar \phi \rpar^2 \\
  &< \infty
\end{split}
\end{equation*}
and thus $C^{1/2}$ is $V$-smooth.

Finally, that $V$ is absolutely continuous on $R(C^{1/2})$ follows directly
from Theorem~\ref{thrm:r_ranA_ac}.\proofend


\section{\label{sec:r_finite}Finite rank perturbations}

Here, I utilise the results of \refsec{sec:r_spectra} to show that
perturbations of the form \refeq{eq:r_rankNpert} lead to a pure point
spectrum for the Floquet operator for a.e.\ perturbation strength $\lambda$.

I use directly the definition of \emph{strongly $H$-finite} from Howland.


\begin{defn}[Strongly H-finite]
\tcaption{\textbf{Definition:} The operator $A$ is strongly $H$-finite}
\label{defn:r_strongHfinite}
Let $H$ be a self-adjoint operator on $\mathcal{H}$ with pure point spectrum,
$\phi_n$ a complete orthonormal set of eigenvectors, and
$H\phi_n = \alpha_n\phi_n$. A bounded operator
$A: \mathcal{H}\to \mathcal{K}$ is strongly $H$-finite if and only if
\begin{equation}
\label{eq:r_strongHfinite}
\sum_{n=1}^\infty |A\phi_n| < \infty\text{.}
\end{equation}
\end{defn}

If $H$ is thought of as a diagonal matrix on $l_2$, i.e.,\ $H = \sum_n
\alpha_n\left|\phi_n\rangle\langle\phi_n\right|$, and $A$ as an infinite
matrix $\{a_{ij}\}$, i.e.,\
$A = \sum_{m,n} a_{mn}\left|\phi_m\rangle\langle\phi_n\right|$, then
\refeq{eq:r_strongHfinite} says
\begin{equation}
\label{eq:r_A-strongHfinite}
\sum_n \left[ \sum_i \left|a_{in}\right|^2 \right]^{\frac{1}{2}} <
  \infty\text{.}
\end{equation}

For our purposes, we need to show that if $A$ is strongly $H$-finite, then it
is $U$-finite. To satisfy the assumption that $Q_\epsilon$ is trace class
in Lemma~\ref{lemma:r_5} (and hence also compact in Theorem~\ref{thrm:r_4})
we also need to show that $A$ is trace class.


\begin{thrm}
\tcaption{\textbf{Theorem:} If $A$ is $H$-finite, then $A$ is $U$-finite}
\label{thrm:r_ufinite}
If $A$ is strongly $H$-finite, then given $U=e^{iTH/\hbar}$ for the period
$T$ in \refeq{eq:r_hamiltonian} and $H\phi_n = \alpha_n\phi_n$,
\begin{enumerate}
\item $A$ is trace class, and
  \label{thrm:r_ufinitea}	
\item $A$ is $U$-finite.
  \label{thrm:r_ufiniteb}
\end{enumerate}
\end{thrm}


\emph{Proof.} (\ref{thrm:r_ufinitea}) Simply consider
\begin{equation}
\label{eq:r_trA}
\tr(A) = \sum_l \langle \phi_l \left| A \right| \phi_l \rangle
  = \sum_l a_{ll} \leq \sum_l \left| a_{ll} \right|\text{.}
\end{equation}
For each term in the sum \refeq{eq:r_trA} we trivially have
\begin{equation*}
\left| a_{ll} \right| \leq \sqrt{\sum_i \left| a_{il} \right|^2}
\end{equation*}
and thus \refeq{eq:r_trA} is finite so $A$ is trace class.

\vspace{\baselineskip}
(\ref{thrm:r_ufiniteb}) Noting that
\begin{equation*}
U|\phi_n\rangle = e^{iTH/\hbar}|\phi_n\rangle
    = e^{iT\alpha_n/\hbar}|\phi_n\rangle
\end{equation*}
we calculate, by insertion of a complete set of states,
\begin{equation*}
\begin{split}
\sum_n &\langle \phi_n | G_\epsilon(\theta;U,A)|\phi_n\rangle \\
  &= \sum_n \langle \phi_n
    |A\frac{1}{\left(1-U^*e^{-i\theta_-}\right)\left(1-Ue^{i\theta_+}\right)}
    A^*|\phi_n\rangle \\
  &= \sum_{n,m} \langle \phi_n | A | \phi_m \rangle \langle \phi_m |
    \frac{1}{\left(1-U^*e^{-i\theta_-}\right)\left(1-Ue^{i\theta_+}\right)}
    A^*|\phi_n\rangle \\
  &= \sum_{n,m} \frac{\langle \phi_n | A | \phi_m \rangle
    \langle \phi_m | A^* | \phi_n\rangle}
    {\left(1-e^{-iT\alpha_m/\hbar}e^{-i\theta_-}\right)
    \left(1-e^{iT\alpha_m/\hbar}e^{i\theta_+}\right)} \\
  &= \sum_{n,m} \frac{\langle \phi_m | A^* | \phi_n \rangle
    \langle \phi_n | A | \phi_m\rangle}
    {\left(1-e^{-iT\alpha_m/\hbar}e^{-i\theta_-}\right)
    \left(1-e^{iT\alpha_m/\hbar}e^{i\theta_+}\right)} \\
  &= \sum_m \frac{\langle \phi_m | A^*A | \phi_m \rangle}
    {\left|1-e^{-\epsilon}e^{iT\alpha_m/\hbar}e^{i\theta}\right|^2} \\
  &= \sum_n \frac{\langle \phi_n | A^*A | \phi_n \rangle}
    {\left|1-e^{-\epsilon}e^{iT\alpha_n/\hbar}e^{i\theta}\right|^2}\text{.}
\end{split}
\end{equation*}
The trace norm is then
\begin{equation*}
\tr G_\epsilon(\theta) = \sum_n \frac{| A\phi_n |^2}
  {\left|1-e^{-\epsilon}e^{iT\alpha_n/\hbar}e^{i\theta}\right|^2}\text{.}
\end{equation*}
If this is bounded for $\epsilon = 0$, then it is trivially bounded for all
$\epsilon > 0$. By \refeq{eq:r_strongHfinite} and a slightly modified version
of (Theorem~3.1, \cite{Howland87}) this is finite a.e.\ for $\epsilon = 0$.
Thus the trace norm of $G_\epsilon$ exists as $\epsilon\rightarrow 0$,
which implies that the strong limit of $G_\epsilon$ exists and we conclude
that $A$ is $U$-finite.\proofend


\begin{thrm}
\tcaption{\textbf{Theorem:} The Floquet operator $V$ is pure point if
$A_1,\ldots,A_N$ are strongly $H$-finite and commute with each other}
\label{thrm:r_sum_pert}
Let $U$ be a pure point unitary operator, and let $A_1,\ldots,A_N$ be
strongly $H$-finite. Assume that the $A_k$s commute with each other. Then
for a.e.\ $\lambda = (\lambda_1,\ldots,\lambda_N)$ in $\mathbb{R}^N$,
\begin{equation*}
V(\lambda) = e^{i\left(\sum_{k=1}^N \lambda_k A_k^*A_k\right)/\hbar}U
\end{equation*}
is pure point.
\end{thrm}


\emph{Proof.} This is a trivial modification of
(Theorem~4.3, \cite{Howland87}). Let
\begin{equation*}
\mathcal{K} = \bigoplus_{k=1}^N \overline{R(A_k)}\text{.}
\end{equation*}
The elements of $\mathcal{K}$ are represented as column vectors. The
operator $A:\mathcal{H}\rightarrow\mathcal{K}$ is defined, for
$y\in\mathcal{H}$, by
\begin{equation*}
Ay =
  \begin{bmatrix}
  A_1y \\ \vdots \\ A_Ny
  \end{bmatrix}
  = \begin{bmatrix}
  x_1 \\ \vdots \\ x_N
  \end{bmatrix}
\end{equation*}
and therefore $A^*: \mathcal{K}\rightarrow\mathcal{H}$ is given by
\begin{equation*}
A^*x = A_1^*x_1 + \cdots + A_N^*x_N\text{.}
\end{equation*}
Accordingly,
$G_\epsilon(\theta): \mathcal{K}\rightarrow\mathcal{K}$, the matrix equivalent
of equation \refeq{eq:r_Geps}, is introduced,
\begin{equation*}
\begin{split}
G_\epsilon(&\theta;U,A) \\
  &= A\bigl[1-U^*e^{-i\theta_-}\bigr]^{-1}
  \bigl[1-Ue^{i\theta_+}\bigr]^{-1}A^* \\
  &= \bigl\{A_i
    \bigl[1-U^*e^{-i\theta_-}\bigr]^{-1}\bigl[1-Ue^{i\theta_+}\bigr]^{-1}
    A_j^*\bigr\}_{1\leq i,j \leq N}\text{.}
\end{split}
\end{equation*}
The diagonal terms are finite a.e.\ because each $A_k$ is
$U$-finite by Theorem~\ref{thrm:r_ufinite}. The off-diagonal terms are
of the form $X_1^*X_2$, and so the Schwartz inequality,
\begin{equation*}
\left|X_1^*X_2\right|^2 \leq \lpar X_1 \rpar^2 \lpar X_2 \rpar^2
\end{equation*}
ensures that they are finite a.e.\ too. Hence, $A$ is $U$-finite as every
term in the matrix $G_\epsilon(\theta;U,A)$ is a.e.\ finite as
$\epsilon\rightarrow 0$.

The Hamiltonian may now be written as
\begin{equation}
\label{eq:r_hamiltonian_N}
H(\lambda) = H_0 + A^*W(\lambda)A \sum_{n=0}^\infty \delta(t-nT)
\end{equation}
and the Floquet operator as
\begin{equation*}
V(\lambda) = e^{iA^*W(\lambda)A/\hbar}U
\end{equation*}
where $W(\lambda) = \text{diag}\{\lambda_k\}$. In this form, the formalism
of \refsec{sec:r_spectra} is essentially fully regained, and we can proceed
to apply Theorems~\ref{thrm:r_2}, \ref{thrm:r_4}, \ref{thrm:r_6} and
\ref{thrm:r_7}.

To establish the absolute continuity of the multiplication operator
$\mathbb{V}$ on the space $L^2(\mathbb{R}^N;\mathcal{H})$ we proceed as in
Proposition~\ref{prop:r_8}. Write
\begin{equation*}
V(t_1,\ldots,t_N) = e^{ic\sum_{k=1}^N t_k A_k^* A_k / \hbar} U\text{,}
\end{equation*}
define
\begin{equation*}
D = -\sum_{k=1}^N \arctan(p_k/2)\text{,}
\end{equation*}
where $p_k = -id/dt_k$, and compute
\begin{equation*}
C = \mathbb{V}[\mathbb{V}^*,D] = \sum_{k=1}^N C_k \geq 0\text{.}
\end{equation*}
In obtaining $C$ as a direct sum of the $C_k$, we have had to assume that
the $A_k$s commute with each other. This complication comes when considering
the term
\begin{equation*}
V(t_1,\ldots,t_N)V^*(t_1,\ldots,t_{k-1},y_k,t_{k+1},\ldots,t_N)
\end{equation*}
in the equivalent of \refeq{eq:r_C}. To obtain the required form of
$e^{ic(t_k-y_k)A^*_kA_k}$ we require that the $A_k$s commute.\footnote{This
restriction is not required in
Howland's self-adjoint work because the summation over $k$ in the
Hamiltonian \refeq{eq:r_hamiltonian_N} enters directly, rather than in the
exponent of $V$.}

Moving on, each $C_k \geq 0$ is equivalent to $C$ in
Proposition~\ref{prop:r_8}
and hence positive. Finally, we must show that $R(C^{1/2})$ is cyclic for
$\mathbb{V}$. This is no longer trivial as, for each $k$, while
$R(C_k^{1/2}) = R(A_k^*)$, the range of $A_k^*$ is not cyclic for $U$, hence
$\mathbb{V}$. To proceed, first note that
\begin{equation*}
R(A^*) = \bigcup_{k} R(A_k^*)\text{.}
\end{equation*}
Now, as argued in Howland, we can assume that $R(A^*)$ is cyclic for $U$.
To elaborate, define $\mathcal{M}(U, R(A^*))$ to be the smallest closed
reducing subspace of $\mathcal{H}$ containing $R(A^*)$. If $R(A^*)$ is not
cyclic for $U$, then $\mathcal{H} \ominus \mathcal{M}$ is not empty. 
However, as shown below, if $y\in\mathcal{H}\ominus\mathcal{M}$, then
$A^*WAy=0$, so in $\mathcal{H}\ominus\mathcal{M}$, $V(t) = U$ and is
therefore pure point trivially. Thus, we can ignore the space
$\mathcal{H}\ominus\mathcal{M}$ and restrict the discussion to
$\mathcal{M}$---i.e.,\ we may assume $R(A^*)$ cyclic for $U$.

The above relied upon showing that $A^*WAy=0$ for
$y\in\mathcal{H}\ominus\mathcal{M}$. I now prove this. If
$y\in\mathcal{H}\ominus\mathcal{M}$ and $y'\in\mathcal{M}$, then
\begin{equation*}
\langle y, y'\rangle = 0\text{.}
\end{equation*}
Given $y'\in\mathcal{M}$, there exists an $x\in\mathcal{K}$ such that
$y'=A^*x$, so
\begin{equation*}
\langle y, A^*x\rangle = 0\text{.}
\end{equation*}
That is
\begin{equation*}
\langle Ay, x\rangle = 0\text{.}
\end{equation*}
This is true for all $x\in\mathcal{K}$.
Suppose $y''\in\mathcal{H}$. Then $WAy''\in\mathcal{K}$ and so
\begin{equation*}
\langle Ay, WAy''\rangle=0\text{.}
\end{equation*}
That is
\begin{equation*}
\langle A^*WAy, y''\rangle=0\text{.}
\end{equation*}
As this is true for any $y''\in\mathcal{H}$, we conclude that $A^*WAy=0$ on
$\mathcal{H}\ominus\mathcal{M}$.

Thus, $R(A^*)$ (with $A$ acting on $L^2(\mathbb{R^N};\mathcal{H})$) may be
assumed cyclic for $U$, hence cyclic for $\mathbb{V}$.

I must finally show that $R(C^{1/2}) = R(A^*)$. We have
\begin{equation*}
R(A^*) = \bigcup_k R(A_k^*) = \bigcup_k R(C_k^{1/2})
\end{equation*}
and
\begin{equation*}
R(C) = \bigcup_k R(C_k)\text{.}
\end{equation*}
As $R(A^*) = R(A^*A)$, $R(C^{1/2}) = R(C)$ and we have shown that
$R(C^{1/2}) = R(A^*)$ as required.
\proofend


Finally, I wish to make the connection with my original aim---to show
that Hamiltonians of the form
\begin{equation}
\label{eq:r_rankNham}
H(t) = H_0 + \sum_{k=1}^N \lambda_k | \psi_k\rangle\langle\psi_k |
  \sum_{n=0}^\infty \delta(t-nT)
\end{equation}
have a pure point quasi-energy spectrum.


\begin{thrm}
\tcaption{\textbf{Theorem:} The Floquet operator $V$ is pure point for
rank-$N$ perturbations}
\label{thrm:r_rankNpp}
Let $H_0$ be pure point, and define our time-dependent Hamiltonian as
in \refeq{eq:r_rankNham}. If $\psi_1,\ldots,\psi_N \in l_1(H_0)$, then
for a.e.\ $\lambda=(\lambda_1,\ldots,\lambda_N)$ in $\mathbb{R}^N$, the
Floquet operator
\begin{equation*}
V = e^{i\left(
  \sum_{k=1}^N \lambda_k |\psi_k\rangle\langle\psi_k|\right)/\hbar}U
\end{equation*}
has pure point spectrum.
\end{thrm}


\emph{Proof.} This theorem is just a special case of
Theorem~\ref{thrm:r_sum_pert} with the $A_k$s given by
$| \psi_k \rangle\langle \psi_k |$. Noting \refeq{eq:r_orth_states},
the $A_k$s clearly commute. As Howland
shows, $|\psi\rangle\langle\psi|$ is strongly $H$-finite if and only if
$\psi \in l_1(H_0)$. Thus Theorem~\ref{thrm:r_sum_pert} applies and the result
follows.\proofend


\section{\label{sec:r_discussion}Discussion of results and potential
applications}

Of fundamental importance in showing that the quasi-energy spectrum remains
pure point for a.e.\ perturbation strength $\lambda$ was the fact that
$\psi_k \in l_1(H_0)$. That is, if we write
\begin{equation*}
|\psi_k\rangle = \sum_{n=0}^\infty (a_k)_n | \phi_n\rangle\text{,}
\end{equation*}
where the $| \phi_n\rangle$ are the basis states of $H_0$, then $\psi_k \in
l_1(H_0)$ if and only if
\begin{equation*}
\sum_{n=0}^\infty |(a_k)_n| < \infty\text{.}
\end{equation*}
If this requirement is dropped, and we only retain $\psi_k \in l_2(H_0)$,
then (Theorem~3.1, \cite{Howland87}) fails and there is the possibility
that $V(\lambda)$ will have a non-empty continuous spectrum. It was this
fact that Milek and Seba \cite{Milek90.1} took advantage of in showing that
the rank-$1$ kicked rotor could contain a singularly continuous spectral
component under certain conditions on the ratio of the kicking frequency
and the fundamental rotor frequency. They analysed two regimes of the
perturbation. One where $\psi \in l_1(H_0)$, in which case the numerical
results clearly showed pure point recurrent behaviour, and the other where
$\psi \in l_2(H_0)$, but $\psi \notin l_1(H_0)$. In the second case, the
authors further proved that the absolutely continuous part of the spectrum
was empty,\footnote{It turns out that Milek and Seba actually made an
assumption in obtaining this result which is as yet is unjustified.
\refchap{chap:comb} investigates this in detail.} and thus the system
contained a singularly continuous spectral component. The numerical results
reflected this, with a diffusive type energy growth being observed.

With the generalisation of Combescure's work here, namely
Theorem~\ref{thrm:r_rankNpp}, it is now possible to investigate the
full class of rank-$N$ kicked Hamiltonians. A sufficient requirement for
recurrent behaviour has been shown to be $\psi_k \in l_1(H_0)$ and so
I must turn my attention to perturbations where this requirement is
no longer satisfied. This is the topic of \refchap{chap:comb}.


\section{\label{sec:r_summary}Summary}

I have shown, in a rigorous and general fashion, that the spectrum of the
Floquet operator remains pure point for perturbations which are constructed
from projection operators that are in turn built from Hilbert space
vectors which are elements of $l_1(H_0)$. In simple terms, with the
Hamiltonian perturbation $W = |\psi\rangle \langle \psi |$, one requires
\begin{equation*}
\sum_n \left| W^{1/2} \phi_n \right| < \infty
\end{equation*}
for the Floquet operator to have a pure point spectrum for almost every
perturbation strength.

As alluded to in \refsec{sec:r_discussion}, to investigate systems that may
display chaotic behaviour I would like to relax, in a
controlled manner, the conditions that lead to a pure point spectrum for
$V$. The emergence of a continuous spectrum is a vital ingredient in making
further progress. \refchap{chap:comb} investigates this question.

\clearemptydoublepage
\chapter[A generalisation of the work of Combescure and Milek \& Seba]
{\label{chap:comb}A Generalisation of the Work of Combescure and
Milek \& Seba}
\minitoc



This chapter extends the results of Combescure \cite{Combescure90} in a
number of ways. Firstly, as an alternative to the work just presented in
\refchap{chap:rankN} I directly extend Combescure's result on rank-$1$
perturbations to rank-$N$ perturbations---pleasingly, the same results as
in \refchap{chap:rankN} are obtained. I will then move on to the important
question of the emergence of a continuous spectral component of the
Floquet operator. I extend all of the results of Combescure and also those
of Milek and Seba \cite{Milek90.1} to rank-$N$ perturbations.

The investigations lead, in a natural way, to a conjecture presented by
Combescure \cite{Combescure90} concerning the dependence of her results on
the particular $H_0$ eigenvalue sequence. The analysis herein leads to a
number-theoretic conjecture that has stood for over fifty years
\cite{Vinogradov}\footnote{The reference is to the 1954 English translation
of Vinogradov's original work, published in 1947. The work in Vinogradov's
1947 monograph incorporates results from a series of papers and a first
monograph from 1937. It is unknown (to me) when the conjecture I refer to
was first presented, but it was at least fifty years ago.} on the
estimation of finite exponential sums. Work already done in this area
\cite{Bourget02.1} will be examined in detail.

In examining the work of Milek and Seba, I highlight a number of
misconceptions and rectify them. Worryingly though, their work is not in
fact fully justified---a point so far missed in the literature. The
resolution is directly linked to the number-theoretic investigations just
mentioned.


\section{\label{sec:c_outline}Outline and summary of results}

In this chapter, I consider Hamiltonians of the form
\begin{equation}
\label{eq:c_hamiltonian}
  H(t) = H_0 + \left(\sum_{k=1}^N \lambda_k|\psi_k\rangle\langle\psi_k|\right)
    \sum_{n=0}^\infty \delta(t-nT)
\end{equation}
where $\lambda_k \in \mathbb{R}$ and each vector $|\psi_k\rangle$ is a
linear combination of the $H_0$ basis states, $|\phi_n\rangle$. See
\refsec{sec:r_outline} for the definitions and properties of these objects.

As already discussed, the basic result is that if every
$|\psi_k\rangle \in l_1(H_0)$ the spectrum will remain pure point
for almost every set of perturbation strengths $\lambda_k$. If this
condition is dropped for any one of the $|\psi_k\rangle$ then
$V_{\lambda_1,\ldots,\lambda_N}$ is no longer pure point. On the
subspace $\mathcal{H}_k$, the space for which $|\psi_k\rangle$ is
a cyclic vector for the operator $U$, the spectrum is purely continuous.

The other key result of this chapter concerns a number-theoretic conjecture
stated by Vinogradov \cite{Vinogradov}. For Milek and Seba's work to
be properly justified, a sufficient condition is for Vinogradov's
conjecture to be true. This observation is linked to the conjecture put
forward by Combescure \cite{Combescure90} and partially addressed by Bourget
\cite{Bourget02.1}.

In \refsec{sec:c_comb1} I extend Combescure's rank-$1$ theorem on the pure
point spectral nature of $V$ to the rank-$N$ case. In \refsec{sec:c_comb2}
I then show the existence of a continuous spectrum for the case where $H_0$
is the harmonic oscillator and the perturbation is rank-$N$. In
\refsec{sec:c_genspec} I investigate Combescure's conjecture, the partial
answer provided by Bourget and the link to number theory and Vinogradov's
conjecture. Finally, in \refsec{sec:c_milek}, I extend Milek and Seba's work
to the rank-$N$ case. A number of conceptual and mathematical errors are
firstly highlighted and then resolved. 


\section{\label{sec:c_comb1}A rank-N generalisation of Combescure's first
theorem}

Consider the measures
\begin{equation*}
m_{k,\lambda_k} = \langle\psi_k | E_{\lambda_k}(S) | \psi_k\rangle\text{.}
\end{equation*}
Each $|\psi_k\rangle$ admits a cyclic subspace of $\mathcal{H}$,
$\mathcal{H}_k$. As argued in the later part of the proof of
Theorem~\ref{thrm:r_sum_pert}, on the space
$\mathcal{H}\ominus \left(\bigoplus_{k=1}^N \mathcal{H}_k\right)$, the
perturbation
\begin{equation*}
\sum_{k=1}^N \lambda_k |\psi_k\rangle\langle\psi_k|
\end{equation*}
is null and thus $V=U$ is trivially pure point. Henceforth, we may safely
restrict the proof to the subspace $\bigoplus_{k=1}^N \mathcal{H}_k$ for
which the vectors $|\psi_k\rangle$ form a cyclic set.

Directly following Combescure, the measure for a point
$x \in [0,2\pi)$ for the operator $V$ acting on the state $|\psi_k\rangle$
is given by
\begin{equation}
\label{eq:c_measure}
m_{k,\lambda_k}(\{x\}) = \frac{-4(1+\mu_k)}{\mu_k^2}B_k(x)\text{,}
\end{equation}
where
\begin{equation*}
\mu_k = e^{i\lambda_k/\hbar} - 1
\end{equation*}
and
\begin{equation*}
B_k(x) = \left[\int_0^{2\pi}
dm_{k,\lambda_k=0}(\theta)\,\left(\sin^2\left[(x-\theta)/2\right]\right)^{-1}
  \right]^{-1}\text{.}
\end{equation*}
This result is the essence of Lemma~1 in Combescure's work. When $H_0$ is
pure point, it is a trivial calculation to show that
\begin{equation}
\label{eq:c_Binv}
B_k^{-1}(x) = \sum_{n=0}^\infty \frac{|(a_k)_n|^2}
  {\sin^2\left[(x-\theta_n)/2\right]}\text{.}
\end{equation}
Corollary~2 in Combescure's work is replaced with the following.

\begin{thrm}
\tcaption{\textbf{Theorem:} $e^{ix}$ is an eigenvalue of the Floquet
operator $V$ if and only if $\prod_{k=1}^N B_k^{-1}(x) < \infty$}
\label{thrm:c_comb1}
Assume $H_0$ is pure point, with $\{\phi_n\}_{n\in\mathbb{N}}$ and
$\{\alpha_n\}_{n\in\mathbb{N}}$ as eigenstates and eigenvalues. Let each
\begin{equation*}
\psi_k = \sum_{n=0}^\infty (a_k)_n\phi_n
\end{equation*}
be cyclic for $H_0$ (hence, cyclic for $U$ and $V$) on $\mathcal{H}_k$ and
$\langle\psi_k | \psi_l\rangle = \delta_{kl}$. Then $e^{ix}$ belongs to the
point spectrum of $V_{\lambda_1,\ldots,\lambda_N}$ if and only if
\begin{equation*}
\prod_{k=1}^N B_k^{-1}(x) < \infty\text{,}
\end{equation*}
where
\begin{equation*}
\theta_n = 2\pi\{\alpha_n/2\pi\hbar\}\text{,}
\end{equation*}
$\{z\}$ being the fractional part of $z$.
\end{thrm}

\emph{Proof.} (\ref{thrm:c_comb1})
The proof follows that in Combescure. By the cyclicity of each
$|\psi_k\rangle$ on $\mathcal{H}_k$ and the argument in
Theorem~\ref{thrm:r_sum_pert}, $e^{ix}$ is an
eigenvalue of $V_{\lambda_1,\ldots,\lambda_N}$ if and only if \emph{every}
$m_{k,\lambda_k}(\{\theta\}) \neq 0$ at $\theta=x$. As already
mentioned, using
\begin{equation*}
dm_{k,\lambda_k=0} = \sum_{n=0}^\infty |(a_k)_n|^2
  \delta(\theta-\theta_n)d\theta
\end{equation*}
we obtain, for each $k$,
\begin{equation*}
B_k^{-1}(x) = \sum_{n=0}^\infty \frac{|(a_k)_n|^2}{\sin^2[(x-\theta_n)/2]}
\text{.}
\end{equation*}
Now consider the eigenvalue $e^{ix}$. If it were to be that for some
$k$, $m_{k,\lambda_k}(\{x\}) = 0$, then we would have found a vector, namely
$|\psi_k\rangle$, such that $V|\psi_k\rangle$ was continuous. We have in
fact found that the whole subspace $\mathcal{H}_k$ is continuous.
Thus, for $V$ to be pure point, every
$m_{k,\lambda_k}(\{\theta\})\neq 0$. Thus, we are lead to consider the
requirement
\begin{equation*}
\prod_{k=1}^N B_k^{-1}(x) < \infty\text{.}
\end{equation*}
\proofend
\longpage

As in Combescure, the relationship
\begin{equation}
\label{eq:c_cotg}
\sum_{n=0}^\infty |(a_k)_n|^2\cotg\left(\frac{x-\theta_n}{2}\right)
  = \cotg\frac{\lambda_k}{2\hbar}
\end{equation}
also holds for each $k$. To show \refeq{eq:c_cotg}, consider each $k$
separately. The proof is the same as for the rank-$1$ case.
See \cite{Combescure90}. Points to consider are that each projection
operator in the rank-$N$ projection is normalised and hence for every
$k$ we have
\begin{equation*}
\sum_{n=0}^N |(a_k)_n|^2 = 1\text{.}
\end{equation*}
In order to complete the generalisation of Combescure's first theorem, we 
require, just as in Combescure, two additional Lemmas.

\begin{lemma}
\tcaption{\textbf{Lemma:} If $\sum_{n=0}^\infty |(a_k)_n| < \infty$, then
$B_k^{-1}(x)<\infty$ for a.e. $x\in\mathbb{R}$}
\label{lemma:c_lemma1}
If $\sum_{n=0}^\infty |(a_k)_n| < \infty$, then $B_k^{-1}(x)<\infty$ for
almost every $x\in\mathbb{R}$.
\end{lemma}

For each $k\in 1,\ldots,N$ the proof is identical to that in Combescure.

\begin{lemma}
\tcaption{\textbf{Lemma:} The statement ``for almost every
$(\lambda_1,\ldots,\lambda_N)$, $V_{\lambda_1,\ldots,\lambda_N}$
has only a point spectrum'' is equivalent to the statement ``for all
$k\in\{1,\ldots,N\}$ and for almost every $x$, $B_k(x) \neq 0$''}
\label{lemma:c_lemma2}
The following two statements are equivalent.
\begin{enumerate}
\item For almost every $(\lambda_1,\ldots,\lambda_N)$,
  $V_{\lambda_1,\ldots,\lambda_N}$ has only a point spectrum.
\item For every $k\in\{1,\ldots,N\}$ and for almost every $x$,
$B_k(x)\neq 0$.
\end{enumerate}
\end{lemma}

The proof is again virtually identical to Combescure's proof. For each $k$,
the continuous part of the spectrum is supported outside the set
$E_k = \{x\in[0,2\pi): B_k(x)\neq 0\}$ and, for $\lambda_k\neq 0$,
the point part of $dm_{k,\lambda_k}$ is supported by the set $E_k$. Thus,
for $V_{\lambda_1,\ldots,\lambda_N}$ to be pure point for almost every
$\lambda_1,\ldots,\lambda_N$ and for every $k$, we require
\begin{equation*}
m_{k,\lambda_k}([0,2\pi)\setminus E_k) = 0\text{.}
\end{equation*}
This in turn implies that for every $k$
\begin{equation*}
\int_0^{2\pi} d\lambda_k'\,h(\lambda_k')m_{k,\lambda_k}([0,2\pi)\setminus E_k)
  = 0\text{,}
\end{equation*}
where $\lambda_k' = \lambda_k/\hbar$ and
\begin{equation*}
h(\lambda) = 2\Re\frac{1}{1-ce^{i\lambda}}
\end{equation*}
for some $|c|<1$.

Combescure's Lemma 5 trivially applies for each $k$. Thus, I have
generalised Combescure's work to obtain the result that the Floquet
operator for the rank-$N$ perturbed Hamiltonian has a pure point spectrum.
As already mentioned, the result matches that obtained in
\refchap{chap:rankN}.


\section{\label{sec:c_comb2}A rank-N generalisation of Combescure's second
theorem}

Having shown that the Floquet operator remains pure point for perturbations
constructed from the vectors $|\psi_k\rangle\in l_1(H_0)$, Combescure relaxes
this condition to allow for the emergence of a continuous spectral
component of the Floquet operator. This result is easily generalised to
the rank-$N$ case. The key point is that Combescure's technique applies
independently for each $k$. I do not discuss the details of the rank-$1$
proof here at all, delaying an analysis to \refsec{sec:c_genspec} where I
will have the opportunity to generalise the results still further. Here, I
simply provide the argument for why each $k$ may be treated independently.
Before proceeding, some subtleties of what Combescure actually
shows are highlighted. They are seemingly overlooked by some in the
literature (e.g.,\ \cite{Milek90.1}).

The cyclicity requirement was essential in the proof that the Floquet
operator spectrum was pure point. Here, we can happily ignore the
cyclicity conditions, as our only goal is to establish the
existence of a state in the continuous subspace $\mathcal{H}_\cont$.
We need not try and ensure the result obtained by
considering $\langle\psi|E(S)|\psi\rangle$ is applicable to all other
vectors in $\mathcal{H}$---the very idea is ill-formed as the perturbation
is null on a subset of $\mathcal{H}$ and thus there is always part of
$\mathcal{H}$ where $V$ has a discrete spectrum. Milek and Seba seem to
have missed this point, restating Combescure's theorem in a way that
implies that all $\psi$ are in $\mathcal{H}_\cont$.

If
\begin{equation*}
\langle\psi_k|E(\{x\})|\psi_k\rangle = 0 
\end{equation*}
then $\mathcal{H}_{ac}$ contains at least the state
$|\psi_k\rangle$. The point to be mindful of is that this does not allow
one to conclude that the Hilbert space for the operator $V$ has
$\mathcal{H}_{pp} = \emptyset$. It is this error that Milek and Seba
\cite{Milek90.1} have made. To draw that conclusion would require an
argument to show that a cyclic vector does in fact exist for $V$. This does
not seem possible in the general context here.

Combescure's proof (Lemma~6 in \cite{Combescure90}) that $\sigma_\cont(V)
\neq \emptyset$ is based on showing
that $B^{-1}(x)\rightarrow\infty$ (equation \refeq{eq:c_Binv}). As the
spectral measure of a single point $x$ is proportional to $B(x)$
(equation \refeq{eq:c_measure}), if $B^{-1}(x)\rightarrow\infty$, then the
contribution of the single point is zero. That is, $e^{ix}$ is in the
continuous spectrum of the Floquet operator. Combescure argues
(see \refsec{sec:c_genspec} for details) that
\begin{equation*}
B^{-1}(x) \geq \#S(x)
\end{equation*}
where $\#S(x)$ is the number of elements of a particular set $S$. She then
shows (the bulk of the proof) that $\#S(x)\rightarrow\infty$ and thus
$B^{-1}(x)\rightarrow\infty$. I generalise the result in a
straightforward manner.

\begin{thrm}
\tcaption{\textbf{Theorem:} Assume $\alpha_n = n\hbar\omega$ with $\omega$
irrational. If $\sum_{n=0}^\infty |(a_k)_n|\rightarrow\infty$ for at least
one $k \in 1,\ldots,N$, then $\sigma_\cont(V) \neq \emptyset$}
\label{thrm:c_comb2}
Assume $\alpha_n = n\hbar\omega$ with $\omega$ irrational.
If $|\psi_k\rangle \notin l_1(H_0)$ for at least one $k\in 1,\ldots,N$,
then $\sigma_\cont(V) \neq \emptyset$.
\end{thrm}

\emph{Proof.} (\ref{thrm:c_comb2})
Following the same argument as for the rank-$1$ case, we take
\begin{equation*}
|(a_k)_n| = n^{-\gamma}2\pi
\end{equation*}
for the state $|\psi_k\rangle$, in such a way that the condition
$\langle\psi_k|\psi_l\rangle=\delta_{kl}$ is preserved.

With this construction, Combescure's proof that the number of
elements in $S(x)$ is infinite applies to each subsequence $S_k(x)$. The
number of elements, $\#S_k(x)$, in each sub-sequence for which
$|\psi_k\rangle \notin l_1(H_0)$, is infinite. The Floquet operator for
the rank-$N$ perturbed harmonic oscillator obtains a continuous spectral
component.
\proofend

\subsection{Discussion}

It must be noted that the proof presented by Combescure (Lemma~6,
\cite{Combescure90}) is only valid for the eigenvalue spectrum,
\begin{equation*}
\alpha_n = n\hbar\omega
\end{equation*}
of the harmonic oscillator. Combescure does however conjecture that the
argument will be valid for more general eigenvalue spectra, including the
rotor,
\begin{equation*}
\alpha_n \propto n^2\text{.}
\end{equation*}
For Milek and Seba's numerical work (using the rotor) to be based on
valid mathematical arguments, a proof of this conjecture is required.
Currently, no such proof exists. In \refsec{sec:c_genspec} I show that if
a conjecture from number theory on the estimation of exponential sums is
true, then Milek and Seba's work can be justified. The rank-$N$
generalisation is straightforward.  Considering the number theory
conjecture has stood for over fifty years, it seems we may have to wait
quite some time for a proof.

For more general eigenvalue spectra (loosely $\alpha_n \propto n^j$) the
situation is significantly better. For $j\geq 3$ Bourget
\cite{Bourget02.1} has made significant progress. A continuous component
of the Floquet operator exists for certain constructions of
$|\psi\rangle$. The conditions are complicated and more restrictive than
the $|\psi\rangle \notin l_1(H_0)$ condition for the harmonic oscillator.
The result is easily extended to the rank-$N$ case due to the independence
of each $k$ as already discussed. Utilising the same number-theoretic
conjecture as in the $j=2$ case will also allow for improvements to the
work of Bourget. See \refsec{sec:c_genspec}.

Returning to the harmonic oscillator case, by applying
Theorem~\ref{thrm:c_comb2} we may conclude that for each
$|\psi_k\rangle \notin l_1(H_0)$, $\mathcal{H}_k$ is purely continuous.
Thus, by dropping the $l_1$ condition for all $|\psi_k\rangle$, I have
shown that $V$ is purely continuous on the subspace of $\mathcal{H}$ where
the perturbation is non-zero. On the subspace of $\mathcal{H}$ where the
perturbation is zero, $V=U$ trivially and thus that portion of the
Hilbert space remains pure point.


\section{\label{sec:c_genspec}Combescure's conjecture and number theory}

Combescure makes a remark (Remark~c., \cite{Combescure90}) that she
believes Theorem~\ref{thrm:c_comb2}
(Lemma~6, Combescure \cite{Combescure90}) is generalisable to include
systems other than the harmonic oscillator. Explicitly, she conjectures
that Hamiltonians, $H_0$, with eigenvalues, $\alpha_n$, of the form
\begin{equation}
\label{eq:c_eigenvalues}
\alpha_n = \hbar \sum_{j=0}^p \beta_j n^j
\end{equation}
with $\beta_j T / 2\pi$ Diophantine for some $j: 1 \leq j \leq p$ will
have the vector $\psi$ in the continuous spectral subspace of $V_\lambda$.

At an intuitive level, one would expect this to be true. The precise nature
of the eigenvalue spectra (proportional to $n$ or a polynomial in $n$)
should not make a significant difference. Milek \cite{Milek89}
argues that Combescure's work can be used in the $n^2$ case based on
evidence from some numerical work that shows that the sequences obtained
are ``almost random''---however, the argument is not entirely convincing to
me.
The 
cited numerical work of Casati \etal\ \cite{Casati85}
discusses
the existence of correlations in the energy levels, rather than the lack of
correlations. While the deviations from a Poisson distribution look small
to the naked eye, Casati \etal\ \cite{Casati85} find deviations from the
expected
Poisson distribution of up to 17 standard deviations. The energy levels are
correlated---it is arguable that they are not characterisable as ``almost
random'' as Milek asserts.

I began to explore the possibility of developing a proof to the conjecture.
A few interesting results have come from this investigation and will be
presented here. While doing this work, I was unaware that in late 2002,
Bourget \cite{Bourget02.1} produced a proof of a slightly modified
conjecture for all but the $p=2$ case in \refeq{eq:c_eigenvalues}. The
techniques used by Bourget are the same as those followed in my work. I
will analyse Bourget's work, and highlight the key breakthrough made.  I
also provide a modified argument to obtain the proof which is, I believe,
significantly easier to follow. Importantly, it also covers the $p=2$ case
missed by Bourget due to technical difficulties. However, it comes at the
expense of relying upon a (quite reasonable) conjecture. I do not claim
that what is presented is adequate on its own, but it does play a
complementary role in understanding, or perhaps appreciating, Bourget's
proof. The reliance on the conjecture simply removes the need for much of
the technical wizardry in Bourget's proof. Use of the conjecture also
strengthens the work. The work also indicates, or highlights, that
Combescure's conjecture is solved by a number-theoretic conjecture that has
stood for over fifty years. What seems a perfectly reasonable conjecture on
physical grounds is shown to be directly related to an abstract
mathematical conjecture.

In what follows, I will rely heavily upon the lemmas and theorems in
Chapter~2 of \cite{Kuipers}. I also use some results on Weyl sums from
\cite{Vinogradov}. Of key importance is an understanding of Combescure's
proof of her Lemma~6 on the emergence of a continuous spectrum for the
kicked harmonic oscillator. This will be discussed at the appropriate time
in this section.

\subsection{Number theory}

To investigate Combescure's conjecture we require two concepts from
number theory---the classification of irrational numbers and the
\emph{discrepancy} of a sequence. I first introduce the concepts and
define the relevant ideas. I then proceed to analyse the conjecture
and the proof provided by Bourget. As the discussion progresses, the new
work that I have done will be presented.

For any number $\beta$, define
\begin{itemize}
\item $[\beta]$, the integer part of $\beta$,
\item $\{\beta\}$, the fractional part of $\beta$, and
\item $\langle \beta \rangle = \min(\{\beta\}, 1-\{\beta\})$.
\end{itemize}
$\langle\beta\rangle$ is simply the ``distance to the nearest integer''.
Definition~\ref{defn:c_type_eta} is taken directly from Kuipers and 
Niederreiter (Definition~3.4, p.~121, \cite{Kuipers}).

\begin{defn}
\tcaption{\textbf{Definition:} The irrational number $\beta$ is of
type $\eta$}
\label{defn:c_type_eta}
Let $\eta$ be a positive real number or infinity. The irrational, $\beta$,
is of type $\eta$ if $\eta$ is the supremum of all $\tau$ for
which
\begin{equation}
\label{eq:c_type_eta}
\varliminf_{n\rightarrow\infty} q^\tau \langle q\beta \rangle = 0\text{,}
\end{equation}
where $q$ runs through the positive integers.
\end{defn}

The idea behind this definition can be seen by considering \emph{rational}
$\beta = p/q'$ for integers $p$ and $q'$. Run through the positive
integers $q$. At $q=q'$, $\langle q\beta \rangle= 0$, and so there is no
supremum $\eta$ for $\tau$ in \refeq{eq:c_type_eta}. In effect,
$\eta\rightarrow\infty$. For irrational $\beta$,
$\langle q\beta \rangle$ is never equal to zero but will approach zero. If
the approach is very slow, then a small $\tau$ is enough to prevent
\refeq{eq:c_type_eta} from approaching zero. $\langle q\beta \rangle$
approaching zero slowly is, in a sense, indicative of $\beta$ being badly
approximated by rational numbers. Even for very large $q'$, $p/q'$ remains
a poor approximation to $\beta$. Thus, the smaller $\eta$, the stronger
the irrationality of $\beta$. This is reasonable in the sense that rational
$\beta$s act like numbers with $\eta\rightarrow\infty$. As stated in
\cite{Kuipers}, all numbers $\beta$ have type $\eta\geq 1$.

I now define the discrepancy of a sequence---a measure of the
non-uniformity of the sequence. Consider a sequence of
numbers\footnote{Equivalently, consider any sequence $x_n$ and consider the
discrepancy of the sequence modulo~1.} $x_n$ in $[0,1)$,
\begin{equation*}
\omega = (x_n)_{n\in\mathbb{N}} \text{ with } x_n\in[0,1)\text{.}
\end{equation*}
For $0\leq a < b \leq 1$ and positive integer $N$, $A([a,b),N)$ counts the
number of terms of the sequence (up to $x_N$) contained in the interval
$[a,b)$,
\begin{equation*}
A([a,b),N) = \#\{n \leq N: x_n\in[a,b)\}\text{.}
\end{equation*}

\begin{defn}
\tcaption{\textbf{Definition:} The sequence $\omega$ has a discrepancy
$D_N(\omega)$}
\label{defn:c_discrepancy}
The discrepancy $D_N$ of the sequence $\omega$ is
\begin{equation}
\label{eq:c_discrepancy}
D_N(\omega) = \sup_{0\leq a < b \leq 1}
  \left| \frac{A([a,b),N)}{N} - (b-a)\right|\text{.}
\end{equation}
\end{defn}

If the sequence $\omega$ is uniformly distributed in $[0,1)$ then
$D_N\rightarrow 0$ as $N\rightarrow\infty$. In this case, every interval
$[a,b]$ in $[0,1)$ gets its  ``fair share'' of terms from the sequence
$\omega$.

Estimating the discrepancy of a sequence will turn out to be vital in the
analysis of Combescure's work. The sequence of interest is basically the
eigenvalue sequence for $H_0$, but I will discuss this in greater detail
later.

The starting point for the estimations that we require is
(equation~(2.42), Chapter~2, \cite{Kuipers}). This is a famous result
obtained by Erd\"os and Tur\'an. It states that
\begin{equation}
\label{eq:c_gendisc}
D_N \leq C\left(\frac{1}{m} +
  \sum_{h=1}^m \frac{1}{h}\left|\frac{1}{N}
  \sum_{n=1}^N e^{2\pi i h x_n} \right| \right)
\end{equation}
for any real numbers $x_1,\ldots,x_N$ and any positive integer $m$.
The sum
\begin{equation*}
S = \sum_{n=1}^N e^{2\pi i h x_n}
\end{equation*}
is an example of a class of exponential sums known as \emph{Weyl sums},
reflecting the pioneering work of Weyl on providing estimations for them.
Vinogradov \cite{Vinogradov} improved on some of the estimations of Weyl.
Weyl and Vinogradov's results concern the modulus of the sum, $|S|$, and
characterise it as
\begin{equation*}
|S| \leq \gamma N\text{,}
\end{equation*}
where $N$ is the number of terms in the sum and $\gamma$ tends to zero as
$N\rightarrow\infty$. The subtle behaviour of $\gamma$ is linked to the
rational/irrational nature of the terms in the sequence.

I will use some basic results from the introductory chapter of
\cite{Vinogradov}. In general, write
\begin{equation*}
S = \sum_{n=1}^N \exp\left(2\pi i F(n) \right)
\end{equation*}
for some function $F(n)$. The application here is when
\begin{equation*}
F(n) = \beta n^j\text{.}
\end{equation*}
For $\beta$ rational (\emph{not} the case I will be interested
in) L. K. Hau proved that $|S|$ was of order
\begin{equation*}
N^{1-(1/j) + \epsilon}
\end{equation*}
(p.~3, \cite{Vinogradov}) and that this estimate could not be much
improved. Here, I am interested in the case where $\beta$ is irrational.
Estimations are much more difficult, and form the major aspect of the work
by Vinogradov. The estimations depend upon making a
rational approximation to $\beta$ and are complicated
functions of $N$ and $j$. Very loosely, he obtains results like
\begin{equation*}
|S| = O(N^{1-\rho'})
\end{equation*}
where
\begin{equation}
\label{eq:c_rho}
\rho' = \frac{1}{3(j-1)^2\log 12j(j-1)}\text{.}
\end{equation}
\shortpage
Vinogradov states
\begin{quote}
It is a plausible conjecture that the estimate in (\ref{eq:c_rho}) holds
with $\rho'$ replaced be $1/j - \epsilon$ \ldots\ A proof or disproof of
this conjecture would be very desirable.
\end{quote}
With the dual aim of extending Bourget's proof to the $p=2$ case (the rotor
considered by Milek and Seba) and ``simplifying'' Bourget's proof, I state
this conjecture formally.

\begin{conj}
\tcaption{\textbf{Conjecture:} For irrational $\beta$, the Weyl sum $S$
is bounded by the optimal rational $\beta$ bound}
\label{conj:c_weyl}
Consider the sum
\begin{equation*}
S = \sum_{n=1}^N \exp{2\pi i n^j\beta_j}\text{.}
\end{equation*}
For all $N$ greater than some critical value,
\begin{equation*}
|S| \leq cN^{1 - (1/j) + \epsilon}
\end{equation*}
for all $\epsilon > 0$ and some constant $c\in\mathbb{R}$.
\end{conj}
I do not attempt to prove Conjecture~\ref{conj:c_weyl}. Given the lengths
gone to by Vinogradov to obtain the results presented above, it seems
rather unlikely that a proof or disproof will be found any time
soon.\footnote{Incremental improvements on the estimations presented by
Vinogradov in \cite{Vinogradov} have been made over time. While Bourget
\cite{Bourget02.1} makes use of these improved results, the
conjecture itself remains unproven which is the only result of any
consequence in this discussion.}

\subsection{New results on discrepancy---upper and lower bounds}

Armed with the estimations on Weyl sums, I now proceed to derive both
upper and lower bounds on the discrepancy for sequences of the type
\begin{equation*}
\omega_j = (n^j\beta)
\end{equation*}
for $\beta$ of any type $\eta\geq 1$. It must be remembered that the upper
bound obtained is contingent upon Conjecture~\ref{conj:c_weyl}. The
lower bound obtained is not dependent upon any unproved conjectures. The
result obtained highlights the ``best possible'' nature of the conjectured
upper bound.

Firstly, (Lemma~3.2, p.~122, \cite{Kuipers}) is generalised to arbitrary
$j$.

\begin{lemma}
\tcaption{\textbf{Lemma:} The discrepancy $D_N(\omega_j)$ of the sequence
$\omega_j = (n^j\beta)$ is bounded by
$D_N(\omega_j) \leq C\left(\frac{1}{m} + N^{1-(1/j)+\epsilon}
  c'\sum_{h=1}^m \frac{1}{h\langle h\beta \rangle}\right)$}
\label{lemma:c_3.2}
The discrepancy $D_N(\omega_j)$ of $\omega_j = (n^j\beta)$ satisfies
\begin{equation*}
D_N(\omega_j) \leq C\left(\frac{1}{m} + N^{1-(1/j)+\epsilon}
  c'\sum_{h=1}^m \frac{1}{h\langle h\beta \rangle}\right)
\end{equation*}
for any positive integer $m$ and $\epsilon>0$, where $C$ and $c'$ are
absolute constants.
\end{lemma}

\emph{Proof.} (\ref{lemma:c_3.2})
Consider equation~\refeq{eq:c_gendisc}. It is
applicable to the first $N$ terms of the sequence $\omega_j$.
We have
\begin{equation}
\label{eq:c_242}
D_N(\omega_j) \leq C\left(\frac{1}{m} + \frac{1}{N}
  \sum_{h=1}^m \frac{1}{h}\left|
  \sum_{n=1}^N e^{2\pi i h n^j \beta} \right| \right)
\end{equation}
for any positive integer $m$. Consider the sum over $n$,
\begin{equation*}
\left| \sum_{n=1}^N e^{2\pi i h n^j \beta} \right|\text{.}
\end{equation*}
Conjecture~\ref{conj:c_weyl} allows this sum to be bounded by
\begin{equation*}
cN^{1-(1/j)+\epsilon}\text{.}
\end{equation*}
We are free to write
\begin{equation*}
c = \frac{c'}{|\sin \pi h \beta|}
\end{equation*}
as $\sin \pi h \beta$ is just some positive real number. Substituting
this result into \refeq{eq:c_242} gives
\begin{equation*}
D_N(\omega_j) \leq C\left(\frac{1}{m} + N^{-(1/j) + \epsilon}
c'\sum_{h=1}^m \frac{1}{h} \frac{1}{|\sin \pi h \beta|}\right)\text{.}
\end{equation*}
Now following the argument at the end of (Lemma~3.2, \cite{Kuipers})
the desired result is obtained.
\proofend

I now give the generalisation of (Theorem~3.2, \cite{Kuipers}). It
provides the ``best'' upper bound one could hope for when estimating
the discrepancy of the sequence $\omega_j = (n^j\beta)$. Again, remember
that the proof relies on Conjecture~\ref{conj:c_weyl}.

\begin{thrm}
\tcaption{\textbf{Theorem:} Assume Conjecture~\ref{conj:c_weyl} is true.
Then the optimal upper bound for the discrepancy of the sequence
$(n^j\beta)$ is given by
$D_N(\omega_j) = O\left(N^{-(1/\eta j) + \epsilon} \right)$}
\label{thrm:c_d_ub}
Assume Conjecture~\ref{conj:c_weyl} is true. Let $\beta$ be of finite
type $\eta$. Let $j$ be a positive integer $j\geq 1$. Then, for every
$\epsilon>0$, the discrepancy $D_N(\omega_j)$ of $\omega_j = (n^j\beta)$
satisfies
\begin{equation*}
D_N(\omega_j) = O\left(N^{-(1/\eta j) + \epsilon} \right)\text{.}
\end{equation*}
\end{thrm}

\emph{Proof.} (\ref{thrm:c_d_ub})
Let $\epsilon>0$ be fixed. By
(Lemma~3.1 and Lemma~3.3, p.~121--3, \cite{Kuipers}),
\begin{equation*}
\sum_{h=1}^m \frac{1}{h\langle h\beta \rangle}
  = O\left(m^{\eta-1+\epsilon'}\right)
\end{equation*}
for a fixed $\epsilon'>0$. Combining this with Lemma~\ref{lemma:c_3.2}
gives
\begin{equation*}
D_N(\omega_j) \leq C\left(\frac{1}{m}
  + N^{-(1/j)+\epsilon''}m^{\eta-1+\epsilon'}\right)
\end{equation*}
for all $m\geq 1$. Now choose $m=\left[N^{1/\eta j}\right]$. We obtain
\begin{equation*}
\begin{split}
D_N(\omega_j) &\leq C\left(N^{-(1/\eta j)}
  + N^{-(1/j) + \epsilon'' + (1/j) - (1/\eta j) + \epsilon'/\eta j}\right) \\
  &= O\left(N^{-(1/\eta j) + \epsilon}\right)
\end{split}
\end{equation*}
where $\epsilon = \epsilon'' + \epsilon'/\eta j$.
\proofend

Theorem~\ref{thrm:c_d_ub} is, in a sense, optimal. For functions $f,g$,
define $f = \Omega(g)$ if $f/g\nrightarrow 0$.
\begin{thrm}
\tcaption{\textbf{Theorem:} The optimal lower bound for discrepancy of the
sequence $(n^j\beta)$ is given by
$D_N(\omega_j) = \Omega\left(N^{-(1/\eta j) - \epsilon} \right)$}
\label{thrm:c_d_lb}
Let $\beta$ be of finite type $\eta$. Let $j$ be a positive integer
$j\geq 1$. Then, for every $\epsilon>0$, the discrepancy
$D_N(\omega_j)$ of $\omega_j = (n^j\beta)$ satisfies
\begin{equation*}
D_N(\omega_j) = \Omega\left(N^{-(1/\eta j) - \epsilon} \right)\text{.}
\end{equation*}
\end{thrm}

\emph{Proof.} (\ref{thrm:c_d_lb})
Let $\epsilon>0$ be fixed. For any given $\epsilon'>0$, there
exists $0<\delta<\eta$ with
$1/(\eta-\delta)=(1/\eta)+\epsilon'$. By
(Definition~3.4, p.~121, \cite{Kuipers}) we have
$\varliminf_{q\rightarrow\infty}
q^{\eta-(\delta/2)}\langle q\beta-j\rangle=0$ and thus
\begin{equation*}
\langle q\beta \rangle < q^{-\eta+(\delta/2)}
\end{equation*}
for an infinite number of positive integers $q$. There are infinitely
many positive integers $q$ and $p$ such that
\begin{equation*}
\left| \beta - p/q \right| < q^{-1-\eta + (\delta/2)}\text{.}
\end{equation*}
That is, by choosing $q$ large enough, we can always find a $p$ such that
$|q\beta - p| = \langle q\beta \rangle$. As $q$ increases $p/q$ is a
better approximation to the irrational $\beta$. For $\theta$ some
irrational with $|\theta|<1$, we have
\begin{equation*}
\beta = p/q + \theta q^{-1-\eta+(\delta/2)}\text{.}
\end{equation*}
Pick a $q$ such that the above relations are valid. Set
\begin{equation*}
N = \left[q^{j(\eta-\delta)}\right]\text{.}
\end{equation*}
Then for $1\leq n^j \leq N^{1/j}$,
\begin{equation*}
n^j \beta = n^j(p/q) + \theta_n\text{,}
\end{equation*}
with
\begin{equation*}
\begin{split}
\left|\theta_n\right| &=  \left|n^j\theta q^{-1-\eta+(\delta/2)}\right| \\
  &< n^j q^{-1-\eta+(\delta/2)} \\
  &\leq q^{[j(\eta-\delta)]^{1/j}-1-\eta+(\delta/2)} \\
  &=q^{-1-(\delta/2)}\text{.}
\end{split}
\end{equation*}
Thus, none of the fractional parts $\{\beta\}$, $\{2^j\beta\}$,
\ldots, $\{\left[N^{1/j}\right]\beta\}$ lie in the interval
$J=\left[q^{-1-(\delta/2)},q^{-1}-q^{-1-(\delta/2)}\right)$, so
\begin{equation*}
D_N(\omega_j) \geq \left|\frac{A(J,N)}{N}-\lambda(J)\right|
  = \lambda(J)\text{,}
\end{equation*}
where $\lambda(J)$ is simply the ``size'' of the set $J$.
For large enough $q$, we have $\lambda(J)\geq 1/2q$. But from the definition
of $N$ it is clear that
\begin{equation*}
N \leq q^{j(\eta-\delta)} \leq N + 1 \leq 2N\text{,}
\end{equation*}
so
\begin{equation*}
q^{-1} \geq cN^{-\left[j(\eta-\delta)\right]^{-1}}\text{.} 
\end{equation*}
Combining these inequalities, we obtain
\begin{equation*}
\begin{split}
D_N(\omega_j) &\geq c'N^{-\left[j(\eta-\delta)\right]^{-1}} \\
 &= c'N^{-(1/j)\left(1/(\eta-\delta)\right)} \\
 &= c'N^{-(1/j)\left((1/\eta)+\epsilon'\right)} \\
 &= c'N^{-(1/\eta j) - \epsilon}
\end{split}
\end{equation*}
where $\epsilon = \epsilon'/j$. That is, we have shown, for all
$\epsilon>0$, that
\begin{equation*}
D_N(\omega_j) = \Omega\left(N^{-(1/\eta j) -\epsilon}\right)\text{.}
\end{equation*}
\proofend

\subsection{Combescure's conjecture, Bourget's work and new results}

Before discussing the conjecture, we must clearly understand Combescure's
proof for the harmonic oscillator case. As stated in \refsec{sec:c_comb2},
the aim is to show that
\begin{equation*}
B^{-1}(x) = \sum_{n=0}^\infty |a_n|^2
  \left(\frac{2}{\sin\left(x-\theta_n\right)}\right)^2
  \rightarrow\infty\text{.}
\end{equation*}
Define the set $S(x)$,
\begin{equation}
\label{eq:c_setS}
S(x) = \{n:|x-\theta_n|\leq |a_n| = n^{-\gamma} 2\pi\}\text{.}
\end{equation}
Each $n$ is an element of $S(x)$ if $x$ is ``close to $\theta_n$''. Note
that $\theta_n = 2\pi \{ \alpha_n / 2\pi \hbar \}$, where $\{.\}$ is
the fractional part, not ``set'' and $\alpha_n$ are the eigenvalues of the
base Hamiltonian $H_0$.

Given that $\sin x\leq x$ for all $x\geq 0$, Combescure obtains a lower
bound for $B^{-1}(x)$,
\begin{equation}
\begin{split}
\label{eq:c_Bestimate}
  B^{-1}(x) &\geq \sum_{n=0}^\infty |a_n|^2
  \left(\frac{2}{x-\theta_n}\right)^2 \\
  &\geq \sum_{n\in S(x)} \frac{4|a_n|^2}{(x-\theta_n)^2}
  \geq 4 \#S(x)\text{.}
\end{split}
\end{equation}
Each $n\in S(x)$ gives a contribution to the sum of greater than one
as $|a_n| / |x-\theta_n | \geq 1$. By only considering $\#S(x)$, we simply
count a ``$1$'' each time.

The results on the discrepancy of sequences are now used, with
the sequence $\omega_\text{HO} = (\theta_n / 2\pi)$. Note that
each element of the sequence $\omega_\text{HO}$ is in $[0,1)$.

Consider the interval, defined for every $x\in (0,2\pi)$ and centred
around $x/2\pi$,
\begin{equation}
\label{eq:c_intervalJ}
J_N(x) = \left[ \frac{x}{2\pi} - N^{-\gamma},
  \frac{x}{2\pi} + N^{-\gamma} \right]\text{.}
\end{equation}
For large enough $N$, $J_N(x) \subset [0,1)$. The size of the interval is
$2N^{-\gamma}$. Using this particular subset and noting that the definition
of discrepancy \refeq{eq:c_discrepancy} involves taking the supremum over
all subsets of $[0,1)$, Combescure obtains
\begin{equation*}
\left| N^{-1}A(J_N(x),N) - 2N^{-\gamma} \right| \leq D_N(\omega_\text{HO})
\text{.}
\end{equation*}
Multiplying through by $N$ gives
\begin{equation}
\label{eq:c_inequality}
\left| A(J_N(x),N) - 2N^{1-\gamma} \right| \leq ND_N(\omega_\text{HO})
\text{.}
\end{equation}
As $|\psi\rangle \notin l_1(H_0)$
\begin{equation*}
\sum |a_n| \rightarrow\infty
\end{equation*}
and thus
\begin{equation*}
1/2 < \gamma \leq 1
\end{equation*}
from simple convergence arguments. Therefore, $N^{1-\gamma}$ grows at a
rate\footnote{Interestingly, it can in fact not grow
at all ($\gamma = 1$) which is a subtle point seemingly missed by 
Combescure and others. The rank-$1$ projection operator from the vector
$|\psi\rangle$ constructed with $\gamma=1$ is not shown to lead to the
emergence of a continuous spectrum. Therefore, the statement that
$|\psi\rangle \notin l_1(H_0)$ implies $|\psi\rangle \in \mathcal{H}_\cont$
is not in fact proved to be true. There are states not in $l_1(H_0)$
that may not be in the continuous spectrum. In practice (numerical,
experimental work) this should not cause any trouble. It is clearly easy to
avoid $\gamma=1$.} less than $N^{1/2}$.
At this stage, Combescure utilises the theorems discussed
above on the discrepancy of sequences. For the eigenvalue sequence,
$\alpha_n = n\hbar\omega$, of the harmonic oscillator\footnote{Do not
confuse $\omega$, the harmonic oscillator frequency, with
$\omega_\text{HO}$, the label for the sequence in $[0,1)$, the discrepancy
of which is being bounded.} the $j=1$ case of
Theorem~\ref{thrm:c_d_ub} applies which
is exactly (Theorem~3.2, \cite{Kuipers}). Combescure obtains the
result\footnote{This is not based on a conjecture as for $j=1$ a direct
proof is possible, bypassing Conjecture~\ref{conj:c_weyl}.
See \cite{Kuipers}.}
\begin{equation*}
D_N(\omega_\text{HO}) = O(N^{-1/\eta + \epsilon})\text{.}
\end{equation*}
For the sequence $\omega_\text{HO}$, $\beta = \omega/2\pi$. If $\beta$ is an
irrational of constant type ($\eta = 1$), the strongest type of irrational,
then
\begin{equation*}
ND_N(\omega_\text{HO}) = O(N^\epsilon)\text{.}
\end{equation*}
As the right-hand side of \refeq{eq:c_inequality} can be made to grow
arbitrarily
slowly, we conclude that the left-hand side must grow slowly too. Thus, to
cancel
the growth of $2N^{1-\gamma}$, $A(J_N(x),N)$ must grow at a
rate arbitrarily close to that of $2N^{1-\gamma}$. We see that
\begin{equation*}
A(J_N(x),N)\rightarrow\infty
\end{equation*}
as $N\rightarrow\infty$. It is now a simple observation \cite{Combescure90}
that this implies that $\#S(x)\rightarrow\infty$ and thus
$B^{-1}(x)\rightarrow\infty$. Thus, $e^{ix}$ is in the continuous spectral
subspace of the Floquet operator $V$.

The importance of the eigenvalue sequence is seen in that if we cannot limit
the right-hand side of \refeq{eq:c_inequality}, then we cannot place a
lower limit on
$A(J_N(x),N)$ and thus we cannot conclude that
$B^{-1}(x)\rightarrow\infty$. Two barriers to limiting the right-hand side
of this
equation exist---$j$ and $\eta$. If, still in the harmonic oscillator case,
we wished for $\beta = \omega/2\pi$ to only be of a weaker type, say
$\eta=2$, we would no longer be able to conclude that
$B^{-1}\rightarrow\infty$. The right-hand side would grow like
$N^{1/2 + \epsilon}$,
which is always faster than $2N^{1-\gamma}$ for $1/2<\gamma\leq 1$ which
grows at a rate of $N^{1/2 - \epsilon}$. Thus, no suitable lower limit for
$A(J_N(x),N)$ can be found. Similarly, if the eigenvalue sequence is
generalised (Combescure's conjecture) then we run into trouble. For $j=2$,
the lowest possible growth rate for the right-hand side we can obtain, taking
Conjecture~\ref{conj:c_weyl} as true, applying Theorem~\ref{thrm:c_d_ub}
and noting Theorem~\ref{thrm:c_d_lb} which says we cannot do any better,
is, once again, $N^{1/2 + \epsilon}$. For larger $j$, the situation only
gets worse.

Given these seemingly significant problems, the natural question to ask is:
``How does one get around this problem?''. The answer is provided in the
work of Bourget \cite{Bourget02.1}. Bourget proves a weaker theorem than
Combescure's conjecture. Where Combescure kept the same requirement on
$|\psi\rangle$, that it be in $l_1(H_0)$, Bourget has a $j$-dependent
requirement. Essentially, for increasing $j$ the $a_n$ terms used to
construct $|\psi\rangle$ must decrease more slowly with $n$. See Bourget's
work \cite{Bourget02.1} for the exact requirement, which depends on the
best estimates available for Weyl sums discussed earlier and thus is
a non-trivial function of $j$.

The key insight in obtaining the proof is to modify the set $S(x)$
(equation \refeq{eq:c_setS}) and the corresponding interval $J_N(x)$
(equation \refeq{eq:c_intervalJ}) that are considered. Importantly, they
become $j$-dependent. Bourget reduces the shrinking rate of the set
$J_N(x)$ as a function of $N$ just enough so as to allow the
weaker limits on the discrepancy to be good enough to force the right-hand
side of
the equivalent to \refeq{eq:c_inequality} to be less than the left-hand
side, while
keeping strong enough control on terms in the new set $S(x)$ to still argue
that $B^{-1}\rightarrow\infty$.

Using the best available estimations on Weyl sums and plugging these into
the upper bound formulas for discrepancy (as discussed earlier when
introducing the work by Vinogradov), Bourget manages to provide a rigorous
proof of the existence of a continuous spectral component of the Floquet
operator (the essence of Combescure's conjecture) for $j\geq 3$, leaving
only the $j=2$ case unresolved. The proof is, unfortunately,
unavoidably clouded by the ``messy'' estimates available for Weyl sums and
thus, the essence of the proof is difficult to see. Here, I will revisit
the proof, but (utilising Conjecture~\ref{conj:c_weyl}) apply
Theorem~\ref{thrm:c_d_ub} which says (using
$2\epsilon$, rather than $\epsilon$ for technical reasons), for all
$\epsilon> 0$
\begin{equation*}
D_N(\omega) = O\left(N^{-(1/\eta j) + 2\epsilon}\right)\text{.}
\end{equation*}
With this very clean estimate, it is far easier to see how Bourget's work
provides a proof that a continuous spectral component of the
Floquet operator exists. It also extends the result to $j=2$. Of course,
the $j=2$ case remains unproved as I have relied upon
Conjecture~\ref{conj:c_weyl}, but I highlight the fact that a solution to
Vinogradov's conjecture would solve Combescure's physics conjecture. I
have also simplified the $j$-dependence of the $a_n$s used to construct
$|\psi\rangle$.

\begin{thrm}
\tcaption{\textbf{Theorem:} Assume Conjecture~\ref{conj:c_weyl} is true.
Then for all $j\geq 1$, the Floquet operator $V$ has
$\sigma_\cont(V) \neq \emptyset$ if $1/2 < \gamma < 1/2 + 1/2 \eta j$}
\label{thrm:c_allj}
Assume Conjecture~\ref{conj:c_weyl} is true and thus
Theorem~\ref{thrm:c_d_ub} follows. Assume $\beta$ is irrational and of
type $\eta$. Then for all positive integers, $j$, the Floquet operator,
$V$, has $\sigma_\cont(V) \neq \emptyset$ if
$1/2 < \gamma < 1/2 + 1/2 \eta j$.
\end{thrm}

\emph{Proof.} (\ref{thrm:c_allj})
The proof relies upon the techniques utilised by Bourget.
In essence, we simply increases the size of the interval
(equation \refeq{eq:c_intervalJ}) from
$2N^{-\gamma}$ to $2N^{2(1/2-\gamma)}(\log N)^{-1/2}$. The important change 
is the first factor. The $\log N$ term is essential for
technical reasons, but has a negligibly small effect on the shrinkage rate
of the interval for large $N$. As $\log N / N^{4\delta} \rightarrow 0$ as
$N\rightarrow\infty$ for all $\delta>0$, for $N$ large enough we have
\begin{equation*}
2N^{2(1/2-\gamma)}(\log N)^{-1/2} > 2N^{2(1/2 - \gamma - \delta)}\text{.}
\end{equation*}
Using this underestimate for the size of the interval, we easily obtain
the equivalent of \refeq{eq:c_inequality},
\begin{equation*}
\left| A(J_N(x),N) - 2N^{2(1 - \gamma - \delta)} \right|
\leq ND_N(\omega_j)\text{,}
\end{equation*}
for the sequence $\omega_j = (n^j\beta)$. Now, using
Theorem~\ref{thrm:c_d_ub}, it is evident that to ensure
$A(J_N(x),N)\rightarrow\infty$, we must have
\begin{equation*}
2(1 - \gamma - \delta) > 1-(1/ \eta j) + 2\epsilon\text{,}
\end{equation*}
or
\begin{equation*}
\gamma < 1/2  + (1/2\eta j) - \epsilon - \delta \text{.}
\end{equation*}
The condition
\begin{equation}
\label{eq:c_gengamma}
1/2 < \gamma < 1/2 + (1/2\eta j)\text{,}
\end{equation}
where the ``$<$'' sign has absorbed the arbitrarily small numbers
$\epsilon$ and $\delta$, must be satisfied to force 
$A(J_N(x),N)\rightarrow\infty$. 

Finally, we must show that $B^{-1}(x)\rightarrow\infty$ when this
larger interval is used. Corresponding to the new interval $J_N(x)$, we
introduce the new set $S(x)$,
\begin{equation*}
S(x) = \left\{n:|x-\theta_n|
  \leq 2\pi N^{2(1/2-\gamma)}\log N^{-1/2}\right\}\text{.}
\end{equation*}
The estimate \refeq{eq:c_Bestimate} is the same, except with the new set
$S(x)$, which no longer has all terms greater than unity. Thus, it is not
enough to simply count the number of terms in $S(x)$. A more subtle
estimate is required. Replacing the numerator, $|a_n|$, with something
smaller, $N^{-\gamma}$, and the denominator, $(x-\theta_n)$, with something
larger, $2\pi N^{2(1/2-\gamma)}\log N ^{-1/2}$, we obtain
\begin{equation*}
B^{-1}(x) \geq \frac{1}{\pi^2} \sum_{n\in S(x)}
  \frac{\log N}{N^{2(1-\gamma)}}\text{,}
\end{equation*}
which is essentially the estimate Bourget obtains. The estimate contained
therein (Lemma~3.5 in \cite{Bourget02.1}) then shows that
$B^{-1}(x)\rightarrow\infty$ and the argument is complete.
\proofend

Examining \refeq{eq:c_gengamma} note that for $j=1$ (for $\eta=1$) we
recover the simple result of Combescure. For all $j\geq 2$, we have a stronger
($j$-dependent) condition on $|\psi\rangle$ than simply
$|\psi\rangle \notin l_1(H_0)$. This complication is the main weakening
of Combescure's conjecture that Bourget and I have been forced to make.
Note that the restriction on $\gamma$ takes into account the end point
subtleties referred to in the preceding discussions.

I have replaced the requirement that $|\psi\rangle\notin l_1(H_0)$
(i.e.,\ $1/2 < \gamma \leq 1$) with the $j$-dependent requirement
$1/2 < \gamma < 1/2 + (1/2j)$. In Bourget's work, the requirement is
stronger---directly related to the replacement of the known limits on Weyl
sums (in terms of $\rho$ in the earlier sections) with the ``best
possible'' estimate from our Conjecture~\ref{conj:c_weyl} of
$(1/j)-\epsilon$.

\subsection{Summary}

Reliance on Conjecture~\ref{conj:c_weyl} and the result of
Theorem~\ref{thrm:c_d_ub} derived from it has allowed me to discuss Bourget's
proof without the complications of the messy estimations on Weyl sums. This
simplified discussion highlights the key aspects of Bourget's proof. It has
also shown that the $j=2$ case for the emergence of a continuous spectral
component of the Floquet operator is solved by Vinogradov's conjecture. A
proof of Vinogradov's conjecture is no longer just of mathematical interest.
It has a direct mathematical physics consequence.

Finally, note that the rank-$N$ equivalent of this work follows in the
same way as presented for the harmonic oscillator case in
\refsec{sec:c_comb2}, providing a complete rank-$N$ generalisation of the
work of Combescure \cite{Combescure90}.


\section{\label{sec:c_milek}Generalising the results of Milek and Seba}

Having established that the continuous subspace of $\mathcal{H}$,
$\mathcal{H}_\cont$, is not empty, I now wish to characterise
it---by identifying the singular and absolutely continuous components. Here,
I extend the result of Milek and Seba to rank-$N$ perturbations. I do not
extend the numerical results of Milek and Seba as they rely on the
assumption that the $j=2$ eigenvalue spectra lead to a continuous Floquet
spectrum---a result I have just shown to be as yet unjustified.

\begin{thrm}
\tcaption{\textbf{Theorem:} The Floquet operator $V$ has
$\sigma_\ac(V) = \emptyset$. Thus, $\sigma_\cs(V)$ is not empty}
\label{thrm:c_ac_empty}
Assume $H(t)$ is given by \refeq{eq:c_hamiltonian} and that
\refeq{eq:c_measure} applies. Assume $B_k^{-1}(x)\rightarrow\infty$ and thus
$\mathcal{H}_\cont \neq \emptyset$. Then $\mathcal{H}_\ac = \emptyset$
and thus $\mathcal{H}_\cs$ is not-empty. The Floquet operator, $V$, has
a non-empty singular continuous spectrum.
\end{thrm}

\emph{Proof.} (\ref{thrm:c_ac_empty})
As shown in the proof of Theorem~\ref{thrm:r_4}\ref{thrm:r_4a} and easily
calculated, the Floquet operator can be written in the form
\begin{equation*}
	V = U+\sum_{k=1}^N R_k\text{,}
\end{equation*}
where
\begin{equation}
\label{eq:c_r_k}
R_k = \left(e^{i\lambda_k/\hbar} - 1\right)|\psi_k\rangle
  \langle\psi_k|U\text{.}
\end{equation}

We can now use either (Theorem~5, Howland \cite{Howland79}) or
(Theorem~1, Birman and Krein \cite{Birman}). The theorem from the paper
of Birman and Krein
is more direct, so we use it here. It states that if we have two unitary
operators, $U$ and $V$, that differ by a trace class operator, then the
wave operators
\begin{equation*}
\Omega_\pm = \text{s-}\underset{\nu\rightarrow\pm\infty}{\lim}
	V^\nu U^{-\nu} P_\ac(U)
\end{equation*}
exist and their range is the absolutely continuous subspace of $V$,
\begin{equation}
\label{eq:c_birman}
R(\Omega_\pm) = \mathcal{H}_\ac(V)\text{.}
\end{equation}
We must show that the difference $V-U$ is finite. With the notation in
\refchap{chap:rankN}, where the perturbation $W$ is given by
$A^*A$ and
\begin{equation*}
A = |\psi\rangle\langle\psi|\text{,}
\end{equation*}
with
\begin{equation*}
|\psi\rangle = \sum_n a_n \phi_n\text{,}
\end{equation*}
we obtain
\begin{equation*}
\begin{split}
\tr A^*A = \tr A &= \sum_l \langle\phi_l|A|\phi_l\rangle \\
  &= \sum_{l,m,n} \langle\phi_l | a_n | \phi_n\rangle
     \langle\phi_m | a_m^* | \phi_l\rangle \\
  &= \sum_{l,m,n} a_n a_m^* \delta_{ln}\delta_{ml} \\
  &= \sum_l |a_l|^2 = 1
\end{split}
\end{equation*}
as $|\psi\rangle\in l_2(H_0)$ and is normalised. The perturbation to the
Hamiltonian is trace class. The difference in unitary operators, $U$ and
$V$, is also trace class. By the triangle inequality for norms,
\begin{equation*}
\lpar R_k \rpar_{\text{tr}} \leq
\lpar \left(e^{i\lambda_k/\hbar}-1\right)\rpar\lpar|\psi_k\rangle
\langle\psi_k|\rpar_{\text{tr}} \lpar U \rpar_{\text{tr}}\text{.}
\end{equation*}
As $\lpar U \rpar_{\text{tr}}=1$,
\begin{equation*}
\begin{split}
\tr \left( \sum_{k=1}^N R_k \right)
  &\leq \sum_k \lpar(e^{i\lambda_k/\hbar}-1\rpar
    \sum_{l,m,n} \langle\phi_l | (a_k)_n | \phi_n \rangle
    \langle\phi_m | (a_k)_m^* | \phi_l \rangle \\
  &= \sum_k \left | e^{i\lambda_k\hbar}-1 \right| \\
  &= \sum_k \sqrt{2(1-\cos \lambda_k/\hbar)}\text{.}
\end{split}
\end{equation*}
Armed with a trace-class perturbation, we conclude that the wave operators
exist. The existence of the operators $\Omega_\pm$ means that they are
defined for all states in the Hilbert Space $\mathcal{H}$. Note (equation
\refeq{eq:c_birman}) that the subspace $\mathcal{H}_\ac(V)$ is
equal to the range of these operators. However, $P_\ac(U)$ gives
zero when acting on any state in $\mathcal{H}$ because $U$ is pure point.
Thus, $\mathcal{H}_\ac(V)$ is empty. As $\mathcal{H}_\cont$
is not empty, $\mathcal{H}_\cs$ must be non-empty, and we have
proved that a singular continuous subspace of the Floquet operator $V$
exists.\proofend

The key assumption in Theorem~\ref{thrm:c_ac_empty} is that
$B_k^{-1}(x)\rightarrow\infty$. This is certainly true for $j=1$ if
$|\psi_k\rangle \neq l_1(H_0)$. For $j\geq 2$ the results were discussed in
detail in \refsec{sec:c_genspec}. For $j\geq 3$, Bourget showed that
one can construct vectors $|\psi_k\rangle$ for which
$B_k^{-1}(x)\rightarrow\infty$. I have shown, in
Conjecture~\ref{thrm:c_allj}, that if Conjecture~\ref{conj:c_weyl} is
true then this result extends to $j\geq 2$ and with improved requirements
on the states $|\psi_k\rangle$.

\subsection{Discussion}

Milek and Seba make a number of incorrect statements in obtaining this
result for the rank-$1$ case. Firstly, they state that
the operator\footnote{As we are dealing with the rank-$1$ case, the
subscript $k$ may be dropped from \refeq{eq:c_r_k}.}
$R = \left[\exp\left(i\lambda/\hbar\right)-1\right]|\psi\rangle\langle\psi|U$
is rank-$1$ which it is not---the presence of the unitary
operator $U$ stops $R$ from being rank-$1$. This is not, however, important.
The applicability of the theorems in \cite{Howland79,Birman} does not rely
upon the rank of the operator $R$, but upon it being of trace-class.
Secondly, they claim that the existence of the wave operators implies that
\begin{equation}
\label{eq:c_ac_spec}
\sigma_\ac{V} \subset \sigma_\ac(U)\text{.}
\end{equation}
This is, again, not true. Given that $\sigma_\ac(U)$ is
empty, it is indeed possible to conclude that $\sigma_\ac(V)$ is
empty, as discussed above, but the relation \refeq{eq:c_ac_spec} does not
follow. Consider the situation where $\sigma_\cont(U)$ is not
empty. Then there is a set of vectors in $\mathcal{H}$ which are continuous
for $U$. These vectors form the domain for the operator $V^\nu$ in the wave
operators. The action with $V^\nu$ does not however keep us in the subspace
$\mathcal{H}_\cont(U)$ as the space we get to (the range for
$V^\nu$) is only invariant for $\mathcal{H}_\cont(V)$, not
$\mathcal{H}_\cont(U)$. Thus, we may obtain a vector, necessarily
in $\mathcal{H}_\cont(V)$ due to invariance, but possibly in
$\mathcal{H}_\sing(U)$, and thus, we cannot conclude that
$\sigma_\ac(V) \subset \sigma_\ac(U)$.
These two points discussed do not make the final results of Milek and Seba
wrong, but ``only'' the proofs.

Of greatest concern is the use of (Lemma~6, Combescure \cite{Combescure90})
without justification. Milek and Seba have assumed that Combescure's
conjecture is true. Bourget's demonstration that a continuous spectral
component of the Floquet operator exists does not cover the $j=2$ case
which is exactly the situation in Milek and Seba's paper. Furthermore, I
have shown that, using the ``best possible'' cases for discrepancy, the
$j=2$ case is covered, but, as I relied on Conjecture~\ref{conj:c_weyl},
I have not actually proved it.  It is worrying that Milek and Seba's work
remains unjustified.


\section{\label{sec:c_summary}Summary}

I have generalised the work of both Combescure \cite{Combescure90} and
Milek and Seba \cite{Milek90.1} from rank-$1$ to rank-$N$. I have also
discussed in detail Combescure's conjecture, my work on estimations of
discrepancy and the demonstration by Bourget \cite{Bourget02.1} that a
continuous spectral component of the Floquet operator does exist for
certain constructions of $|\psi\rangle$. This covers the essential aim of
Combescure's conjecture on the existence of a continuous spectral
component. A clear view of the essence of Bourget's proof has been
provided by taking a reasonable number-theoretic conjecture to be true.
With this clear view, the work of Bourget becomes more accessible. I
also demonstrated that reliance on Vinogradov's conjecture allows one to
extend the work to the $j=2$ case, showing that a proof of
Vinogradov's conjecture would have direct implications in mathematical
physics.

An in depth critical analysis of the work of Milek and Seba was also
undertaken; I highlighted a number of misconceptions and errors in the
work. I have also highlighted that, even with Bourget's demonstration of
the existence of a continuous spectral component of the Floquet operator
for $j\geq 3$, Milek and Seba's work \emph{still} is not fully justified.
Using Vinogradov's conjecture, one can then justify Milek and Seba's
result. A direct proof of Combescure's conjecture or a proof of
Vinogradov's conjecture, allowing my work to bridge the gap, remains
desirable.

\clearemptydoublepage
\chapter[Conclusions and further work]
{\label{chap:conc}Conclusions and Further Work}
\minitoc

In this thesis I have presented new results on the classification of the
spectrum of the Floquet operator for a class of kicked Hamiltonian systems.

Before presenting the main results, I reviewed the fields of classical and
quantum chaos (\refchap{chap:chaos}) and presented a detailed investigation
of the links between quantum dynamics and spectral analysis
(\refchap{chap:spectrum}). A point of ambiguity in the physics literature
was also identified and commented on.

In \refchap{chap:maths} I provided a conceptual introduction to the
mathematical field of spectral and functional analysis, highlighting the
physical meaning behind much of the basic mathematical building blocks. The
aim was not only to lay the groundwork for the material in
\refchap{chap:rankN}, but to make this important field of mathematics more
accessible to physicists. An appreciation of functional and spectral
analysis provides one with a deeper understanding of basic quantum
mechanics.

In \refchap{chap:rankN} I developed a number of unitary equivalent theorems
to those well known in the self-adjoint theory (e.g.,\ the Putnam--Kato
theorem) and successfully applied them to show that the spectrum of the
Floquet operator remains pure point (given that $U= e^{-iH_0 T}$ has pure
point spectrum) when the perturbation is suitably constrained. It should be
stressed that the result is non-perturbative. I also obtained this result
in a more straight forward, but less general, way by extending the work of
Combescure. This was presented in the early parts of \refchap{chap:comb}.

In the remaining sections of \refchap{chap:comb} I extended the results on
the emergence of a continuous spectrum of the Floquet operator to
rank-$N$ perturbations and investigated Combescure's conjecture that the
eigenvalue sequence for the unperturbed Hamiltonian, $H_0$, should not affect
the results in a significant way. Reviewing the work of Bourget and linking
it to a number-theoretic conjecture put forward by Vinogradov, I showed
that if one could solve Vinogradov's conjecture, then the essence of
Combescure's conjecture would follow. For any eigenvalue sequence for $H_0$
described by a polynomial, a continuous component will emerge under
realisable conditions for the rank-$N$ operator perturbation.

\section{Future directions}

The clearest loose end in the work presented in this thesis would be to
find a proof of Vinogradov's conjecture on the estimation of Weyl sums.
Without a proof, it seems reasonable to conclude that the work of Milek and
Seba on kicked rotors and the existence of a continuous component cannot be
properly justified. Having stood for over fifty years, a proof or disproof
seems unlikely to turn up any time soon.

When considering open questions of physical, rather than mathematical,
interest, a different direction is seen for extending this work.  It
is broadly acknowledged that moving beyond the closed system dynamics
governed by the Schr\"odinger equation is of great use in analysing chaos
in quantum systems and investigating the quantum--classical
correspondence.  Thus, it would be interesting to allow for environmental
interactions and to study the behaviour of kicked systems in so called
``open quantum systems''. There is already a great deal of research effort
in open systems and specifically, the analysis of kicked systems in such
environments. Many of the results look promising. Applying the techniques
developed in this work would be interesting.

Returning to the closed system dynamics, an extension of the work presented
in \refchap{chap:rankN} to infinite $N$ would be valuable. Similar work for
the Hamiltonian system was done in the later part of Howland's paper
\cite{Howland87}. As $N\rightarrow\infty$ a number of the tools and
arguments used for finite $N$ clearly collapse and one must be rather
cautious. The physical implications of infinite $N$ systems are also open
to investigation.

\clearemptydoublepage


\addcontentsline{toc}{chapter}{\numberline{}References}
\bibliographystyle{hplain}
\bibliography{bibliography}
\clearemptydoublepage

\nocite{p_Ames,p_Casati93,p_Cohen86,p_Garbaczewski}


\end{document}